\documentclass[11pt]{article}% добавляется twoside для двусторонней печати

\usepackage[T1]{fontenc}
\usepackage[cp1251]{inputenc}
\usepackage{textcomp}
\usepackage[centertags]{amsmath}
\usepackage{amsfonts}
\usepackage{amssymb}
\usepackage[pdftex]{hyperref}
\usepackage{graphicx}
\usepackage{graphbox}
\usepackage[numbers,sort&compress]{natbib}% упорядочивает ссылки

\usepackage{paperinitial}% Установка параметров страницы

\paperinitialization{15mm}{15mm}{15mm}{15mm}{2pt}{10pt}

\DeclareMathOperator{\re}{Re}
\DeclareMathOperator{\im}{Im}
\DeclareMathOperator{\sgn}{sgn}
\DeclareMathOperator{\Ai}{Ai}
\DeclareMathOperator{\Sif}{Si}

\newcommand{\lan}{\langle}
\newcommand{\ran}{\rangle}

\newcommand{\e}{\varepsilon}
\newcommand{\vf}{\varphi}

\newcommand{\s}{\sigma}

\newcommand{\al}{\alpha}
\newcommand{\be}{\beta}
\newcommand{\ga}{\gamma}
\newcommand{\Ga}{\Gamma}
\newcommand{\de}{\delta}
\newcommand{\De}{\Delta}

\newcommand{\la}{\lambda}

\newcommand{\ups}{\upsilon}

\newcommand{\spx}{\mathbf{x}}
\newcommand{\spy}{\mathbf{y}}

\newcommand{\spk}{\mathbf{k}}
\newcommand{\spe}{\mathbf{e}}
\newcommand{\R}{\mathbb{R}}

\begin{document}
\allowdisplaybreaks[4]% позволяет переносить многострочные формулы
\frenchspacing% уменьшение пробелов после запятых
\setlength{\unitlength}{1pt}% устанавливает единицу длины в окружении picture

\title{{\Large\textbf{Probability of radiation of twisted photons by classical currents}}}

\date{}

\author{O.V. Bogdanov${}^{1),2)}$\thanks{E-mail: \texttt{bov@tpu.ru}},\; P.O. Kazinski${}^{1)}$\thanks{E-mail: \texttt{kpo@phys.tsu.ru}},\; and G.Yu. Lazarenko${}^{1)}$\thanks{E-mail: \texttt{lazarenko.georgijj@icloud.com}}\\[0.5em]
{\normalsize ${}^{1)}$ Physics Faculty, Tomsk State University, Tomsk 634050, Russia}\\
{\normalsize ${}^{2)}$ Tomsk Polytechnic University, Tomsk 634050, Russia}}

\maketitle

\begin{abstract}

The general formula for the probability of radiation of a twisted photon by a classical current is derived. The general theory of generation of twisted photons by undulators is developed. It is proved that the probability to record a twisted photon produced by a classical current is equal to the average number of twisted photons in a given state. The general formula for the projection of the total angular momentum of twisted photons with given the energy, the longitudinal projection of momentum, and the helicity is obtained. The symmetry property of the average number of twisted photons produced by a charged particle moving along a planar trajectory is found. The explicit formulas for the average number of twisted photons generated by undulators both in the dipole and wiggler regimes are obtained. It is established that, for the forward radiation of an ideal right-handed helical undulator, the harmonic number $n$ of the twisted photon coincides with its projection of the total angular momentum $m$. As for the ideal left-handed helical undulator, we obtain that $m=-n$. It is found that the forward radiation of twisted photons by a planar undulator obeys the selection rule that $n+m$ is an even number. It turns out that the average number of twisted photons produced by the undulator and detected off the undulator axis is a periodic function of $m$ in a certain spectral band of the quantum numbers $m$.

\end{abstract}

\section{Introduction}

At the present moment, there are rather well developed techniques to produce and detect the vortex electromagnetic radiation (see, for review, \cite{PadgOAM25,AndBabAML,TorTorTw,AndrewsSLIA}). This type of radiation is loosely treated as the electromagnetic waves carrying an orbital angular momentum\footnote{The first theoretical studies \cite{Sadovs,Poynting} of the angular momentum of electromagnetic waves date back to the turn of the $20$th century.}. The rigorous quantum definition of such a radiation is coined as a twisted photon \cite{TorTorTw,MolTerTorTor,Ivanov11,JenSerprl,JenSerepj} and refers to the states of a free electromagnetic field with the definite energy, the longitudinal projection of momentum, the projection of the total angular momentum, and the helicity \cite{GottfYan,JaurHac,BiaBirBiaBir,JenSerprl,JenSerepj}. Similar states of electrons were also produced experimentally (see, for review, \cite{BliokhVErev,LBThY}). A number of quantum electrodynamic processes involving the twisted photons and electrons had been already described in the literature (see, e.g., \cite{Ivanov11,JenSerprl,JenSerepj}). However, to our knowledge, the general formula for the probability of radiation of twisted photons by classical currents was not presented. This formula is an analog of the well known expression for the spectral angular distribution of radiation of plane-wave photons \cite{LandLifshCTF.2,JacksonCE} in classical electrodynamics. The primary aim of this paper is to fill this gap and to provide the examples of how to use this formula.

The model of a classical source producing the twisted photons is a good approximation to reality when the quantum recoil experienced by the source can be neglected. For ultrarelativistic electrons moving in the external electromagnetic field, the latter limitation is rather weak. For example, for the laser radiation with the photon energies of the order $1$ eV and the intensity $I\approx10^{20}$ W/cm${}^2$, this restriction says, roughly, that the energies of electrons evolving in the laser radiation field must be less than $2.5$ GeV. Such electrons may produce the photons with the energies of the order $250$ MeV, and this radiation is still described by formulas of classical electrodynamics fairly well. Furthermore, the classical currents radiating the twisted photons were already used in theoretical investigations \cite{HMRR,HeMaRo,SasMcNu,TaHaKa,AfanMikh,BordKN}, and the efficiency of generation of twisted photons by such sources was confirmed experimentally \cite{HKDXMHR,BHKMSS}. Of course, there are other more traditional ways to produce the twisted photons in the optical range using various optical devices \cite{PadgOAM25,AndBabAML,TorTorTw,AndrewsSLIA,Liuperfvort,ABSW,BAVW,SSZZWX}, but from the theoretical point of view all these means can be reduced to the production of the twisted photons by classical currents in the formalism of electrodynamics of continuous media (see, e.g., \cite{LandLifshECM}). Not to overestimate this approximation, mention should be made that the classical currents generate photons in a coherent state (see Sec. \ref{Prob_Rad}). Therefore, this approximation cannot reproduce the non-trivial quantum correlations of photons (see, e.g., \cite{MVWZ,MolTerTorTor,MolTerZeil,fickler12}).

To date, the undulators or undulator-type devices (the free-electron lasers, for example) are the most investigated systems that generate the twisted photons and are based on free electrons \cite{HMRR,HeMaRo,SasMcNu,TaHaKa,AfanMikh,BordKN,HKDXMHR,BHKMSS}. Therefore, we apply the general formula for the probability of radiation of twisted photons to the analysis of the undulator radiation and develop a theory of radiation of twisted photons by undulators. We consider the undulator radiation both in the dipole and non-dipole (wiggler) approximations and give its complete description in terms of twisted photons. For a particular case of the forward radiation of a helical undulator, the general theory reproduces the results known in this case \cite{SasMcNu,TaHaKa}.

In particular, we establish that the $n$th harmonic of the forward radiation of an ideal right-handed helical undulator consists of the twisted photons with the projection of the total angular momentum $m=n$, irrespective of the photon helicity and longitudinal momentum (an ideal left-handed helical undulator produces the twisted photons with $m=-n$). This result is in agreement with the direct analysis of the Li\'{e}nard-Wiechert potentials \cite{SasMcNu,TaHaKa}, where it was found that such an undulator radiates the photons in the Laguerre-Gaussian modes with the orbital angular momentum. We derive the explicit expression for the average number of twisted photons produced in a given state. The forward radiation of the planar undulator can also be used to generate the twisted photons. It was found in \cite{SasMcNu} that this radiation is described by the Hermite-Gaussian modes without the orbital angular momentum, but these modes can be converted to the Laguerre-Gaussian modes with the aid of cylindrical lenses \cite{SasMcNu,ABSW,BAVW}. We show that, without any mode conversion or any special helical modulation of the electron bunch \cite{HeMaRo,HKDXMHR}, the average number of twisted photons with the fixed helicity is not symmetric under $m\rightarrow-m$, and the projection of the total angular momentum per one photon with fixed the energy, the longitudinal momentum, and the helicity can be as large as in the case of the helical undulator, at least, for small harmonic numbers. We also establish the property of the forward radiation of the planar undulator that $n+m$ must be an even number, otherwise the number of twisted photons produced is zero.

As for the radiation at an angle to the undulator axis, the spectacular result is that the average number of twisted photons is a periodic function of $m$ in a certain spectral band of the quantum numbers $m$. This is a robust result in the sense that this property holds for different types of undulators both in the dipole and wiggler regimes. We give the explicit formulas for the period of oscillations and the width of the spectral band. The signal of such a type can be employed for the high density information transfer (see Sec. \ref{Undulator_Dip}). We also analyze the spectrum of twisted photons and derive the formulas for the average number of radiated twisted photons. These formulas are obtained under the assumption that the number of the undulator sections $N$ is finite but large. So the analytical results become more and more accurate when $N$ is increased.

We start in Sec. \ref{Field_Operat} with the derivation of the mode functions of the electromagnetic field describing the twisted photons. Using these mode functions, we construct quantum electrodynamics with the twisted photons. The subject matter of this section is known in the literature (see, e.g., \cite{GottfYan,JaurHac,BiaBirBiaBir,JenSerprl,JenSerepj,TamVic,photonw,AndBabAMLch,AkhBerQED,Ivanov11}), and we include it into this article for the reader convenience in order to assign the notation and conventions. Sec. \ref{Prob_Rad} is devoted to the derivation of a general formula for the probability to detect a twisted photon radiated by a classical current. We show that this probability is, in fact, the average number of photons produced by the current. We also provide a pictorial representation of the general formula in terms of the usual plane-wave amplitudes of radiation of photons by a classical current. Sec. \ref{Prob_Rad} is concluded by the discussion of some properties of the wave packets composed of the twisted photons. In Sec. \ref{Angul_Mom}, we introduce the quantities that characterize the twist of the electromagnetic radiation and derive the general formulas for them. In particular, we establish the general symmetry property for the average number of twisted photons radiated by a charged particle moving along a planar trajectory. In Sec. \ref{Undulator}, the radiation of twisted photons by undulators is studied. Sec. \ref{Undulator_Dip} is devoted to the dipole case, while Sec. \ref{Wiggler} is for the the wiggler radiation. We obtain the average number of twisted photons produced by undulators in these cases and reveal some of its general properties. The useful formulas for the special functions appearing in the course of our study are collected in Appendix \ref{Bessel_Prop}.

We shall use the system of units such that $\hbar=c=1$ and $e^2=4\pi\al$, where $\al$ is the fine structure constant.

\section{Field operators}\label{Field_Operat}

Let us consider a quantum electromagnetic field interacting with a classical current in the Coulomb gauge. A thorough description of the quantization procedure in this gauge can be found, for example, in \cite{WeinbergB.12}. In the absence of source, the electromagnetic potential $A_i(t,\spx)$, $i=\overline{1,3}$, obeys the equations
\begin{equation}\label{Max_eqns}
    \ddot{A}_i-\Delta A_i=0,\qquad\partial_iA_i=0.
\end{equation}
In order to construct the field operators and quantum field theory, one needs to find the mode functions (a complete set of the solutions to \eqref{Max_eqns}) and partition them into positive- and negative-frequency modes.

To this aim it is useful to consider the eigenvalue problem for a self-adjoint Maxwell Hamiltonian operator (the curl operator)
\begin{equation}\label{Max_statio}
    h_M\psi_i(\spx):=\e_{ijk}\partial_j\psi_k(\spx)=sk_0\psi_i(\spx),\qquad k_0>0,\;s=\pm1.
\end{equation}
As we shall see, $k_0$ characterizes the energy of a state and $s$ is its helicity. It is assumed also that the complex vector fields $\psi_i$ obey the boundary conditions such that $k_0\neq0$. In this case, Eqs. \eqref{Max_statio} imply
\begin{equation}\label{div_free}
    \partial_i\psi_i=0.
\end{equation}
If one puts $\psi_i=E_i+iH_i$, then \eqref{Max_statio} is the system of free Maxwell equations Fourier-transformed with respect to time. The complete orthonormal set of eigenfunctions of the Maxwell Hamiltonian constitutes the basis in the Hilbert space of divergence-free complex vector fields $\psi_i(\spx)$ with the scalar product
\begin{equation}\label{scal_prod}
    \lan\phi,\psi\ran=\int d\spx\phi^*_i(\spx)\psi_i(\spx).
\end{equation}
It follows form \eqref{Max_statio}, \eqref{div_free} that
\begin{equation}\label{Laplace_eig}
    \De \psi=-k_{0}^2\psi,
\end{equation}
i.e., the general solution of \eqref{Max_eqns} can readily be found with the aid of the eigenfunctions \eqref{Max_statio}.

The Hamiltonian $h_M$ commutes with the operator of the total angular momentum (see, e.g., \cite{AkhBerQED})
\begin{equation}\label{ang_mom}
    J_{l ij}:=\e_{lmn}x_m k_n\de_{ij}-i\e_{l ij},\qquad k_n:=-i\partial_n.
\end{equation}
The index $l$ in $J_{l ij}$ marks the components of the angular momentum operator. The last term in \eqref{ang_mom} is the photon spin operator. The helicity operator,
\begin{equation}\label{helicity}
    S_{ij}=-ik_n\e_{n ij}/|\spk|=J_{l ij}k_l/|\spk|=(h_M)_{ij}/|\spk|,
\end{equation}
commutes with the Maxwell Hamiltonian and with $J_l$. As a result, we can construct the complete set of eigenfunctions of $h_M$ with definite values of the projection of the total angular momentum onto the $z$ axis, the helicity, and the projection of the momentum onto the $z$ axis:
\begin{equation}\label{compl_set}
    \hat{h}_{M}\psi_\al=sk_0\psi_\al,\qquad \hat{k}_3\psi_\al=k_3\psi_\al,\qquad \hat{J}_3\psi_\al=m\psi_\al,\qquad\hat{S}\psi_\al=s\psi_\al,
\end{equation}
where $\al\equiv(s,m,k_3,k_0)$, $m\in \mathbb{Z}$.

In solving system \eqref{compl_set}, it is convenient to introduce the basis spanned on the eigenvectors of the projection of the photon spin operator onto the $z$ axis:
\begin{equation}\label{spin_eigv}
    \mathbf{e}_\pm:=\mathbf{e}_1\pm i\mathbf{e}_2,\quad \mathbf{e}_3,
\end{equation}
where $\spe_i$ are the standard basis vectors. It is clear that
\begin{equation}
    (\spe_{\pm},\spe_{\pm})=0,\qquad (\spe_{\pm},\spe_{\mp})=2,\qquad\spe^*_\pm=\spe_\mp,
\end{equation}
and any vector can be decomposed in this basis as
\begin{equation}\label{psi_decomp}
    \psi=\frac12(\psi_-\spe_++\psi_+\spe_-)+\psi_3\spe_3.
\end{equation}
The scalar product \eqref{scal_prod} becomes in this basis:
\begin{equation}
    \lan\phi,\psi\ran=\int d\spx\big[\frac12(\phi^*_+\psi_+ +\phi^*_-\psi_-)+\phi^*_3\psi_3\big].
\end{equation}
Then the complete orthonormal set satisfying \eqref{compl_set} takes the form \eqref{psi_decomp} with \cite{GottfYan,JaurHac,BiaBirBiaBir,JenSerprl,JenSerepj}
\begin{equation}\label{mode_func_1}
    \psi_3(m,k_3,k_\perp)=\frac{1}{\sqrt{RL_z}}\frac{k_\perp^{3/2}}{2k_0}J_m(k_\perp r)e^{im\vf+ik_3z},\qquad \psi_\s(s,m,k_3,k_\perp)=\frac{i^\s k_\perp}{\s sk_0+k_3}\psi_3(m+\s,k_3,k_\perp),
\end{equation}
where $\s=\pm1$, $L_z$ is the size of a system along the $z$ axis, $R$ is the radius of a system counted from the $z$ axis, and $r:=\sqrt{x^2+y^2}$. To characterize the complete set of eigenfunctions, we use the quantum number $k_\perp:=\sqrt{k_0^2-k_3^2}\geq0$ instead of the energy $k_0$. It is supposed that $k_\perp R\gg1$ and $k_3L_z\gg1$. The completeness relation reads
\begin{equation}
\begin{gathered}
    \sum_\al\psi_{\al i}(\spx) \psi^*_{\al j}(\spy)=(\de_{ij}-\partial^x_i\partial^x_j\De^{-1})\de(\spx-\spy) =\de_{ij}\de(\spx-\spy) +\partial^x_i\partial^x_j \frac1{4\pi|\spx-\spy|}=:\de^\perp_{ij}(\spx-\spy),\\
    \sum_\al\equiv\sum_{s=\pm1}\sum_{m=-\infty}^\infty\int_{-\infty}^\infty\frac{L_z dk_3}{2\pi}\int_0^\infty\frac{Rdk_\perp}{\pi}.
\end{gathered}
\end{equation}
The following useful relations hold
\begin{equation}\label{compl_conj}
\begin{gathered}
    \psi_3^*(m,k_3,k_\perp)=(-1)^m\psi_3(-m,-k_3,k_\perp),\qquad \psi^*_\s(s,m,k_3,k_\perp)=(-1)^{m}\psi_{-\s}(s,-m,-k_3,k_\perp),\\
    \psi^*_i(s,m,k_3,k_\perp)=(-1)^m\psi_i(s,-m,-k_3,k_\perp),\qquad\psi_\s(s,m,k_3,k_\perp)=-\psi_{-\s}(-s,m+2\s,k_3,k_\perp).
\end{gathered}
\end{equation}
By construction, these mode functions are divergence-free. The mode functions \eqref{mode_func_1}, \eqref{mode_func_an} and their linear combinations corresponding to the same energy $k_0$ are the stationary solutions of \eqref{Max_statio}. Therefore, they do not spread with time (see, e.g., \cite{Durnin87,PampEnd,ChChChEf,MillEber}).

In quantum field theory, the mode functions of a boson field should be normalized by $(2k_0)^{-1/2}$. Furthermore, we have to write the mode functions in an arbitrary frame. Let us given the system of coordinates $\spx\equiv(x,y,z)$. Choose the unit vector $\spe_3$ that defines the projection direction of the angular momentum of a photon measured by the detector, and take the other two orthonormal vectors $\spe_{1,2}$ that are orthogonal to $\spe_3$ and constitute a right-handed system with it. Then we should substitute
\begin{equation}\label{replacmnt}
    z\rightarrow(\spe_3,\spx)=:x_3,\qquad r\rightarrow|x_+|=\sqrt{x_+x_-},\qquad \vf\rightarrow\arg x_+=\tfrac12\arg(x_+/x_-),
\end{equation}
where $x_\pm=(\spe_\pm,\spx)$ and, henceforward, $x_3$ is understood as \eqref{replacmnt}. In this case, the mode functions take the form \eqref{psi_decomp}, \eqref{mode_func_1} with
\begin{equation}\label{mode_func_an}
\begin{split}
    \psi_3(m,k_3,k_\perp)&=\frac{1}{\sqrt{RL_z}}\Big(\frac{k_\perp}{2k_0}\Big)^{3/2}J_m(k_\perp|x_+|)e^{im\arg x_+ +ik_3x_3}=\\
    &=\frac{1}{\sqrt{RL_z}}\Big(\frac{k_\perp}{2k_0}\Big)^{3/2}\frac{x_+^{m/2}}{x_-^{m/2}}J_m(k_\perp x_+^{1/2}x_-^{1/2})e^{ik_3x_3}=\\
    &=\frac{1}{\sqrt{RL_z}}\Big(\frac{k_\perp}{2k_0}\Big)^{3/2}j_m(k_\perp x_+, k_\perp x_-)e^{ik_3x_3},
\end{split}
\end{equation}
where we have introduced the shorthand notation for the Bessel functions (see Appendix \ref{Bessel_Prop}) and have assumed the principal branches of the multi-valued functions. The last representation in \eqref{mode_func_an} is more convenient for analytical calculations since, in this representation, the mode functions are entire analytic functions of the complex coordinates $\spx$ and the components $x_\pm$, $x_3$. The relations \eqref{compl_conj} are valid in an arbitrary frame. The quantum number $s$ defines the helicity of the detected photon, $m$ is the projection of the total angular momentum onto $\spe_3$, $k_3$ characterizes the projection of the photon momentum onto $\spe_3$, and $k_\perp$ is the absolute value of the momentum projection orthogonal to $\spe_3$.

Decomposing $A_i(t,\spx)$ in the complete set  $\psi_\al(\spx)$ with the coefficients depending on $t$ and taking into account \eqref{Laplace_eig}, we find the general solution to the wave equation \eqref{Max_eqns}. As a result, the self-adjoint field operator is written as
\begin{equation}\label{field_oper}
    \hat{A}(t,\spx)=\sum_\al \hat{c}_\al\psi_\al(\spx) e^{-ik_{0_\al}t}+ \sum_\al \hat{c}^\dag_\al\psi^*_\al(\spx) e^{ik_{0_\al}t},\qquad k_0=\sqrt{k_\perp^2+k_3^2},
\end{equation}
where $\hat{c}^\dag_\al$, $\hat{c}_\al$ are the creation-annihilation operators
\begin{equation}
    [\hat{c}_\al, \hat{c}^\dag_\be]=\de_{\al\be},
\end{equation}
and the standard partition of the field operator onto positive- and negative-frequency parts on a stationary background has been used (see, e.g., \cite{GFSh.3}). The canonical momentum operator is
\begin{equation}\label{field_oper_1}
\begin{gathered}
    \hat{E}(t,\spx)=-i\sum_\al k_{0\al}\hat{c}_\al\psi_\al(\spx) e^{-ik_{0_\al}t}+ i\sum_\al k_{0\al}\hat{c}^\dag_\al\psi^*_\al(\spx) e^{ik_{0_\al}t},\\
    [\hat{A}_i(t,\spx),\hat{E}_j(t,\spy)]=i\de^\perp_{ij}(\spx-\spy).
\end{gathered}
\end{equation}
Expressing the creation-annihilation operators from \eqref{field_oper} and \eqref{field_oper_1}, it is easy to obtain the secondary quantized operators corresponding to the one-particle operators \eqref{Max_statio}, \eqref{ang_mom}, \eqref{helicity} in terms of $\hat{A}$ and $\hat{E}$. For example,
\begin{equation}
    \hat{H}=\frac{1}{2}\int d\spx(\hat{E}\hat{E}+\hat{A}h^2_M\hat{A}),\qquad \hat{S}=\frac{1}{2}:\int d\spx(\hat{E}h_M^{-1}\hat{E}+\hat{A}h_M\hat{A}):.
\end{equation}
The helicity and total angular momentum operators commute with the Hamiltonian $\hat{H}$, and their averages do not depend on time for a free electromagnetic field. The classical limit or, which is the same, the average over the coherent state of the total angular momentum operator coincides with the corresponding expression following from the Noether theorem. The explicit expression for the classical limit of the helicity operator is given, for example, in \cite{DesTeit,AfanStep,BiaBirBiaBir,AndBabAMLch}.

Let us notice an interesting feature of the mode functions \eqref{mode_func_1}. If one considers the evolution of a localized wave packet composed of these mode functions, which is sufficiently narrow in $k_3$, $k_0$ spaces, then its group velocity along the $z$ axis equals $n_3=k_3/k_0<1$, i.e., it is less than the speed of light \cite{GRPFSBFP,ZZLLSS}. This fact is well known for the electromagnetic field modes in fibers (see, e.g., \cite{Hunsperger,MillEber}). Indeed, let the wave packet be
\begin{equation}\label{wave_pckt}
    \int dk_3dk_\perp e^{-i(k_0t-k_3z)}\vf_i(k_3,k_\perp;x,y),\qquad k_0=\sqrt{k_\perp^2+k_3^2},
\end{equation}
where $\vf_i(k_3,k_\perp;x,y)$ is a linear combination of the mode functions \eqref{mode_func_1} that is a slowly varying function of $k_3$. Its characteristic scale of variation should satisfy the estimate $\De k_3\gg 2\pi/L_z$, where $L_z$ is a distance from the source of radiation to the detector. For large $z$ and $t$, the WKB method applied to the integral over $k_3$ in \eqref{wave_pckt} gives
\begin{equation}
    z=n_3t,
\end{equation}
with the applicability conditions
\begin{equation}\label{nperp_cond}
    k_0T\gg1,\qquad n_\perp(k_0T)^{1/2}\gg n_3,
\end{equation}
where $T$ is the registration time of a photon and $n_\perp=k_\perp/k_0$. For $n_\perp\ll1$, these conditions reduce to
\begin{equation}
    n_\perp(k_0L_z)^{1/2}\gg 1.
\end{equation}
If the rest integral over $k_\perp$ is saturated on the modes with $n_\perp$ satisfying \eqref{nperp_cond}, then the group velocity of the wave packet along the $z$ axis is equal to $n_3$ provided $\De k_3$ complies with the estimate given above. Rather recently, this property of the Bessel beams was confirmed experimentally \cite{GRPFSBFP,ZZLLSS}.

\section{Probability of radiation of twisted photons}\label{Prob_Rad}

Let us consider now the theory of a quantum electromagnetic field interacting with a classical current $j_\mu(x)$. Then, the process of a one-photon radiation is possible
\begin{equation}\label{0g}
    0\rightarrow \ga.
\end{equation}
This process gives the leading in the fine-structure constant contribution to the probability of radiation of photons. We assume that the system is in the vacuum state $|0\ran$ at the initial moment $x^0=-T/2$, and the escaping photon with quantum numbers $(s,m,k_3,k_\perp)$ is detected at the instant $x^0=T/2$. The observation period $T$ is supposed to be very large, and we shall take the limit $T\rightarrow+\infty$ in the final answer. Then, keeping in mind that
\begin{equation}
    \hat{U}_{0,t}^0\hat{c}_\al \hat{U}_{t,0}^0=e^{-ik_{0\al} t} \hat{c}_\al,
\end{equation}
the transition amplitude of the process \eqref{0g} is written as
\begin{equation}
    e^{-iT(E_{vac}+k_{0_\al}/2)}\lan0|\hat{c}_\al\hat{S}_{T/2,-T/2}|0\ran,\qquad \hat{U}_{T/2,-T/2}=\hat{U}^0_{T/2,0}\hat{S}_{T/2,-T/2}\hat{U}^0_{0,-T/2},
\end{equation}
where $\hat{U}_{T/2,-T/2}$ is the evolution operator, $\hat{S}_{T/2,-T/2}$ is the $S$ matrix, and $\hat{U}^0_{T/2,0}$ is the free evolution operator that does not take into account the interaction with the classical current. In this expression, we also assume that $\hat{H}_0|0\ran=E_{vac}|0\ran$. In the first Born approximation, the transition amplitude of the process  \eqref{0g} becomes
\begin{equation}\label{vac_decay}
    -i e^{-iT(E_{vac}+k_{0\al}/2)}\lan0|\hat{c}_\al \int_{-T/2}^{T/2} dx\hat{A}_i(x) j^i(x)|0\ran,
\end{equation}
where the integration over $x^0$ is confined within the limits $[-T/2,T/2]$.

Substituting \eqref{field_oper} into \eqref{vac_decay}, we obtain the amplitude
\begin{equation}\label{amplitude}
    S(\al;0)=-ie^{-iT(E_{vac}+k_{0\al}/2)}\int_{-T/2}^{T/2} dxe^{ik_{0\al} x^0}\psi^*_{\al i}(\spx)j^i(x).
\end{equation}
For the theory of quantum electromagnetic fields to be self-consistent, the $4$-divergence of the current density must be identically zero. The current density of a point charge meeting this requirement has the form
\begin{equation}\label{current_class}
\begin{split}
    j^\mu(x)=\,&e\Big\{\int_{\tau_1}^{\tau_2}d\tau\dot{x}^\mu(\tau)\de^4(x-x(\tau)) +\frac{\dot{x}^\mu(\tau_2)}{\dot{x}^0(\tau_2)}\theta(x^0-x^0(\tau_2))\de\big[\spx-\spx(\tau_2)-(x^0-x^0(\tau_2))\dot{\spx}(\tau_2)/\dot{x}^0(\tau_2)\big]+\\
    &+\frac{\dot{x}^\mu(\tau_1)}{\dot{x}^0(\tau_1)}\theta(x^0(\tau_1)-x^0)\de\big[\spx-\spx(\tau_1)-(x^0-x^0(\tau_1))\dot{\spx}(\tau_1)/\dot{x}^0(\tau_1)\big] \Big\},
\end{split}
\end{equation}
where $x^0(\tau_1)=-\tau_0/2$ and $x^0(\tau_2)=\tau_0/2$, and $\tau_0$ is the time period when the particle moves with acceleration. The expression \eqref{current_class} can be cast into the standard form
\begin{equation}
    j^\mu(x)=e\int_{-\infty}^{\infty}d\tau\dot{x}^\mu(\tau)\de^4(x-x(\tau)),
\end{equation}
where it is assumed that, for $\tau<\tau_1$, the particle moves with the constant velocity $\dot{x}^\mu(\tau_1)$, while, for $\tau>\tau_2$, it moves with the constant velocity $\dot{x}^\mu(\tau_2)$. Having performed the Fourier transform,
\begin{equation}\label{current_fourier_gen}
    j^\mu(x)=:\int\frac{d^4k}{(2\pi)^4}e^{ik_\nu x^\nu}j^\mu(k),
\end{equation}
the last two terms in \eqref{current_class} correspond to the boundary terms in
\begin{equation}\label{current_fourier}
    j_\mu(k)=e\Big(\int_{\tau_1}^{\tau_2}d\tau \dot{x}_\mu e^{-ik_\nu x^\nu(\tau)}-\frac{i\dot{x}_\mu}{k_\la\dot{x}^\la}e^{-ik_\nu x^\nu}\Big|_{\tau_1}^{\tau_2}\Big).
\end{equation}
These boundary contributions are responsible for the radiation created by a particle when it enters to and exits from the external field (see, e.g., \cite{EdRadTh,SynchRad2015,Bord.1,Coisson,BKL,DiHeIld,AkhShul}).

It is useful to factor out the common multiple from the mode functions
\begin{equation}
    \psi_{\al i}(\spx)=:\frac{1}{\sqrt{RL_z}}\Big(\frac{k_\perp}{2k_{0\al}}\Big)^{3/2}e^{ik_3x_3}a_{\al i}(\spx).
\end{equation}
Then, substituting \eqref{current_class} into \eqref{amplitude}, squaring the absolute value of the outcome, and tending $T$ to infinity, we deduce that the probability of the process \eqref{0g} is given by
\begin{multline}\label{probabil}
    dP(s,m,k_3,k_\perp)=e^2\bigg|\int d\tau e^{-i[k_0x^0(\tau)-k_3x_3(\tau)]}\Big\{\frac12\big[\dot{x}_+(\tau)a_-(s,m,k_3,k_\perp;\spx(\tau))+\\
    +\dot{x}_-(\tau)a_+(s,m,k_3,k_\perp;\spx(\tau)) \big]
    +\dot{x}_3(\tau)a_3(m,k_\perp;\spx(\tau))\Big\} \bigg|^2 \Big(\frac{k_\perp}{2k_{0}}\Big)^{3}\frac{dk_3dk_\perp}{2\pi^2},\quad k_0:=\sqrt{k_3^2+k_\perp^2},
\end{multline}
where recall that $e^2=4\pi\al$, $\al\approx1/137$ is the fine-structure constant. As far as we know, this formula for the probability of radiation of twisted photons by a classical current has not been presented in the literature. This formula has the same status as the well-known expression for the spectral angular distribution of radiation created by a charged particle \cite{LandLifshCTF.2,JacksonCE}.

Some comments about this formula are in order. If the charged particle is moving uniformly and rectilinearly in the distant past and future, then the integrand of \eqref{probabil} behaves as
\begin{equation}
    \tau^{-1/2}e^{-i(k_0\ups_0-k_3\ups_3\pm k_\perp\sqrt{\ups_+\ups_-})\tau},\qquad |\tau|\rightarrow\infty,
\end{equation}
where $\ups^\mu$ is the asymptote of $\dot{x}^\mu(\tau)$ for large $\tau$'s. Hence, the integral \eqref{probabil} converges. As long as
\begin{equation}\label{positiveness}
    k_0\ups_0-k_3\ups_3\pm k_\perp\sqrt{\ups_+\ups_-}>0,
\end{equation}
one can speed up convergence of the integral by deforming the integration contour to the lower half-plane of the complex $\tau$ plane for such $\tau$'s where the motion of the particle becomes uniform and rectilinear. As for the complex conjugate integral entering into \eqref{probabil}, the integration contour ought to be deformed to the upper half-plane. For such a deformation to be justified, one needs to employ the last two representations of the mode functions in formula \eqref{mode_func_an}. In the infrared limit, when $k_0\rightarrow0$ and $s$, $m$, and $n_\perp$ are fixed, $dP$ tends to zero for $|m|\geq2$. In the ultraviolet limit, $k_0\rightarrow\infty$, the probability \eqref{probabil} also tends to zero, the asymptote being controlled by either the singular points of $x^\mu(\tau)$, or by the stationary points of
\begin{equation}
    k_0 x^0(\tau)-k_3 x_3(\tau)\pm k_\perp\sqrt{x_+(\tau)x_-(\tau)}
\end{equation}
in the $\tau$ plane, $\im\tau\leq0$, that are nearest to the real axis.

It is assumed in \eqref{probabil} that the Cartesian system of coordinates is chosen. Its origin lies on the ray emanating from the detector along $\spe_3$ (the projection direction of the angular momentum of a photon). For brevity, we call this ray as the $\spe_3$ axis. The specific choice of the reference point on the $\spe_3$ axis is inessential. Recall that the vectors $\spe_1$, $\spe_2$, and $\spe_3$ constitute a right-handed system, which, for example, can be taken to be the basis vectors $\spe_\vf$, $\spe_\theta$, and $\spe_r$ of the spherical system of coordinates. In a general case, $\spe_3\neq \mathbf{n}$, where $\mathbf{n}$ is a unit vector directed from the radiation point to the detector. The expression \eqref{probabil} is invariant under
\begin{equation}
    x_\pm \rightarrow e^{\pm i\vf}x_\pm,\qquad y_\pm \rightarrow e^{\pm i\vf}y_\pm,\qquad j_m\rightarrow e^{im\vf}j_m,\qquad\forall\vf\in\R.
\end{equation}
This property is a consequence of the symmetry of the expression \eqref{probabil} under rotations around the $\spe_3$ axis.

Formula \eqref{probabil} can be interpreted pictorially in terms of the usual plane-wave amplitudes of radiation of photons with the fixed helicity produced by the current $j^i$ (see Figs. \ref{plane_symm_plots}, \ref{cylind_plots}). In order to obtain the amplitude entering into \eqref{probabil}, one needs to rotate the trajectory of a charge around the $\spe_3$ axis and to add up the plane-wave contributions with fixed $s$, $k_3$, and $k_0$ coming from the rotating trajectory at the location of the detector with the ``weight''
\begin{equation}
     e^{im\vf}d\vf/(2\pi),
\end{equation}
where $\vf\in[0,2\pi)$ is the rotation angle. Such a picture is valid for distributed currents too. This interpretation allows one to predict the properties of \eqref{probabil} without making any detailed calculations (see Secs. \ref{Angul_Mom}, \ref{Undulator}). The mathematical proof of this interpretation comes from the integral representation of the Bessel functions (see also \eqref{Bessel_int1}),
\begin{equation}\label{Bessel_int}
    J_m(x)=i^{-m}\int_0^{2\pi}\frac{d\vf}{2\pi}e^{-im\vf+ix\cos\vf},\quad m\in \mathbb{Z},
\end{equation}
substituted into \eqref{probabil} instead of the Bessel functions, or from the way of derivation of the mode functions \eqref{mode_func_1} proposed in \cite{GottfYan,JenSerprl,JenSerepj}.

In the approximation scheme we consider, where the quantum electromagnetic field is supposed to interact with a classical current, the model is linear and can be solved exactly. In particular, one can find the exact amplitude of the process \eqref{0g}. For such a model, the following relation holds
\begin{equation}\label{coher_stt}
    \lan\be|\hat{c}_\al\hat{U}_{T/2,-T/2}|0\ran =\s(\al;0)\lan\be|\hat{U}_{T/2,-T/2}|0\ran,\qquad \s(\al;0):=S(\al;0)|_{E_{vac}=0},
\end{equation}
for any state $|\be\ran$ of the Fock space. Therefore, the exact expression for the probability of the process \eqref{0g}, when $T\rightarrow\infty$, is
\begin{equation}\label{probabil_exct}
    w(\al;0)=|\s(\al;0)|^2_{T\rightarrow\infty} \Big|\exp\Big[-\frac{i}{2}\int dxdy j^\mu(x) G_F(x-y) j_\mu(y)\Big]\Big|^2,
\end{equation}
where
\begin{equation}\label{Feyn_prop}
    G_F(x)=-\int\frac{d^4k}{(2\pi)^4}\frac{e^{-ik_\mu x^\mu}}{k^2+i0}.
\end{equation}
The last factor in \eqref{probabil_exct} describes the vacuum to vacuum transition probability in the model at hand. The simplest means of how to obtain formula \eqref{probabil_exct} is to integrate the Gaussian functional integral with respect to the fields $A_i$ and to employ the charge conservation law $\partial_\mu j^\mu=0$ in order to bring the answer to the explicitly Lorentz-invariant form. Introducing the Fourier-transform of the current density \eqref{current_fourier_gen} and using the Sokhotski formula in \eqref{Feyn_prop}, we deduce
\begin{equation}\label{probabil_exct1}
    w(\al;0)=|\s(\al;0)|^2_{T\rightarrow\infty}\exp\Big[\int\frac{d^4k}{16\pi^3}\de(k^2)j^\mu(k)j^*_\mu(k)\Big].
\end{equation}
The first factor in this formula is \eqref{probabil}. The expression standing in the exponent in the second factor is negative and, with a reversed sign, equals the total number of photons created by the current $j_\mu(x)$ during the whole observation period (see, e.g., \cite{LandLifshCTF.2}). It follows from \eqref{current_fourier} that, for $k_0\rightarrow0$ and $\dot{x}^\mu(\tau_1)\neq\dot{x}^\mu(\tau_2)$,
\begin{equation}
    j^\mu(k)j^*_\mu(k)\sim k_0^{-2}.
\end{equation}
Taking into account that $\de(k^2)$ removes one integration in \eqref{probabil_exct1} and $\de(k^2)\sim k_0^{-1}$, we see that the integral in \eqref{probabil_exct1} diverges logarithmically at the small photon energies. This is the standard infrared divergence of quantum electrodynamics. One can get rid of it by introducing the infrared energy cutoff \cite{WeinbergB.12}.

When $\dot{x}^\mu(\tau_1)=\dot{x}^\mu(\tau_2)$, the infrared divergence does not arise. However, in any case, the probability of the process \eqref{0g} is extremely small. From the physical point of view, it is more appropriate to consider the average number of photons in the state $\al$ created by the current $j_\mu(x)$ during the whole observation period
\begin{equation}\label{average_nph}
    n(\al;0)=\sum_{\be}\lan 0|\hat{U}_{-\infty,\infty}\hat{c}^\dag_\al|\be\ran \lan\be|\hat{c}_\al\hat{U}_{\infty,-\infty}|0\ran=\lan 0|\hat{U}_{-\infty,\infty}\hat{c}^\dag_\al\hat{c}_\al\hat{U}_{\infty,-\infty}|0\ran.
\end{equation}
Using \eqref{coher_stt}, we obtain
\begin{equation}
    n(\al;0)=|\s(\al;0)|^2_{T\rightarrow\infty},
\end{equation}
i.e., formula \eqref{probabil} describes the average number of photons with the given quantum numbers. The probability of the inclusive process
\begin{equation}\label{0gX}
    0\rightarrow\ga+X,
\end{equation}
where $X$ are the photons that are not recorded by the detector, is given by
\begin{equation}
    w_{incl}(\al;0)=\lan 0|\hat{U}_{-\infty,\infty}(1-:e^{-\hat{c}^\dag_\al \hat{c}_\al}:)\hat{U}_{\infty,-\infty}|0\ran=1-e^{-n(\al;0)}.
\end{equation}
Recall that $:\exp(-\hat{c}^\dag_\al \hat{c}_\al):$ is the projector to the state without the photons characterized by the quantum number $\al$.

Notice also that the relation \eqref{coher_stt} implies that the state
\begin{equation}
    \hat{U}_{T/2,-T/2}|0\ran
\end{equation}
is an eigenvector of the annihilation operators of photons and, consequently, is a coherent state (see, e.g., \cite{GottfYan,Glaub2,KlauSud}). In this state, the non-trivial quantum correlations (the entanglement) are absent, viz., the logarithm of the generating functional of the correlation functions \cite{KlauSud,Glaub1} is linear in the sources. Of course, this is a consequence of the approximation we made when the current operator $\hat{j}^i$ had been replaced by the classical quantity $j^i$. Formula \eqref{probabil} can be improved by taking into account the matrix structure of the current and the effects of quantum recoil caused by the radiation of a photon (see, e.g., \cite{Bord.1,BaKaStrbook,AkhShul,BBTq1,BBTq2}), the notion of a classical trajectory still being applicable in this case. It is also easy to generalize this formula to the case of a source possessing the higher multipole moments (see, e.g., \cite{TeitVilWee,GrooSutt,rrmm,JacksonCE}). Formula \eqref{probabil} for the average number of photons created in a given state is obviously written for any other complete set of the photon mode functions.

Concluding this section, we give comments on how to take into account in \eqref{probabil} the finite sizes of the detector recording a photon. The states \eqref{mode_func_1} are not localized in space, and any detector with a finite spatial extension is not able to detect the photon in this state. Let $f_\al$ be the form-factor of the wave function of a photon recorded by the detector, viz., the detector detects the photon in the state
\begin{equation}
    \sum_\al f_\al\hat{c}_\al^\dag|0\ran,\qquad\sum_\al f^*_\al f_\al=1.
\end{equation}
As for the functions $f_\al$, they can be taken, for example, in the form of wave packets with a Gaussian envelope studied in \cite{KarlJHEP} at length. Then the transition amplitude \eqref{amplitude} takes the form
\begin{equation}
    \sum_\al f^*_\al S(\al;0).
\end{equation}
On squaring the module of this expression, the dependence on $T$ does not disappear provided $f_\al$ correspond to the modes with different energies. We can formally remove this dependence by introducing
\begin{equation}\label{wp_disp}
    f_\al=:\tilde{f}_\al e^{-ik_{0\al}T/2},
\end{equation}
and interpreting \eqref{probabil} as the probability of the photon production in the state
\begin{equation}
    \sum_\al\tilde{f}_\al\hat{c}_\al^\dag|0\ran
\end{equation}
at the instant of time $t=0$, the photon propagating freely to the detector after its creation. The instant $t=0$ corresponds to the moment in time when the reaction is passing, i.e., in our case, to that instant of time when the acceleration of a charged particle is being different from zero. It is assumed that the acceleration does not vanish when $t\in[-\tau_0/2,\tau_0/2]$, and $\tau_0\ll T$. Obviously, these considerations are valid for the average number of photons \eqref{average_nph} as well.

\section{Angular momentum}\label{Angul_Mom}

The properties of radiation related to its twist can be characterized by the differential asymmetry and by the projection of the total angular momentum per one photon with given the energy, the momentum projection $k_3$, and the helicity:
\begin{equation}
    A(s,m,k_3,k_0):=\frac{dP(s,m,k_3,k_0)-dP(s,-m,k_3,k_0)}{dP(s,m,k_3,k_0)+dP(s,-m,k_3,k_0)},\qquad \ell(s,k_3,k_0):=\frac{dJ_3(s,k_3,k_0)}{dP(s,k_3,k_0)},
\end{equation}
where
\begin{equation}\label{ang_mom_expl}
    dJ_3(s,k_3,k_0):=\sum_{m=-\infty}^\infty mdP(s,m,k_3,k_0),\qquad dP(s,k_3,k_0)=\sum_{m=-\infty}^\infty dP(s,m,k_3,k_0).
\end{equation}
The integral characteristic,
\begin{equation}\label{ell}
\begin{split}
    \ell(s,k_0):=\frac{dJ_3(s,k_0)}{dP(s,k_0)},\qquad dJ_3(s,k_0)=\int_0^{k_0} dk_3 \frac{dJ_3(s,k_3,k_0)}{dk_3},\quad dP(s,k_0)=\int_0^{k_0} dk_3 \frac{dP(s,k_3,k_0)}{dk_3},
\end{split}
\end{equation}
is also of interest. It specifies the total angular momentum of radiation projected to the $\spe_3$ axis recorded by the detector per one radiated photon with given the energy and the helicity. Notice that the projection onto the $\spe_3$ axis of the total angular momentum of \emph{all} radiated photons with given $k_0$ and $s$ is obtained by integrating $dJ_3(s,k_3,k_0)/dk_3$ with respect to $k_3$ over the symmetric interval $[-k_0,k_0]$ (cf. \eqref{ell}).

One can expect by the symmetry reasons that, in the case when the trajectory of a charged particle is planar and the detector lies in the orbit plane and projects the angular momentum onto the axis also lying in the orbit plane, the distribution of the detected photons over the angular momentum projection should be symmetric with respect to the sign change of the angular momentum. More precisely, the following relation holds for this configuration:
\begin{equation}\label{plane_symm}
    dP(s,m,k_3,k_0)=dP(-s,-m,k_3,k_0).
\end{equation}
Indeed, for the configuration considered, in virtue of the symmetry of the expression \eqref{probabil} with respect to the rotations around the $\spe_3$ axis, one can always bring the trajectory to the position that $x_+=x_-$ for all the points of the trajectory. After that, one performs the rotation by the angle $\pi$ around the $\spe_3$ axis. Then the equality $x_+=x_-$ does not change for all the points of the trajectory, $dP$ remains the same, and
\begin{equation}
    \dot{x}_\pm\rightarrow-\dot{x}_\pm,\qquad j_m\rightarrow j_{-m},\qquad j_{m-1}\rightarrow j_{-m+1},\qquad j_{m+1}\rightarrow j_{-m-1}.
\end{equation}
Since
\begin{equation}
    \frac{ik_\perp}{sk_0-k_3}=-\frac{ik_\perp}{-sk_0+k_3},
\end{equation}
we deduce \eqref{plane_symm} from \eqref{probabil}.

\begin{figure}[!t]
\centering
i)\;\includegraphics*[align=c,width=0.35\linewidth]{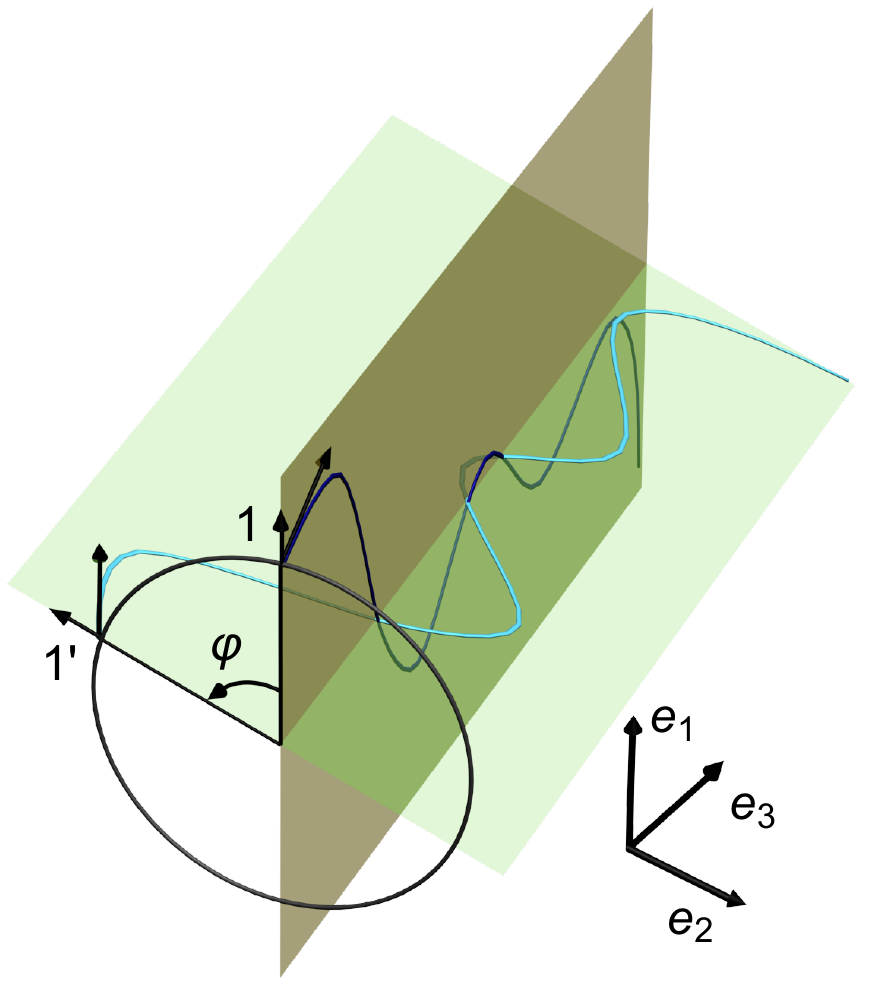}\qquad\quad
ii)\;\includegraphics*[align=c,width=0.35\linewidth]{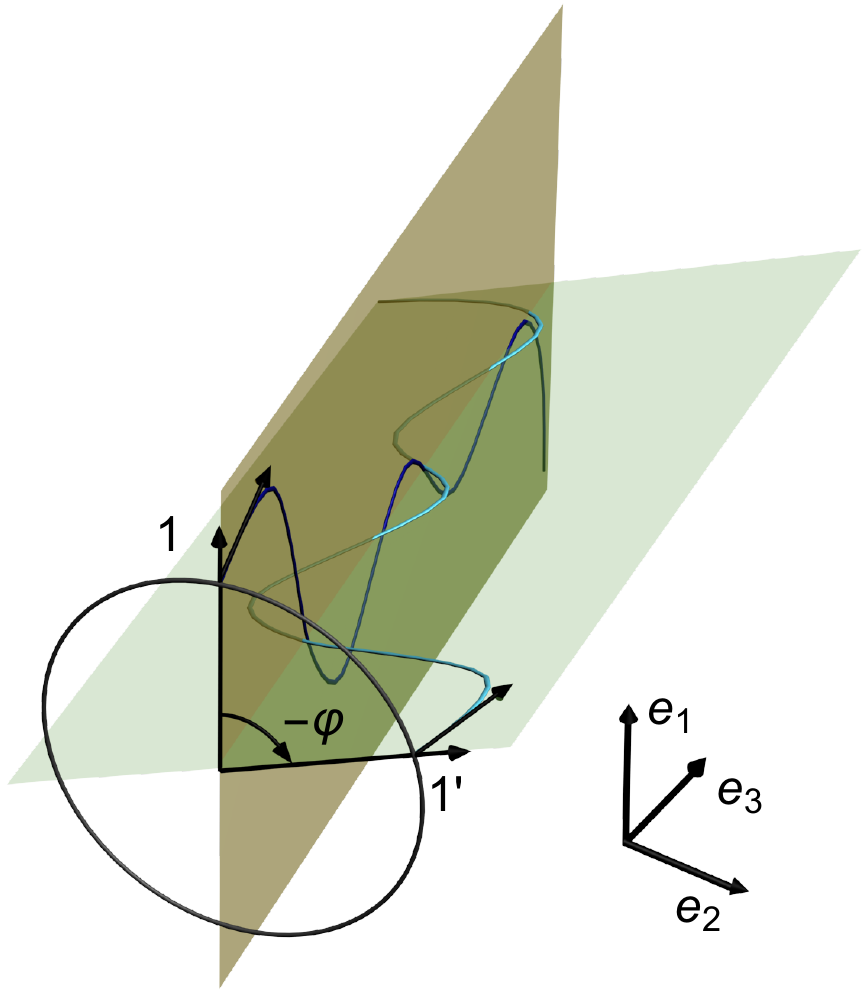}
\caption{{\footnotesize On the left panel: The pictorial representation of the transition amplitude \eqref{probabil} corresponding to the family (i) of the trajectories (see the main text) with the prescribed additional phase factor $e^{im\vf}$. The initial blue trajectory is rotated around $\spe_3$ by an angle of $\vf$ so that the point $1$ is shifted to $1'$ and the blue trajectory goes to the azure one. The vectors tangent to the trajectories at the points $1$ and $1'$ are the velocity vectors. The vectors lying in the plane normal to the rotation axis are the projections of the velocity vectors onto this plane. On the right panel: The pictorial representation of the transition amplitude \eqref{probabil} corresponding to the family (ii) of the trajectories with the prescribed additional phase factor $e^{-im\vf}$. The initial blue trajectory is rotated around $\spe_3$ by an angle of $-\vf$ so that the point $1$ is shifted to $1'$ and the blue trajectory goes to the azure one. The vectors tangent to the trajectories at the points $1$ and $1'$ are the velocity vectors. The vectors lying in the plane normal to the rotation axis are the projections of the velocity vectors onto this plane.}}
\label{plane_symm_plots}
\end{figure}

The relation \eqref{plane_symm} for the configuration at issue can also be proved with the help of the pictorial representation of \eqref{probabil} in terms of the radiation of the plane-wave photons from a family of trajectories (see Sec. \ref{Prob_Rad} and Fig. \ref{plane_symm_plots}). Indeed, let us rotate the trajectory around the $\spe_3$ axis and bring the trajectory to the plane with the basis vectors $\spe_1$, $\spe_3$. Then we consider the contributions of the families of the trajectories (i) and (ii) corresponding to $m$ and $-m$, respectively. For an arbitrary trajectory from the family (i) with the rotation angle $\vf$ around the aforementioned axis there exists the trajectory from the family (ii) with the rotation angle $-\vf$ and the same phase factor $e^{im\vf}$. The contributions of these trajectories to the transition amplitude differs only by the sign of the $y$ component of the current vector (see Fig. \ref{plane_symm_plots}). The sign change of $s$ corresponds to the sign change of the basis vector $\spe_2$ (see \eqref{spin_eigv}) or, which is the same, to the sign change of the $y$ component of the current vector. Consequently, we deduce the property \eqref{plane_symm}.

It follows from the symmetry relation \eqref{plane_symm} that, in the case considered, the average helicity and the average projection of the total angular momentum of the electromagnetic field onto $\spe_3$ are zero. Indeed,
\begin{equation}
\begin{split}
    dJ_3(k_3,k_0)&=\sum_{s,m}m dP(s,m,k_3,k_0)=-\sum_{s,m}m dP(s,m,k_3,k_0)=0,\\
    dS(k_3,k_0)&=\sum_{s,m}s dP(s,m,k_3,k_0)=-\sum_{s,m}s dP(s,m,k_3,k_0)=0.
\end{split}
\end{equation}
This property is in agreement with the estimates following from the classical formulas \cite{TeitVilWee,IvSokCFT,BordKN} for the radiation of the angular and spin momenta by a point charge moving along a planar trajectory.

Now let us obtain the general formulas for the quantities  \eqref{ang_mom_expl}. The sums over $m$ in \eqref{ang_mom_expl} can be found explicitly with the help of the addition theorem \eqref{add_thm}. These formulas are useful not only for the analytical calculations but for the numerical simulations as well. In the case when the distribution over $m$ in \eqref{probabil} is wide, the immediate summation over $m$ in \eqref{probabil} can be highly time-consuming. Let us introduce the notation
\begin{equation}
    \De_\pm:=x_\pm-y_\pm,\qquad\De_{0,3}:=x_{0,3}-y_{0,3},
\end{equation}
where $x_\pm:=x_\pm(\tau)$, $x_{0,3}:=x_{0,3}(\tau)$, $y_\pm:=x_\pm(\s)$, and $y_{0,3}:=x_{0,3}(\s)$. Then
\begin{multline}\label{phot_prob_s}
    dP(s,k_3,k_0)=e^2\int d\tau d\s e^{-i(k_0\De_0-k_3\De_3)}\Big[\Big(\dot{x}_3\dot{y}_3 +\frac{\dot{x}_+\dot{y}_-}{4}\frac{k_\perp^2}{(sk_0-k_3)^2} +\frac{\dot{x}_-\dot{y}_+}{4}\frac{k_\perp^2}{(sk_0+k_3)^2}\Big)j_0 +\\
    + \Big(\frac{\dot{x}_+\dot{y}_3}{sk_0-k_3} -\frac{\dot{x}_3\dot{y}_+}{sk_0+k_3}\Big)\frac{ik_\perp}{2}j_{-1} + \Big(\frac{\dot{x}_-\dot{y}_3}{sk_0+k_3} -\frac{\dot{x}_3\dot{y}_-}{sk_0-k_3}\Big)\frac{ik_\perp}{2}j_{1} +\frac{\dot{x}_+\dot{y}_+}{4}j_{-2} +\frac{\dot{x}_-\dot{y}_-}{4}j_{2}\Big] \Big(\frac{k_\perp}{2k_{0}}\Big)^{3}\frac{dk_3dk_\perp}{2\pi^2},
\end{multline}
where $j_m\equiv j_m(k_\perp\De_+,k_\perp\De_-)$. It is clear that the expression obtained does not depend on the choice of the origin of the system of coordinates. We can single out explicitly the dependence on the photon polarization in \eqref{phot_prob_s}:
\begin{multline}
    dP(s,k_3,k_0)=\frac12dP(k_3,k_0)+
    se^2\int d\tau d\s e^{-i(k_0\De_0-k_3\De_3)}\times\\
    \times\Big[(\dot{x}_+\dot{y}_--\dot{x}_-\dot{y}_+)\frac{k_3}{k_\perp}j_0 +i(\dot{x}_+\dot{y}_3-\dot{x}_3\dot{y}_+)j_{-1} +i(\dot{x}_-\dot{y}_3-\dot{x}_3\dot{y}_-)j_{1} \Big]\Big(\frac{k_\perp}{2k_{0}}\Big)^{2}\frac{dk_3dk_\perp}{8\pi^2},
\end{multline}
where
\begin{multline}%\label{prob_k3_k0}
    dP(k_3,k_0)=\sum_{s=\pm1}dP(s,k_3,k_0)=e^2\int d\tau d\s e^{-i(k_0\De_0-k_3\De_3)}
    \Big[\Big(2\dot{x}_3\dot{y}_3 +(\dot{x}_+\dot{y}_-+\dot{x}_-\dot{y}_+)\frac{k_0^2+k_3^2}{2k_\perp^2}\Big)j_0+\\
    +(\dot{x}_+\dot{y}_3+\dot{x}_3\dot{y}_+)\frac{ik_3}{k_\perp}j_{-1} -(\dot{x}_-\dot{y}_3+\dot{x}_3\dot{y}_-)\frac{ik_3}{k_\perp}j_{1} +\frac{\dot{x}_+\dot{y}_+}{2}j_{-2} +\frac{\dot{x}_-\dot{y}_-}{2}j_{2}\Big]\Big(\frac{k_\perp}{2k_{0}}\Big)^{3}\frac{dk_3dk_\perp}{2\pi^2}.
\end{multline}

Using the addition theorem \eqref{add_thm} and the recurrence relations \eqref{recurr_rels}, we derive a similar expression for the angular momentum projection
\begin{multline}%\label{ang_mom_s_k3_k0}
    dJ_3(s,k_3,k_0)=e^2\int d\tau d\s e^{-i(k_0\De_0-k_3\De_3)}%\times\\
    \Big[\Big(\dot{x}_3\dot{y}_3 +\frac{k_\perp^2}{4}\Big(\frac{\dot{x}_+\dot{y}_-}{(sk_0-k_3)^2} +\frac{\dot{x}_-\dot{y}_+}{(sk_0+k_3)^2}\Big)\Big)\frac{k_\perp}{2}(c_+j_{-1}+c_-j_1) +\\
    +\Big(\frac{\dot{x}_+\dot{y}_3}{sk_0-k_3} -\frac{\dot{x}_3\dot{y}_+}{sk_0+k_3}\Big)\frac{ik_\perp^2}{4}(c_-j_0+c_+j_{-2}) + \Big(\frac{\dot{x}_-\dot{y}_3}{sk_0+k_3} -\frac{\dot{x}_3\dot{y}_-}{sk_0-k_3}\Big)\frac{ik_\perp^2}{4}(c_+j_0+c_-j_2)+\\
    +\frac{\dot{x}_+\dot{y}_+}{8}k_\perp (c_+j_{-3}+c_-j_{-1}) +\frac{\dot{x}_-\dot{y}_-}{8}k_\perp(c_+j_{1}+c_-j_3) +\Big(\frac{\dot{x}_+\dot{y}_3}{sk_0-k_3}+\frac{\dot{x}_3\dot{y}_+}{sk_0+k_3}\Big)\frac{ik_\perp}{4}j_{-1}-\\ -\Big(\frac{\dot{x}_-\dot{y}_3}{sk_0+k_3}+\frac{\dot{x}_3\dot{y}_-}{sk_0-k_3}\Big)\frac{ik_\perp}{4}j_{1} +\Big(\frac{\dot{x}_+\dot{y}_-}{(sk_0-k_3)^2}-\frac{\dot{x}_-\dot{y}_+}{(sk_0+k_3)^2}\Big)\frac{k^2_\perp}{4}j_0 \Big] \Big(\frac{k_\perp}{2k_{0}}\Big)^{3}\frac{dk_3dk_\perp}{2\pi^2},
\end{multline}
where $c_\pm:=(x_\pm+y_\pm)/2$. As one can expect, this expression is not invariant under the translations perpendicular to the $\spe_3$ axis. Isolating the dependence on $s$, we have
\begin{multline}\label{ang_mom_s_k3_k01}
    dJ_3(s,k_3,k_0)=\frac12dJ_3(k_3,k_0) +se^2\int d\tau d\s e^{-i(k_0\De_0-k_3\De_3)}
    \Big[k_3(\dot{x}_+\dot{y}_--\dot{x}_-\dot{y}_+)(c_+j_{-1}+c_-j_1)+\\
    +ik_\perp(\dot{x}_+\dot{y}_3-\dot{x}_3\dot{y}_+)(c_-j_0+c_+j_{-2}) +ik_\perp(\dot{x}_-\dot{y}_3-\dot{x}_3\dot{y}_-)(c_+j_0+c_-j_{2})+\\
    +i(\dot{x}_+\dot{y}_3+\dot{x}_3\dot{y}_+)j_{-1} -i(\dot{x}_-\dot{y}_3+\dot{x}_3\dot{y}_-)j_{1}+\frac{2k_3}{k_\perp}(\dot{x}_+\dot{y}_-+\dot{x}_-\dot{y}_+)j_0\Big] \Big(\frac{k_\perp}{2k_{0}}\Big)^{2}\frac{dk_3dk_\perp}{16\pi^2},
\end{multline}
where
\begin{multline}\label{ang_mom_s_k3_k02}
    dJ_3(k_3,k_0)=e^2\int d\tau d\s e^{-i(k_0\De_0-k_3\De_3)}%\times\\
    \Big[\Big(2\dot{x}_3\dot{y}_3 +(\dot{x}_+\dot{y}_- +\dot{x}_-\dot{y}_+)\frac{k_0^2+k_3^2}{2k_\perp^2} \Big)\frac{k_\perp}{2}(c_+j_{-1}+c_-j_1) +\\
    +(\dot{x}_+\dot{y}_3 +\dot{x}_3\dot{y}_+)\frac{ik_3}{2}(c_-j_0+c_+j_{-2}) -(\dot{x}_-\dot{y}_3 +\dot{x}_3\dot{y}_-)\frac{ik_3}{2}(c_+j_0+c_-j_2)+\\
    +\frac{\dot{x}_+\dot{y}_+}{4}k_\perp (c_+j_{-3}+c_-j_{-1}) +\frac{\dot{x}_-\dot{y}_-}{4}k_\perp(c_+j_{1}+c_-j_3) +(\dot{x}_+\dot{y}_3 -\dot{x}_3\dot{y}_+) \frac{ik_3}{2k_\perp}j_{-1}+\\ +(\dot{x}_-\dot{y}_3 -\dot{x}_3\dot{y}_-)\frac{ik_3}{2k_\perp}j_{1} +(\dot{x}_+\dot{y}_--\dot{x}_-\dot{y}_+) \frac{k_0^2+k_3^2}{2k_\perp^2}j_0 \Big] \Big(\frac{k_\perp}{2k_{0}}\Big)^{3}\frac{dk_3dk_\perp}{2\pi^2}.
\end{multline}
In the expressions above, one has to deform the integration contours in the $\tau$ and $\s$ planes for $|\tau|\rightarrow\infty$, $|\s|\rightarrow\infty$ in the same way as it was pointed out in discussing formula \eqref{probabil}.

The average number of photons \eqref{phot_prob_s} with given the helicity $s$, the energy $k_0$, and the momentum projection $k_3$ can be integrated over $k_3$, the result being expressed in terms of elementary functions. In the expressions \eqref{ang_mom_s_k3_k01}, \eqref{ang_mom_s_k3_k02}, this cannot be done. The arising integrals seem not to be expressible in terms of the known special functions.

\section{Undulator}\label{Undulator}
\subsection{Dipole approximation}\label{Undulator_Dip}

As an example of application of the general formulas derived in the previous sections, we investigate the radiation of twisted photons by undulators in the dipole and non-dipole (wiggler) approximations. A thorough exposition of the general theory of undulator radiation can be found, for example, in \cite{AlfBashCher,Bord.1}. In our study of the undulator radiation, we shall mainly rely on \cite{Bord.1}.

Let a charged particle move along the $z$ axis with the velocity $\be_\parallel\approx1$ for $t<-TN/2$ and $t>TN/2$, while for $t\in[-TN/2,TN/2]$ its trajectory has the form
\begin{equation}\label{traj_undul}
    x^i(t)=r^i(t)+\ups^i t,\qquad \ups^i=(0,0,\be_\parallel t),
\end{equation}
where $t$ is time of the laboratory frame, $r^i(t)$ is a periodic function of $t$ with the period $T=:2\pi\omega^{-1}$, $N\gg1$ is the number of sections of the undulator. The trajectories of the particle are joined continuously at the instants $t=\pm TN/2$. In the dipole approximation, we suppose that
\begin{equation}\label{dipole_appr}
    \beta^2_\parallel=1-\frac{1+K^2}{\gamma^2},\qquad r^2_{x,y}\approx\frac{K^2}{\omega^2\ga^2},\qquad |r_z|\approx\frac{K^2}{2\pi\omega\ga^2},
\end{equation}
where $K\ll1$ is the undulator strength parameter and $\ga=(1-\be^2)^{-1/2}$. The trajectory of such a type can be realized, for example, in the helical magnetic field (see, for details, \cite{Bord.1}, Chap. 5). Then the coordinates of the charged particle take the form
\begin{equation}\label{circular_ondul}
    r_z=a_z\sin(2\omega t),\qquad r_x=a_x\cos(\omega t),\qquad r_y=-a_y\sin(\omega t),
\end{equation}
and
\begin{equation}\label{axyz}
    a_{x,y}=\frac{\la_0^2 H_{y,x}}{4\pi^2\ga},\qquad a_z=\frac{\la_0^3(H_y^2-H_x^2)}{64\pi^3\ga^2},
\end{equation}
where $\la_0:=2\pi\be_\parallel\omega^{-1}$, the magnetic field strength $H_i$ is measured in the units of the critical field
\begin{equation}\label{crit_field}
    H_0=\frac{m^2}{|e|\hbar}\approx 4.41\times 10^{13}\;\text{G},
\end{equation}
and the lengths are measured in the units of the Compton wavelength of the electron, $l_C:=\hbar/m=3.86\times 10^{-11}$ cm. The undulator strength parameter for the trajectory \eqref{circular_ondul} is written as
\begin{equation}
    K=\la_0\frac{\sqrt{H_x^2+H_y^2}}{2\sqrt{2}\pi}.
\end{equation}
One can verify that the relations \eqref{dipole_appr} are fulfilled for the trajectory \eqref{circular_ondul}.

Further, we have to evaluate the integrals entering into \eqref{probabil} keeping in mind that $K\ll1$ and $\ga\gg1$. Let us choose the basis
\begin{equation}\label{basis1}
    \spe_3=(\sin\theta \cos\vf,\sin\theta \sin\vf,\cos\theta),\qquad \spe_1=(\cos\theta \cos\vf,\cos\theta \sin\vf,-\sin\theta),\qquad \spe_2=(-\sin\vf,\cos\vf,0),
\end{equation}
i.e., we assume that the detector measures the total angular momentum projection onto the axis emanated from the source of radiation to the detector. In the case at issue, the most part of the radiation is concentrated in the cone with the opening angle $1/\ga$. Therefore, we may suppose that
\begin{equation}
    \theta\ga\lesssim1,\qquad n_\perp\ga\lesssim1,\qquad n_3=\sqrt{1-n_\perp^2}\approx 1-\frac{n_\perp^2}{2},
\end{equation}
where $n_\perp:=k_\perp/k_0$ and $n_3:=k_3/k_0$. It is also known from the theory of undulator radiation, and we shall ascertain this fact below, that
\begin{equation}
    k_0\approx 2n\omega\ga^2,
\end{equation}
where $n$ is the harmonic number. The main contribution to the radiation comes from the lowest harmonics $n$ provided that $K\ll1$. Then the following estimates are valid:
\begin{equation}\label{dip_est}
\begin{gathered}
    k_\perp|r_\pm|\approx2nK,\qquad k_3|r_3|\approx \frac{nK^2}{\pi},\qquad|\dot{r}_\pm|\approx\frac{K}{\ga},\qquad|\dot{r}_3|\approx\frac{K^2}{2\pi\ga^2},\\
    \ups_\pm\approx-\theta\Big(1-\frac{1+K^2}{2\ga^2}\Big)\sim\ga^{-1},\qquad \ups_3\approx1-\frac{1+K^2+\theta^2\ga^2}{2\ga^2}\approx1,
\end{gathered}
\end{equation}
Recall that, for example, $r_3=(\spe_3,\mathbf{r})$. The estimates \eqref{dip_est} imply that both the exponent in \eqref{probabil} and the functions $j_m$ can be developed as a Taylor series in $r_i$ keeping only the terms of the leading order in $K$.

Let us introduce the notation
\begin{equation}\label{I_n0}
\begin{split}
    I_3&:=\int dt\dot{x}_3e^{-ik_0[t(1-n_3\ups_3)-n_3r_3]}j_m\big(k_\perp(\ups_+t+r_+),k_\perp(\ups_-t+r_-)\big),\\
    I_\pm &:=\frac{in_\perp}{s\mp n_3}\int dt\dot{x}_\pm e^{-ik_0[t(1-n_3\ups_3)-n_3r_3]}j_{m\mp 1}\big(k_\perp(\ups_+t+r_+),k_\perp(\ups_-t+r_-)\big).
\end{split}
\end{equation}
Then, in the leading order in $K$, we obtain
\begin{equation}\label{I_n}
\begin{split}
    I_3&=\int dte^{-ik_0t(1-n_3\ups_3)}\Big[\ups_3j_m+\frac{k_\perp\ups_3}{2}(r_+j_{m-1}-r_-j_{m+1})\Big],\\
    I_\pm &=\frac{in_\perp}{s\mp n_3}\int dte^{-ik_0t(1-n_3\ups_3)}\Big[\ups_\pm j_{m\mp 1}+\dot{r}_\pm j_{m\mp 1} \mp\frac{k_\perp\ups_\pm}{2}(r_\mp j_{m} -r_\pm j_{m\mp 2})\Big],
\end{split}
\end{equation}
where $j_m\equiv j_m(k_\perp\ups_+t,k_\perp\ups_-t)$ and, in expanding in the Taylor series, we have employed the recurrence relations \eqref{recurr_rels}. Substituting the representation \eqref{Bessel_int1} of the Bessel function into the integrals \eqref{I_n} and taking into account the condition \eqref{positiveness}, we see that the contribution of the first term in the square brackets in these integrals vanishes. As a result, the integrals \eqref{I_n} are reduced to the integrals over $t\in[-TN/2,TN/2]$.

It is useful to represent $r_i$ as the Fourier series
\begin{equation}\label{traj_Fourie}
    r_3=\sum_{n=-\infty}^\infty r_3(n)e^{i\omega nt},\qquad r_\pm=\sum_{n=-\infty}^\infty r_\pm(n)e^{i\omega nt},
\end{equation}
and substitute them into \eqref{I_n}. Let
\begin{equation}\label{omega_pm}
    \omega_\pm:=\frac{\omega}{1-n_3\ups_3\mp n_\perp|\ups_+|}\approx\frac{2\omega\ga^2}{1+K^2+(n_\perp\mp\theta)^2\ga^2},
\end{equation}
and
\begin{equation}\label{GmN}
    G_N^m(a,b):=\int_{-N}^N\frac{dt}{4}e^{-i\pi t(b-a)/2}J_m(\pi(b+a)t/2).
\end{equation}
Some properties of the functions $G_N^m(a,b)$ are collected in the Appendix \ref{Bessel_Prop}. Then
\begin{equation}\label{GmN1}
\begin{split}
    i^{-m}&\int_{-\pi}^\pi\frac{d\psi}{2\pi}e^{-im\psi}\de_N\big(k_0(1-n_3\ups_3-n_\perp|\ups_+|\cos\psi)-\omega n\big)=\\
    =&\int_{-TN/2}^{TN/2}\frac{dt}{2\pi}e^{-i[k_0(1-\ups_3n_3)-\omega n]t} J_m(k_\perp|\ups_+|t)
    =\frac{2}{\omega}G_N^m(a_n,b_n),
\end{split}
\end{equation}
where
\begin{equation}
    \de_N(x):=\int_{-TN/2}^{TN/2}\frac{dt}{2\pi}e^{-ixt}=\frac{\sin(TNx/2)}{\pi x},
\end{equation}
and
\begin{equation}\label{anbn}
    a_n=n-k_0\omega_+^{-1},\qquad b_n=k_0\omega_-^{-1}-n.
\end{equation}

With the aid of the above functions, we can write
\begin{equation}\label{I_n1}
\begin{split}
    I_3&=\frac{2\pi}{\omega}k_\perp\ups_3(-1)^{m+1}\sum_{n=-\infty}^\infty\Big[G^{m-1}_Nr_+(n)-G^{m+1}_Nr_-(n)\Big],\\
    I_\pm &=\frac{2\pi}{\omega}\frac{in_\perp}{s\mp n_3}(-1)^{m+1}\sum_{n=-\infty}^\infty\Big[\dot{r}_\pm(n)G_{N}^{m\mp1} \pm\frac{k_\perp\ups_\pm}{2}\big(r_\mp(n)G_{N}^{m} -r_\pm(n)G_{N}^{m\mp 2}\big)\Big],
\end{split}
\end{equation}
where $G_{N}^{m}\equiv G_N^m(a_n,b_n)$.

For large $N$ the function $G_{N}^{m}$ is well approximated by the expression \eqref{GmN_approx}. Hence, $a_n\geq0$, $b_n\geq0$, and the photon energy should belong to the intervals
\begin{equation}\label{energy_ints}
    k_0\in n[\omega_-,\omega_+],\quad n\in \mathbb{N}.
\end{equation}
Therefore, the terms with $n\geq1$ only survive in the sums over $n$ in \eqref{I_n1}. The intensity of radiation at the harmonic $n$ is proportional to the modulus squared of the Fourier series coefficient \eqref{traj_Fourie} with the number $n$. This implies that, in the dipole approximation, the radiation is concentrated at the lowest harmonics. The intervals \eqref{energy_ints} overlap starting with the number
\begin{equation}\label{n0over}
    n_0=\frac{\omega_-}{\omega_+-\omega_-}\approx\frac{1+K^2+(n_\perp-\theta)^2\ga^2}{4n_\perp\theta\ga^2}.
\end{equation}
In order to proceed, we distinguish the three cases: (i) the weakly degenerate case when $k_0$ is close to $n\omega_\pm$; (ii) the strongly degenerate case when $\omega_-$ tends to $\omega_+$; and the regular case (iii) when the photon energy $k_0$ is taken sufficiently far from the boundaries of the intervals \eqref{energy_ints}, and $\de_N(x)$ in \eqref{GmN1} can be replaced by the delta-function.

\begin{figure}[!t]
\centering
a)\;\includegraphics*[align=c,width=0.6\linewidth]{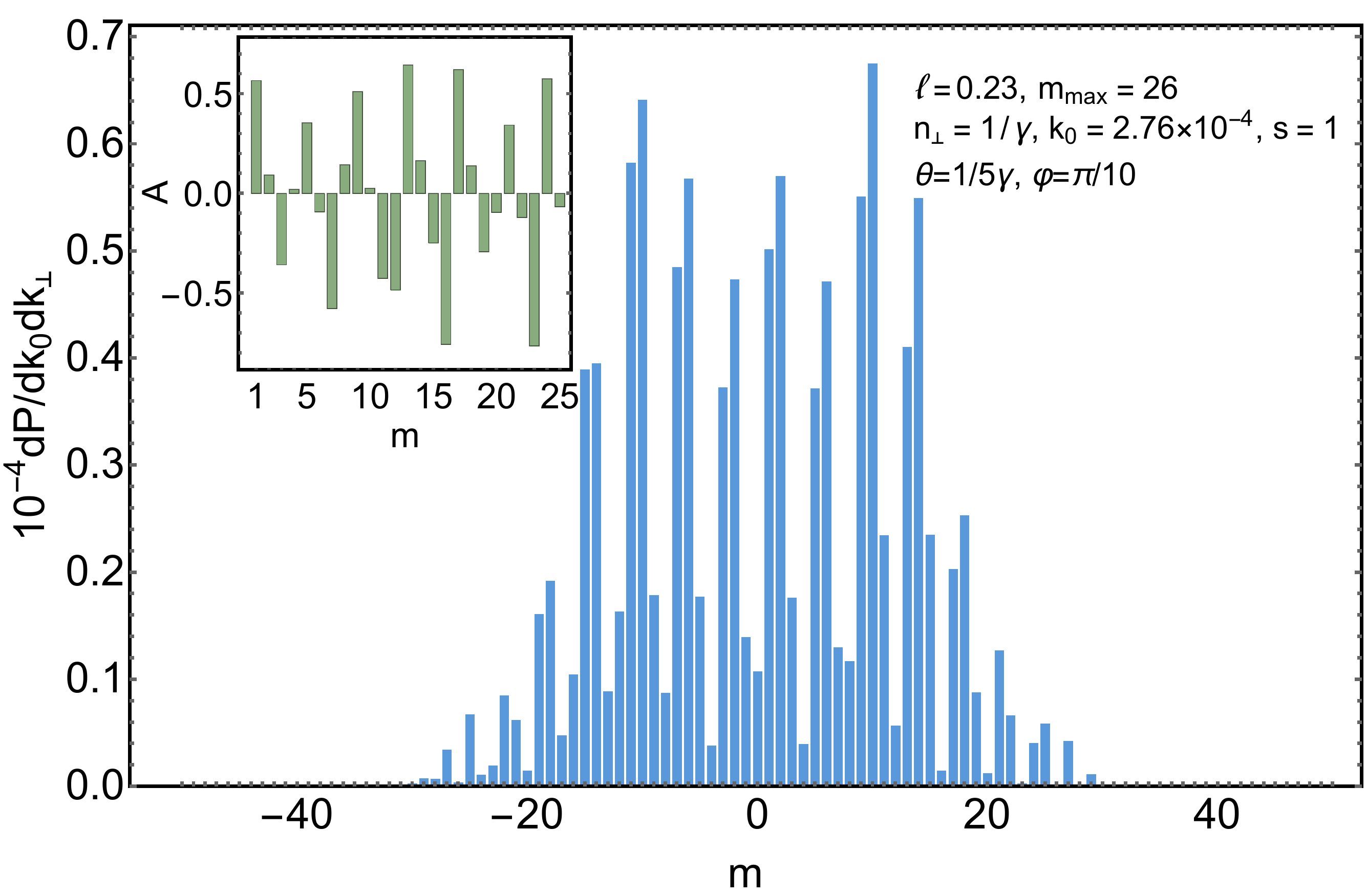}\\
b)\;\includegraphics*[align=c,width=0.4\linewidth]{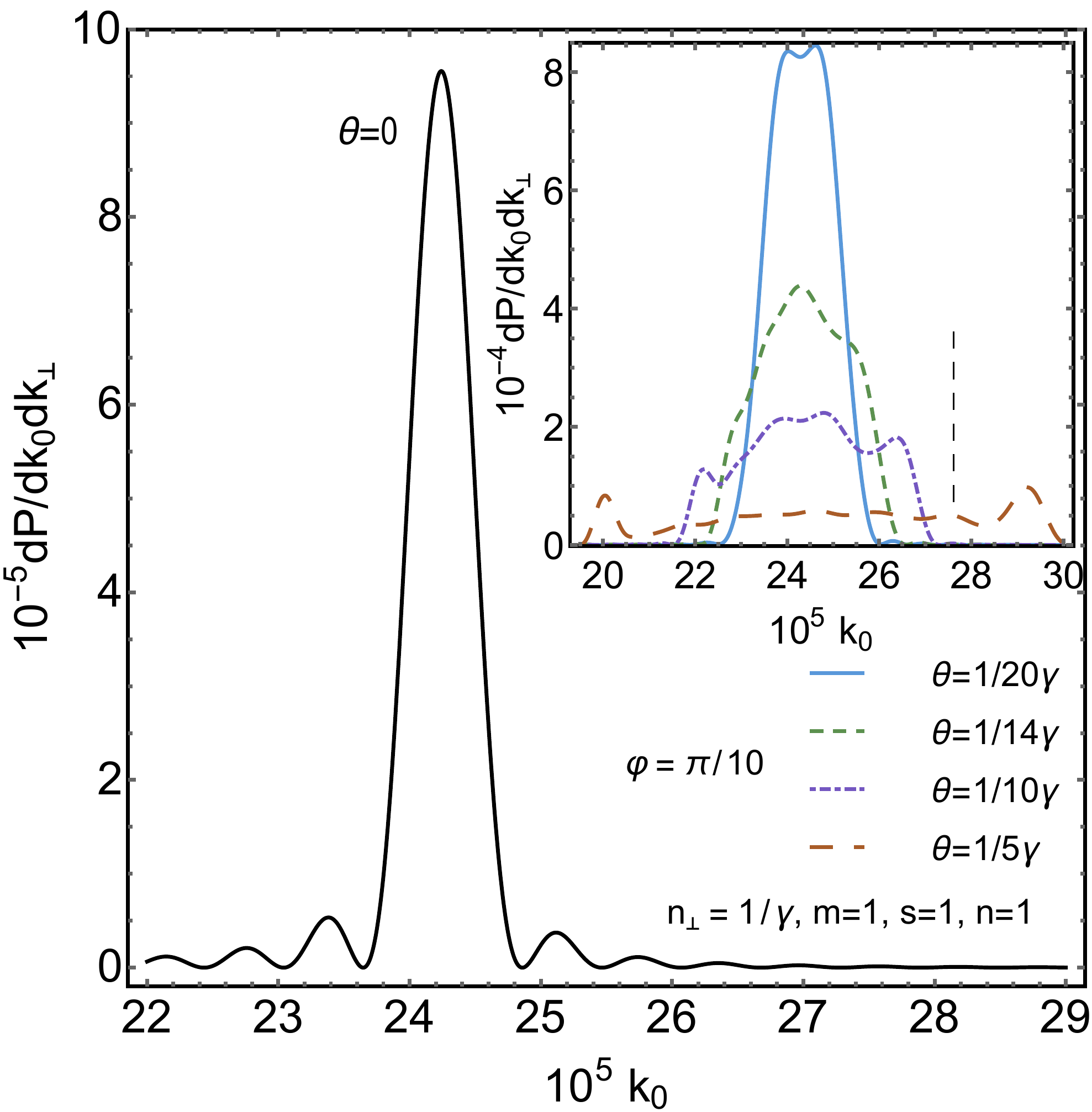}\;\;
c)\;\includegraphics*[align=c,width=0.39\linewidth]{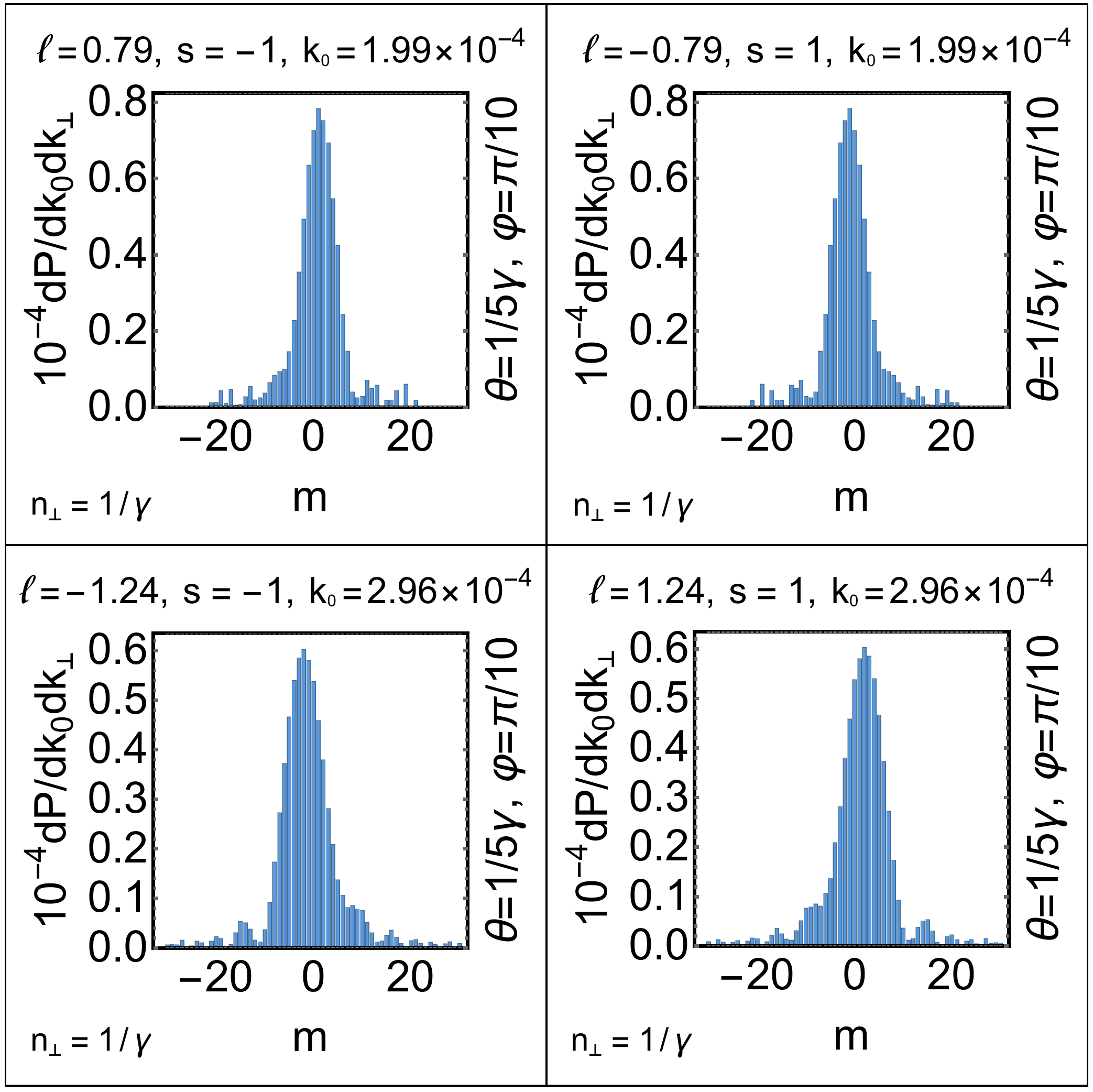}
\caption{{\footnotesize The radiation of twisted photons by the planar undulator in the dipole regime at the first harmonic. The trajectory of the electron is taken in the form \eqref{circular_ondul} with $a_z=a_y=0$, $a_x=K\la_0/(\sqrt{2}\pi\ga)$, where $K=0.03$ is the undulator strength parameter, $\la_0=1$ cm is the length of the undulator section, and $\ga=10^3$ is the Lorentz-factor of the electron. The number of the undulator sections $N=40$. The energy of photons is measured in the units of the rest energy of the electron $0.511$ MeV. (a) The distribution over $m$, the asymmetry, and the angular momentum projection per one photon. In accordance with \eqref{m_period}, the period of oscillations $T_m=4$. (b) The density of the average number of twisted photons against $k_0$ for the different observation angles. The position of the peak in the forward radiation and the boundaries of the spectral band in the inset are well described by \eqref{energy_ints}. The dashed vertical line in the inset depicts the photon energy used in the plot (a). (c) The density of the average number of twisted photons against $m$ and the angular momentum projection per one photon at the left, $\xi_n=\pi$, and right, $\xi_n=0$, peaks appearing in the distribution over the photon energy for $\theta=1/(5\ga)$ (see the inset in the plot (b)). The average number of photons obeys the symmetry relation \eqref{plane_symm}.}}
\label{plan_dip_plots}
\end{figure}

\begin{figure}[!t]
\centering
a)\;\includegraphics*[align=c,width=0.6\linewidth]{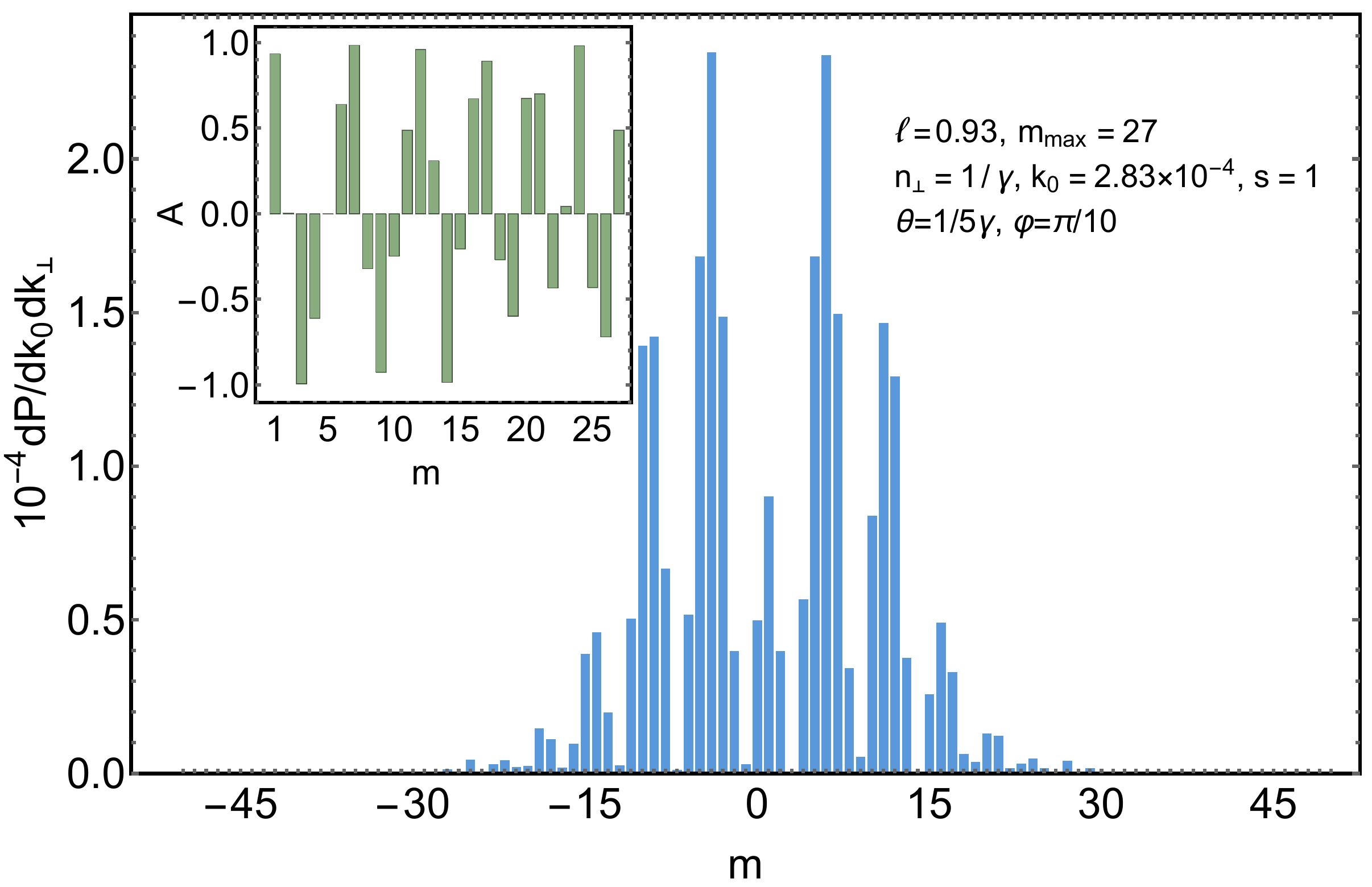}\\
b)\;\includegraphics*[align=c,width=0.4\linewidth]{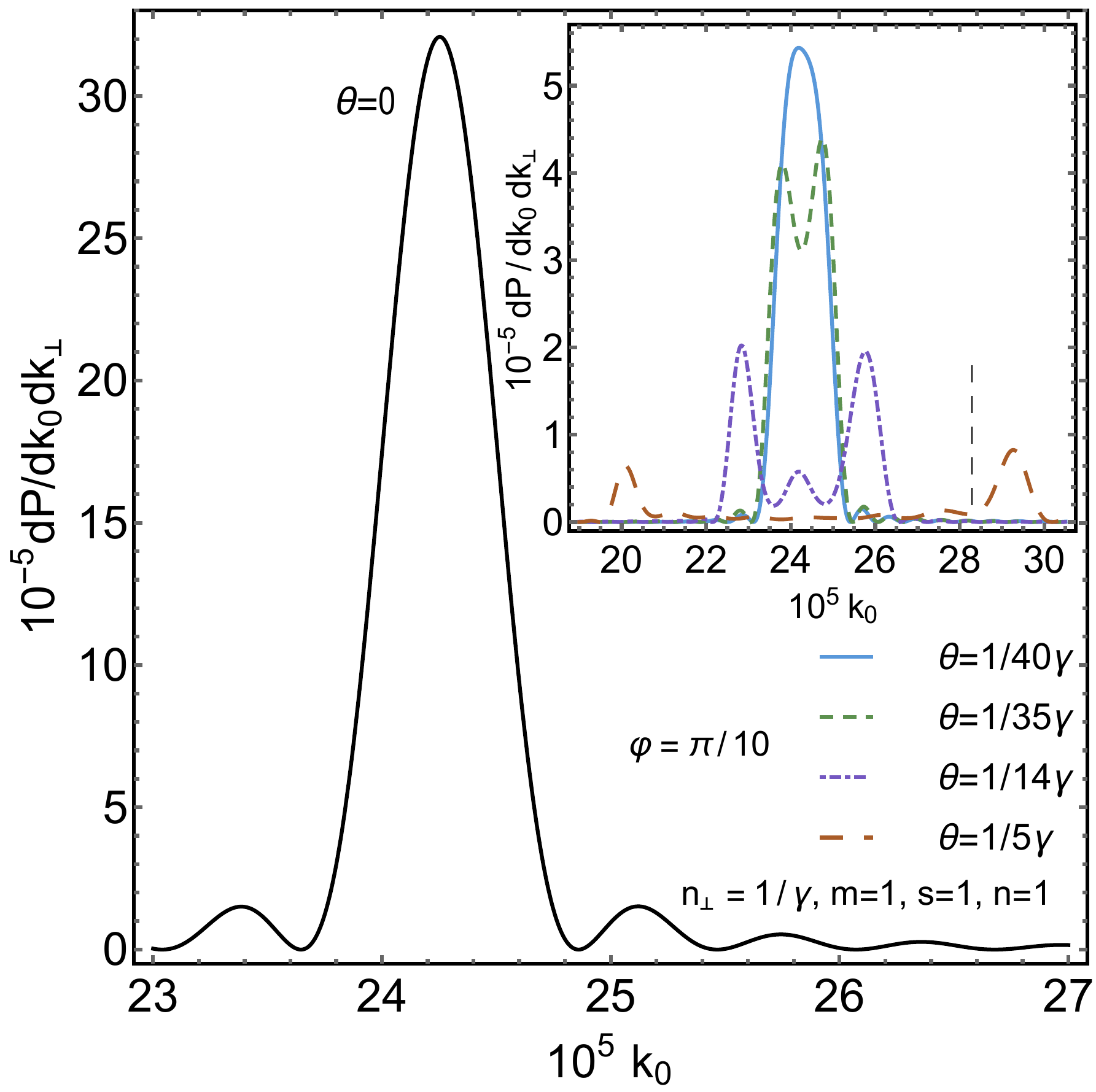}\;\;
c)\;\includegraphics*[align=c,width=0.39\linewidth]{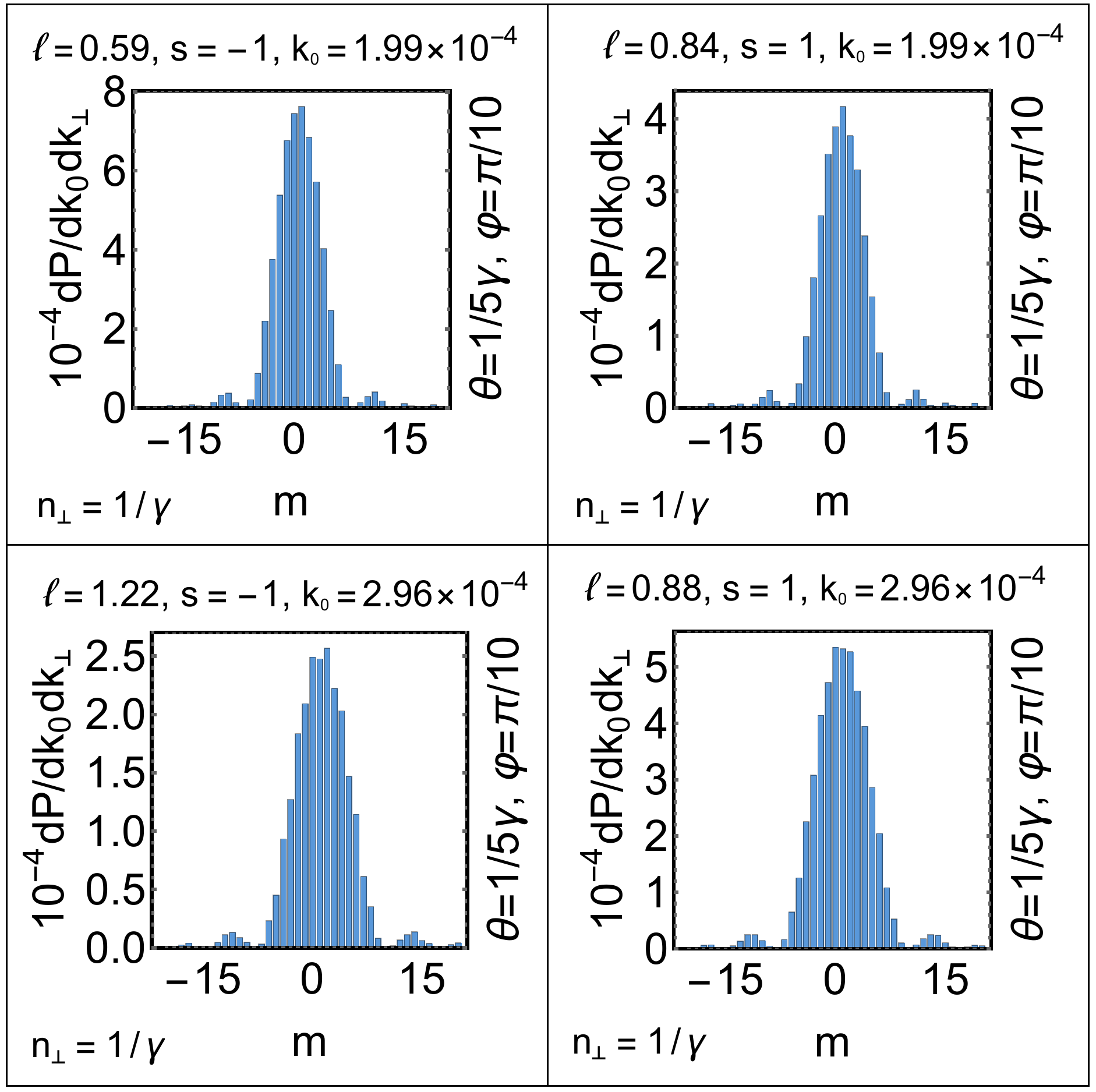}
\caption{{\footnotesize The radiation of twisted photons by the helical undulator in the dipole regime at the first harmonic. The trajectory of the electron is taken in the form \eqref{circular_ondul}, \eqref{axyz}, where $H_x=10^{-12}$, $H_y=1.2\times 10^{-12}$ in the units of the critical field \eqref{crit_field}, $\la_0=1$ cm is the length of the undulator section, and $\ga=10^3$ is the Lorentz-factor of the electron. These data correspond to $K=4.6\times10^{-3}$. The number of the undulator sections $N=40$. The energy of photons is measured in the units of the rest energy of the electron $0.511$ MeV. (a) The distribution over $m$, the asymmetry, and the angular momentum projection per one photon. In accordance with \eqref{m_period}, the period of oscillations $T_m=5$. (b) The density of the average number of twisted photons against $k_0$ for the different observation angles. The position of the peak in the forward radiation and the boundaries of the spectral band in the inset are well described by \eqref{energy_ints}. The dashed vertical line in the inset depicts the photon energy used in the plot (a). (c) The density of the average number of twisted photons against $m$ and the angular momentum projection per one photon at the left, $\xi_n=\pi$, and right, $\xi_n=0$, peaks appearing in the distribution over the photon energy for $\theta=1/(5\ga)$ (see the inset in the plot (b)).}}
\label{hel_dip_plots}
\end{figure}

Let us begin with the case (i). The absolute values of the integrals \eqref{I_n1} and, consequently, the average number of photons \eqref{probabil} increase rapidly near the boundaries of the intervals \eqref{energy_ints}. In the vicinity of these peaks, $a_n\lesssim1/N$ and $b_n\gtrsim 5/N$, i.e.,
\begin{equation}
    n\omega_+>k_0\gtrsim\omega_+(n-1/N),\qquad k_0\gtrsim\omega_-(n+5/N),
\end{equation}
we have from \eqref{I_n1} and \eqref{GmN_bound},
\begin{equation}
    dP=\frac{e^2Nn\omega^{-1}}{\omega^{-1}_--\omega^{-1}_+}\Big|isn_3\sqrt{\frac{\omega}{\omega_+}}r_2(n)+\sqrt{\frac{\omega_+}{\omega}}\Big[\frac{\omega}2(\omega_+^{-1}+\omega_-^{-1}) -n_\perp^2\Big] r_1(n) \Big|^2\frac{n_\perp}{n_3^2}\frac{dk_3dk_\perp}{2\pi^2}.
\end{equation}
The dependence on $m$ is absent, but $|m|$ must satisfy the estimate \eqref{m_est_a0}. Outside of this spectral band of $m$, the average number of photons $dP$ tends exponentially to zero (see Figs. \ref{plan_dip_plots}, \ref{hel_dip_plots}). Analogously, for $b_n\lesssim1/N$ and $a_n\gtrsim 5/N$, i.e., for
\begin{equation}
    n\omega_-<k_0\lesssim\omega_-(n+1/N),\qquad k_0\lesssim\omega_+(n-5/N),
\end{equation}
the average number of photons becomes
\begin{equation}
    dP=\frac{e^2Nn\omega^{-1}}{\omega^{-1}_--\omega^{-1}_+}\Big|isn_3\sqrt{\frac{\omega}{\omega_-}}r_2(n)+\sqrt{\frac{\omega_-}{\omega}}\Big[\frac{\omega}2(\omega_+^{-1}+\omega_-^{-1}) -n_\perp^2\Big] r_1(n) \Big|^2\frac{n_\perp}{n_3^2}\frac{dk_3dk_\perp}{2\pi^2},
\end{equation}
where $|m|$ should satisfy the estimate \eqref{m_est_a0}.

\begin{figure}[!t]
\centering
i)\;\includegraphics*[align=c,width=0.45\linewidth]{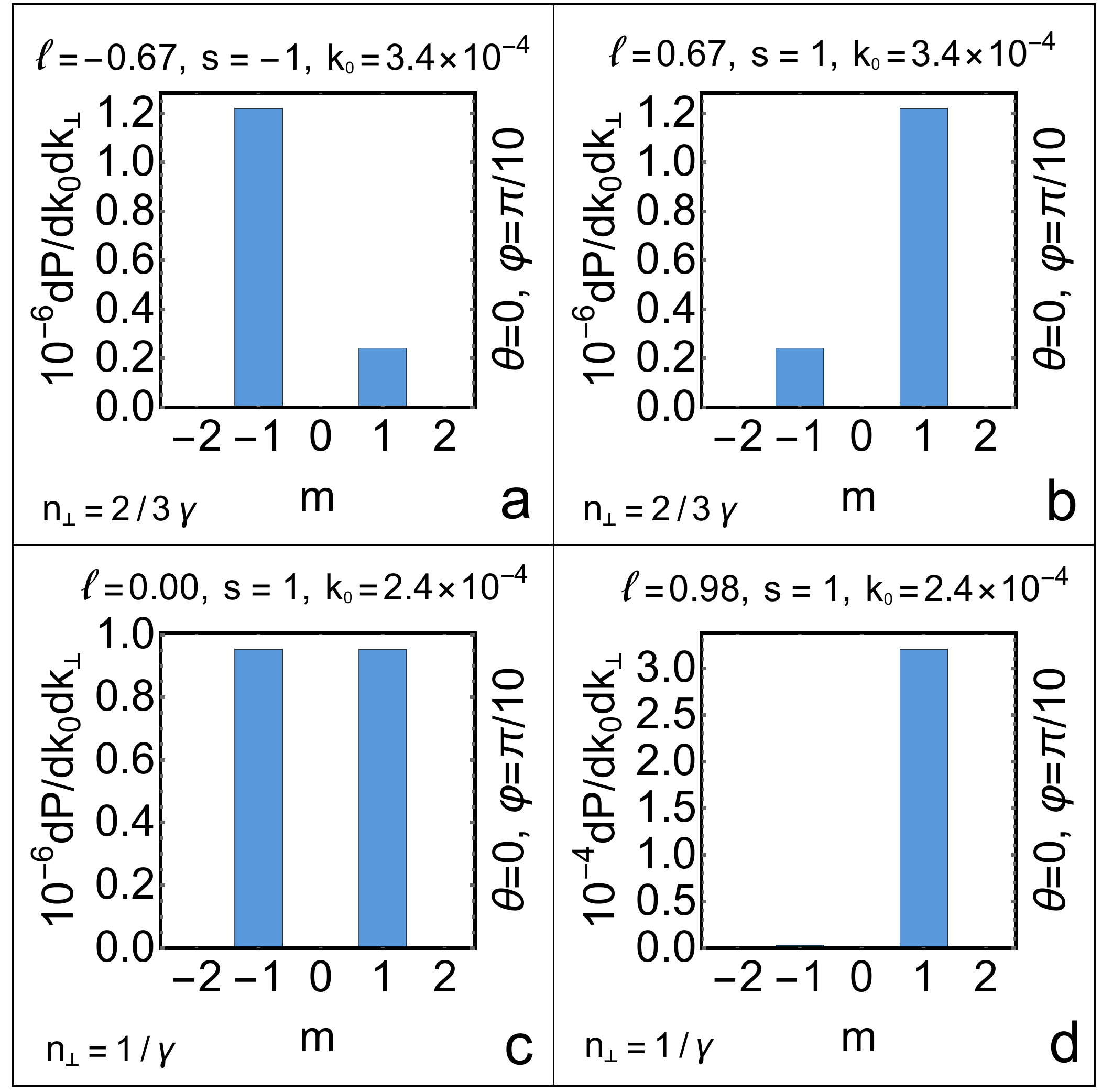}\;\;
ii)\;\includegraphics*[align=c,width=0.45\linewidth]{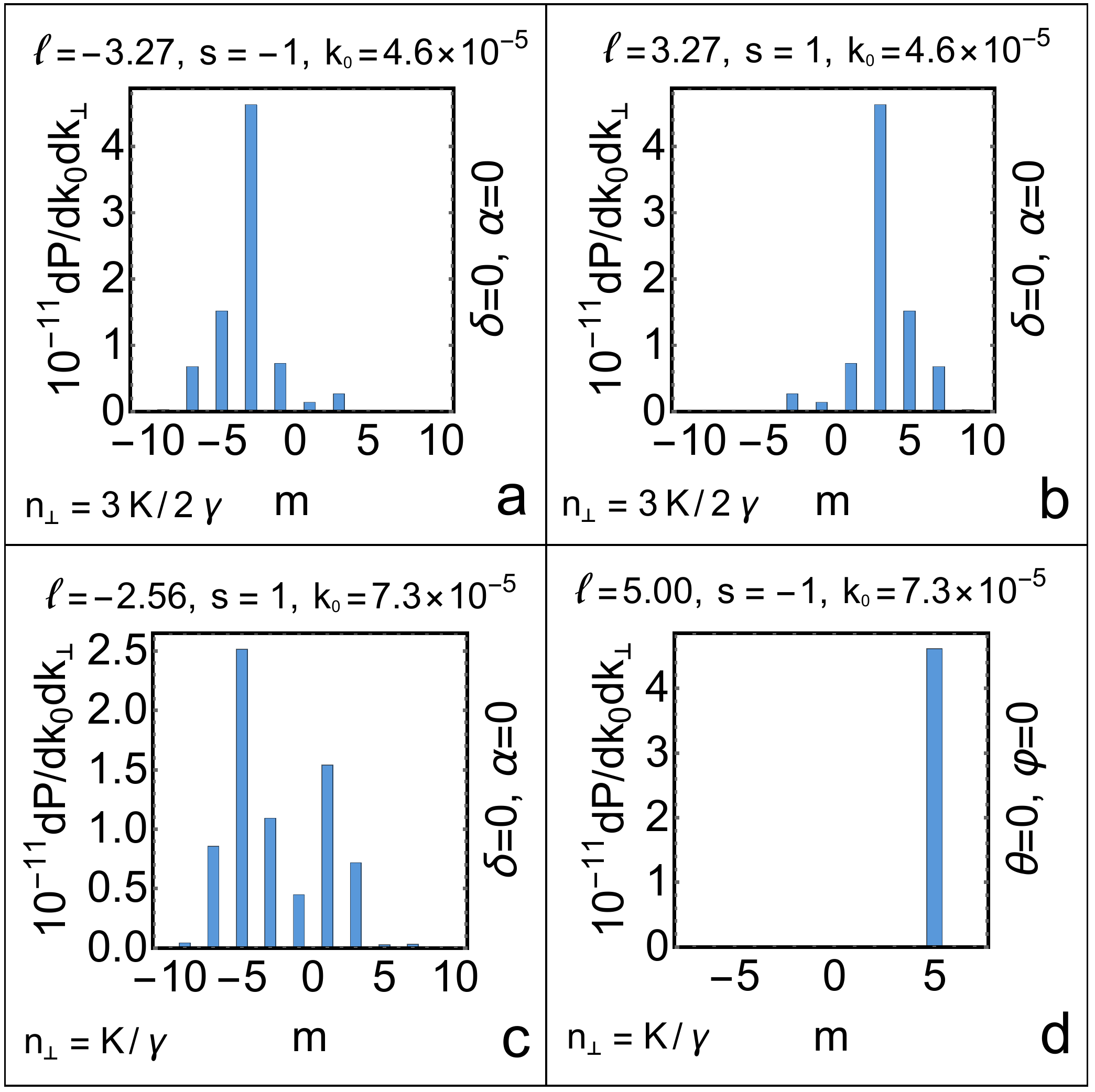}
\caption{{\footnotesize The distribution over $m$ and the angular momentum projection per one photon of the forward radiation of twisted photons produced by the undulators. The energy of photons is measured in the units of the rest energy of the electron $0.511$ MeV. The forward radiation of the planar undulator is subject to the the symmetry relation \eqref{plane_symm} and the selection rule that $m+n$ is an even number. The forward radiation of the right-handed helical undulator obeys the rule $m=n$. On the left panel: the undulator forward radiation in the dipole regime at the first harmonic. The insets (a), (b), (c) are for the planar undulator, and (d) is for the helical one. The trajectories of the electron are the same as in Figs. \ref{plan_dip_plots}, \ref{hel_dip_plots}. The small contribution at $m=-1$ for the helical undulator is a consequence of deviation of the trajectory from an ideal right-handed helix. The density of the average number of twisted photons is well described by \eqref{forward_dip}. On the right panel: the wiggler forward radiation at the fifth harmonic. The insets (a), (b), (c) are for the planar wiggler, and (d) is for the helical one. The trajectories of the electron are the same as in Figs. \ref{hel_wig_plots}, \ref{plan_wig_plots}.}}
\label{forw_dip_plots}
\end{figure}

The largest value of $dP$ is achieved in the case (ii) when the two peaks merge into one (see Figs. \ref{plan_dip_plots}, \ref{hel_dip_plots}). This happens when $a_n\lesssim1/N$ and $b_n\lesssim1/N$, which is equivalent to
\begin{equation}\label{peak_m}
    \omega_+(n-1/N)\lesssim k_0\lesssim \omega_-(n+1/N).
\end{equation}
These inequalities imply
\begin{equation}\label{delta}
    \omega_+-\omega_-<\frac{\omega_++\omega_-}{nN},\qquad \frac{2n_\perp\theta\ga^2}{1+K^2+(n_\perp^2+\theta^2)\ga^2}<\frac{1}{nN}.
\end{equation}
In this case, the functions $G^m_N$ can be replaced by \eqref{GmN00} in \eqref{I_n1}. Moreover, it follows from \eqref{omega_pm} and \eqref{delta} that
\begin{equation}
    |\ups_\pm|=\frac{\omega}{2n_\perp}(\omega_-^{-1}-\omega_+^{-1})<\frac{\omega}{n_\perp\bar{\omega}N},
\end{equation}
where $\bar{\omega}:=(\omega_++\omega_-)/2$. Therefore, the contributions standing at $\ups_\pm$ in \eqref{I_n1} can be neglected in this case. As a result,
\begin{equation}\label{forward_dip}
    dP=\frac{e^2N^2}{16}\sum_{n=1}^\infty n^2\Big[\de_{m1}\Big(\frac{n_\perp^2\bar{\omega}-\omega}{n_3\omega}-s \Big)^2|r_+(n)|^2 +\de_{m,-1}\Big(\frac{n_\perp^2\bar{\omega}-\omega}{n_3\omega}+s \Big)^2|r_-(n)|^2 \Big]n_\perp dk_3dk_\perp.
\end{equation}
This formula describes, in particular, the forward radiation of twisted photons when $\theta\ga\lesssim1/N$ and $n_\perp\ga\lesssim1$. As we see, the whole radiation consists mainly of the photons with the quantum numbers $m=\pm1$. If the trajectory of a charged particle is a right-handed helix, then $|r_-(n)|$, $n=\overline{1,\infty}$, are small or equal to zero. In this case, the whole radiation at the peak \eqref{peak_m} consists of the twisted photons with $m=1$ (see Fig. \ref{forw_dip_plots}). As for the left-handed helical trajectory, the radiation at the peak \eqref{peak_m} consists of the photons with $m=-1$. These properties are independent of the helicity $s$, which can take the both values $\pm1$.

Now we turn to the regular case (iii) when $a_n\gtrsim 5/N$ and $b_n\gtrsim 5/N$, i.e.,
\begin{equation}\label{regul_case_dip}
    \omega_-(n+5/N)\lesssim k_0\lesssim\omega_+(n-5/N),
\end{equation}
and $m$ satisfies \eqref{a_b_m_ests}:
\begin{equation}\label{m_max}
    |m|\lesssim m_{max},\qquad m_{max}:= \frac{10N}{7}k_0(\omega_-^{-1}-\omega_+^{-1})\approx\frac{20N}{7}\frac{k_0}{\omega}n_\perp\theta.
\end{equation}
In this domain, the functions $G^m_N$ in \eqref{I_n1} can be replaced by the expressions \eqref{GmN_approx} with a good accuracy. In a general case, the explicit expression for the average number of photons resulting from such a substitution is rather cumbersome. That is why we do not present it here. However, the dependence of $dP$ on $m$ can readily be found. Let
\begin{equation}\label{xin}
    \xi_n:=\arccos\frac{b_n-a_n}{b_n+a_n}=\arccos\frac{\omega_+^{-1}+\omega_-^{-1}-2nk_0^{-1}}{\omega^{-1}_--\omega^{-1}_+}.
\end{equation}
Then the integrals \eqref{I_n1} are the linear combinations of $\cos(m\xi_n)$ and $\sin(m\xi_n)$ with the coefficients independent of $m$. Adding up these expressions and squaring the modulus of the result, we see that, for those modes where the intervals \eqref{energy_ints} do not overlap or one can neglect this overlapping, the dependence of the average number of photons \eqref{probabil} on $m$ for every mode $n$ can be cast into the form
\begin{equation}\label{m_depend_reg}
    A_n^2\cos^2(m\xi_n+\de_n),
\end{equation}
where $A_n$ and $\de_n$ do not depend on $m$. Consequently, in the domain of parameters \eqref{regul_case_dip}, \eqref{m_max}, the average number of photons $dP$ is a periodic function of $m$ with the period
\begin{equation}\label{m_period}
    T_m=\pi/\xi_n,\quad\xi_n\in(0,\pi/2);\qquad T_m=\pi/(\pi-\xi_n),\quad\xi_n\in[\pi/2,\pi).
\end{equation}
For $|m|>m_{max}$, the approximation \eqref{GmN_approx} does not hold, and the average number of photons \eqref{probabil} tends exponentially to zero (see the representative dependence of $dP$ against $m$ in Figs. \ref{plan_dip_plots}, \ref{hel_dip_plots}). The difference of the peak heights in the plots is an artefact of a finite $N$ ($N=40$). It disappears with increasing $N$.

As an example, we consider the cases of the planar and helical undulators in detail. Let $r_1(n)=0$, $n=\overline{1,\infty}$. This is valid when, for example,
\begin{equation}
    \mathbf{r}(t)\sim\spe_2.
\end{equation}
Thus we have a planar trajectory of the charged particle. In the domain \eqref{regul_case_dip}, \eqref{m_max}, we obtain
\begin{equation}
\begin{split}
    I_3&=\frac{2}{\omega} \frac{i^m}{\sqrt{a_nb_n}} k_\perp\ups_3 \sin\xi_n\sin(m\xi_n) r_2(n),\\
    I_\pm &=-\frac{1}{\omega} \frac{i^m}{\sqrt{a_nb_n}}\big\{\omega n\cos[(m\mp 1)\xi_n]\mp k_\perp\ups_\pm \sin\xi_n\sin[(m\mp 1)\xi_n]\big\}\frac{n_\perp r_2(n)}{s\mp n_3}.
\end{split}
\end{equation}
Then we add up these expressions taking into account $1/2$ in \eqref{probabil} and square the modulus of the outcome. The result is
\begin{equation}\label{probabil_dippl}
    dP=e^2\frac{k_0^2\cos^2(m\xi_n+\de_n)}{(\omega_+-\omega_-)^2}\Big[4\frac{a_nb_n}{n_3^2}\frac{\omega_+^2\omega_-^2}{k_0^2\omega^2} \Big(\frac{\omega}{2}(\omega_+^{-1}+\omega_-^{-1})-n_\perp^2 \Big)^2+\Big(1-n\frac{\omega_++\omega_-}{k_0}\Big)^2 \Big] \frac{|r_2(n)|^2n_\perp}{a_nb_n}\frac{dk_3dk_\perp}{4\pi^2},
\end{equation}
where
\begin{equation}
    \tan\de_n=\frac{2s}{n_3\omega}\frac{\omega_+\omega_-\sqrt{a_nb_n}}{k_0-n(\omega_++\omega_-)} \Big(\frac{\omega}{2}(\omega_+^{-1}+\omega_-^{-1})-n_\perp^2 \Big).
\end{equation}
When $s$, $n_\perp$, and $\theta$ are fixed, the absolute value of the phase $|\de_n|$ reaches its maximal value at
\begin{equation}
    k_0=n\frac{\omega_+^2+\omega_-^2}{\omega_++\omega_-}\in n[\omega_-,\omega_+].
\end{equation}
Notice that, in the domain of applicability of the approximations made, the average number of photons \eqref{probabil_dippl} obeys the relation \eqref{plane_symm} despite the fact that the detector and the axis used to define the projection of the angular momentum do not lie in the orbit plane of a particle. In particular, on summing over the helicities, the average number of photons is symmetric with respect to $m\rightarrow-m$ (see Fig. \ref{plan_dip_plots}).

As for the helical undulator, we suppose that $r_-(n)=0$, $n=\overline{1,\infty}$. Then, if the conditions \eqref{regul_case_dip}, \eqref{m_max} are satisfied, we obtain
\begin{equation}
\begin{split}
    I_3&=\frac{1}{\omega} \frac{i^{m-1}}{\sqrt{a_nb_n}} k_\perp\ups_3 \cos[(m-1)\xi_n] r_+(n),\\
    I_+&=-\frac{1}{\omega} \frac{i^{m-1}}{\sqrt{a_nb_n}}\big\{\omega n\cos[(m-1)\xi_n]-\frac{k_\perp\ups_+}{2}\cos[(m-2)\xi_n]\big\}\frac{n_\perp r_+(n)}{s-n_3},\\
    I_-&=-\frac{1}{\omega} \frac{i^{m-1}}{\sqrt{a_nb_n}}\frac{k_\perp\ups_-}{2}\cos(m\xi_n)\frac{n_\perp r_+(n)}{s+n_3}.
\end{split}
\end{equation}
The average number of radiated photons is written as
\begin{equation}
    dP=e^2\Big[\Big(sn+\frac{n_\perp k_0}{2}(\omega_+^{-1}+\omega_-^{-1})\Big)\cos[(m-1)\xi_n] +s\sqrt{a_nb_n}\sin[(m-1)\xi_n]\Big]^2 \frac{|r_+(n)|^2}{a_nb_n}n_\perp\frac{dk_3dk_\perp}{16\pi^2},
\end{equation}
in the domain of parameters \eqref{regul_case_dip}, \eqref{m_max}. Since $n_\perp\sim\ga^{-1}$ and $k_0\lesssim n\omega_+$, the second term in the round parenthesis can be neglected in comparison with the first one. Then
\begin{equation}
    dP=e^2\Big(n\cos[(m-1)\xi_n] +\sqrt{a_nb_n}\sin[(m-1)\xi_n]\Big)^2 \frac{|r_+(n)|^2}{a_nb_n}n_\perp\frac{dk_3dk_\perp}{16\pi^2}.
\end{equation}
The dependence of the average number of the radiated twisted photons on $s$ disappears. A typical dependence of $dP$ on $k_0$, $n_\perp$, and $m$ is presented in Fig. \ref{hel_dip_plots} in this case.

As it was mentioned in many paper and books (see, e.g., \cite{fickler12,Bozinov13,PadgOAM25,AndBabAML,TorTorTw,AndrewsSLIA,SSDFGCY}), the quantum number $m$ can be considered as a new degree of freedom of photons that can be employed to transmit information. The signals of the form presented in Figs. \ref{plan_dip_plots}, \ref{hel_dip_plots} can be used for information transfer by the means analogous to the frequency modulation in radio engineering. In our case, however, the frequency of radiation is fixed, but the period of the dependence of radiation on $m$ is changed. This allows one to transmit the number with the base approximately equal to $m_{max}/2$ per one signal, which is considerably more efficient than the binary code used in digital communication. Of course, there are many technical issues in realizing this way of information transfer that are related to detection of the twisted photons. We will not discuss these problems here (for the possible techniques of detection of the twisted photons see, e.g., \cite{LPBFAC,BLCBP,SSDFGCY,LavCourPad}). In particular, it is necessary to discriminate the photons with $k_\perp$ belonging to the narrow interval of momenta. Roughly, this can be achieved by employing the resolution of the signal by the arrival time, i.e., by the different values of $n_3$ \cite{GRPFSBFP,ZZLLSS}. More accurate and efficient method to single out the photons with the narrow interval of $k_\perp$ can be based on the use of the phase masks or fibers with circular cross-section (see, e.g., \cite{MillEber}).

\begin{figure}[!t]
\centering
\includegraphics*[align=c,width=0.4\linewidth]{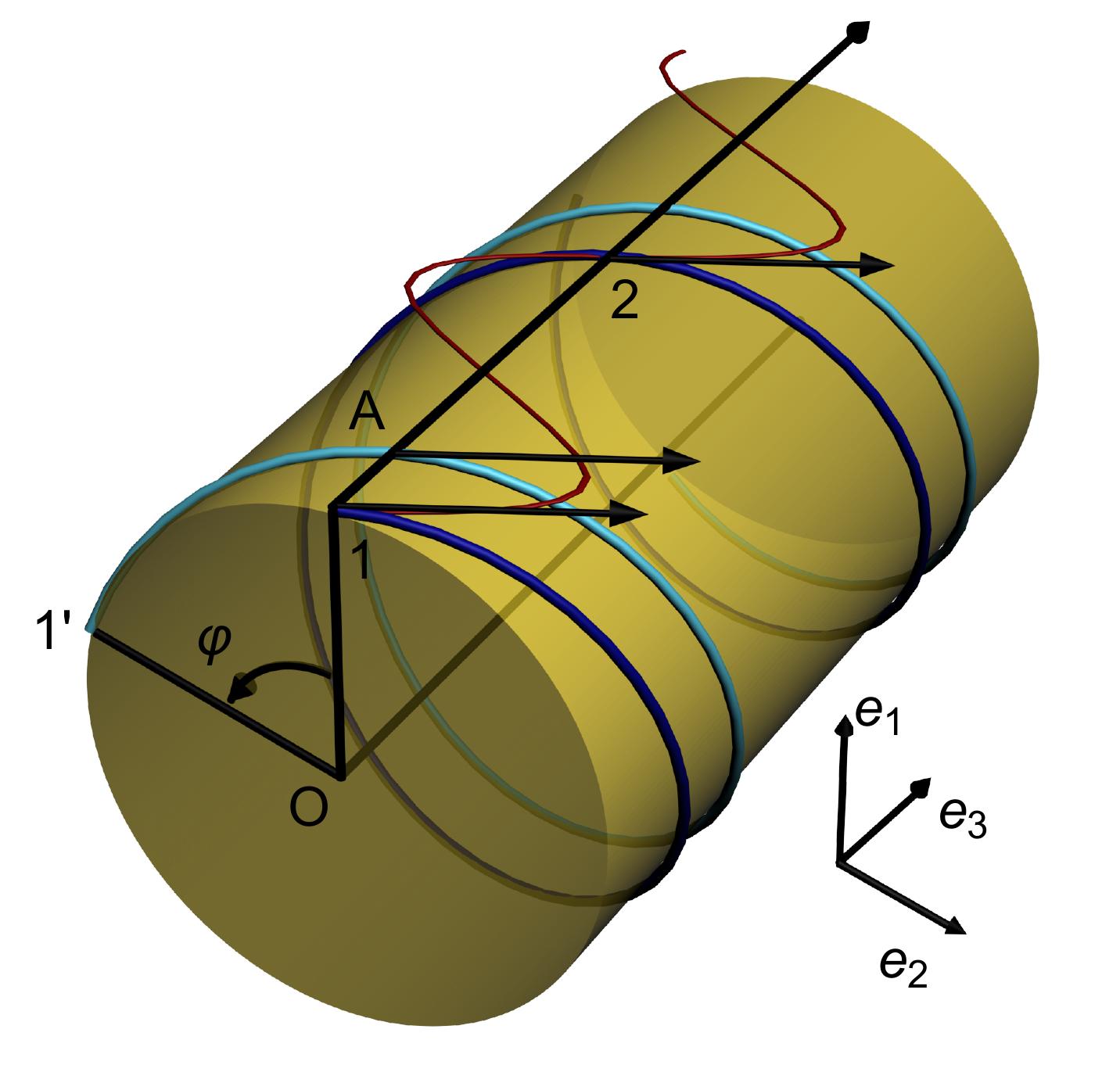}
\caption{{\footnotesize The pictorial representation of the radiation amplitude of twisted photons produced by a charged particle moving along a helical trajectory (see the detailed description at the end of Sec. \ref{Undulator_Dip}). The red sinusoid depicts the undulator radiation at the first harmonic. The vectors attached to the points $1$, $2$, and $A$ are the velocity vectors. The phase of the electromagnetic wave is $\Phi(1,2)=d_{1,2}(k_0\be_\parallel^{-1}-k_3)$, where $d_{1,2}$ is the distance between the points $1$ and $2$. The resonance occurs when $\Phi(1,2)=2\pi n$, where $n$ is the harmonic number. The azure trajectory is obtained from the blue one by the rotation by an angle of $\vf$ so that the point $1$ is shifted to the point $1'$. This trajectory contributes to the amplitude with the additional phase factor $e^{im\vf}$. Therefore, the radiation from the point $A$ adds up constructively with the radiation from the points $1$ and $2$ when $\Phi(1,2)(1-d_{A,1}/d_{1,2})+m\vf=2\pi n$. Inasmuch as $d_{A,1}/d_{1,2}=\vf/(2\pi)$, we infer that $(m-n)\vf=0$, i.e., $m=n$. As for the ideal left-handed helical trajectory, one can easily deduce that $m=-n$.}}
\label{cylind_plots}
\end{figure}

We see that, in the dipole approximation, it is hard to achieve the large angular momentum projection per one photon in the radiation from undulators. In the optimal case, it is of the order $\pm1$ when the (almost) forward radiation of the undulator is used (see Fig. \ref{forw_dip_plots}). This fact can be explained with the help of the pictorial interpretation of the transition amplitude \eqref{amplitude} in terms of the plane-wave photon radiation amplitude (see Sec. \ref{Prob_Rad}). Let us consider the forward radiation from the helical undulator for a charged particle with the trajectory in the form of a right-handed helix. The maximum of the radiation arises in that case when the points $1$ and $2$ of the trajectory radiates in-phase, i.e., the resonance occurs (see Fig. \ref{cylind_plots}). In the dipole approximation, the main contribution to the radiation comes from the first harmonic. Considering the family of trajectories resulting from the rotation of the initial trajectory of a charged particle around the $\spe_3$ axis, we see that the radiation from these trajectories at the first harmonic adds up constructively provided $m=1$ and destructively for all the other quantum numbers $m$. Therefore, in the dipole approximation, the right-handed helical undulator radiates mainly the twisted photons with $m=1$.

If one abandons the dipole approximation, i.e., one considers the case when the charged particle is relativistic in the reference frame moving with velocity $\be_\parallel$ along the $z$ axis, then the radiation of the undulator (the wiggler) is no longer concentrated at the first harmonic. As is known \cite{Bord.1}, the wiggler radiation spectrum extends to the harmonic number $n_{ext}\approx K^3$, and the intensity of radiation drops exponentially above this number. Therefore, as it follows from the considerations above and Fig. \ref{cylind_plots}, the wiggler radiates the twisted photons with the total angular momentum projection $m$ up to $n_{ext}$. Moreover, as for the forward radiation, the quantum number $m$ coincides with the harmonic number $n$, i.e., the forward radiation of an ideal right-handed helical wiggler at the $n$th harmonic consists of the twisted photons with the projection of the total angular momentum $m=n$ (cf. \cite{SasMcNu,BHKMSS,TaHaKa}). In the next section, we shall show this explicitly.

\subsection{Wiggler}\label{Wiggler}
\subsubsection{Helical wiggler}

Now we suppose that $K\gtrsim1$. Let us start with the forward radiation of a helical wiggler assuming that the electron moves exactly along a circle in the $(x,y)$ plane. Then, using the notation \eqref{traj_undul}, we have
\begin{equation}\label{helic_wigg}
    r_\pm=r_0e^{\pm i\omega t},\qquad r_3=0,\qquad\ups_\pm=0,
\end{equation}
where $r_0>0$ and
\begin{equation}\label{dip_par}
    K=\ga\omega r_0.
\end{equation}
Neglecting the radiation created at the entrance and exit points of the undulator, we can write the integrals \eqref{I_n0} as
\begin{equation}
\begin{split}
    I_3&:=\int_{-TN/2}^{TN/2}dt\ups_3 e^{-ik_0t(1-n_3\ups_3-\omega m)}J_m(k_\perp r_0),\\
    I_\pm&:=\mp\frac{n_\perp}{s\mp n_3}\int_{-TN/2}^{TN/2} dt\omega r_0e^{-ik_0t(1-n_3\ups_3-\omega m)}J_{m\mp1}(k_\perp r_0).
\end{split}
\end{equation}
Substituting these expressions into \eqref{probabil}, we arrive at
\begin{equation}\label{wigg_forw}
    dP=\frac{e^2}{4}\de^2_N(k_0(1-n_3\ups_3)-\omega m)\Big[\frac{n_3-\ups_3}{n_\perp}J_m\Big(\frac{mn_\perp K}{\ga(1-n_3\ups_3)}\Big) +s\omega r_0 J'_m\Big(\frac{mn_\perp K}{\ga(1-n_3\ups_3)}\Big) \Big]^2n_\perp dk_3dk_\perp.
\end{equation}
As a function of $k_0$, the average number of photons with given $n_\perp$ and $m$ possesses the sharp maximum at
\begin{equation}\label{spectr_wigg}
    k_0=\frac{m\omega}{1-n_3\ups_3}\approx \frac{2m\omega\ga^2}{1+K^2+n_\perp^2\ga^2},
\end{equation}
i.e., in this case, the principal quantum number (the harmonic number) coincides with the quantum number $m$ (see Fig. \ref{forw_dip_plots}).

Formula \eqref{wigg_forw} resembles the Schott formula for the spectral angular distribution of synchrotron radiation (see, e.g., \cite{Bord.1,LandLifshCTF.2,JacksonCE}). It is not surprising since the radiation at the wiggler axis can be considered as synchrotron one in the reference frame moving with velocity $\be_\parallel$ normally to the plane of motion of a charged particle (see, e.g., \cite{Bord.1,AlfBashCher,AfanMikh}). The estimates for the angular momentum of radiated electromagnetic waves in this case can be found, for example, in \cite{BordKN}. It is evident that a similar radiation is produced by the ultrarelativistic electrons moving helically along the magnetic flux lines provided that $\spe_3$ is in line with the direction of the magnetic field strength vector. Such electrons can be the electrons in the magnetic field of the Earth, the Sun, or the neutron stars. However, at a large distance from the radiation point, it seems impossible to record the twisted photons created in this way by the detector with a short base. In accordance with \eqref{wave_pckt}, \eqref{wp_disp}, the corresponding wave packets spread in the direction normal to the photon propagation direction with the velocity $n_\perp\sim\ga^{-1}$.

Large $|m|$ can be reached when the quantity
\begin{equation}\label{wig_largem}
    x:=1-\frac{n^2_\perp K^2}{\ga^2(1-n_3\ups_3)^2}\approx1-\frac{4n_\perp^2\ga^2K^2}{(1+K^2+n_\perp^2\ga^2)^2}=\frac{(1+(K+n_\perp\ga)^2)(1+(K-n_\perp\ga)^2)}{(1+K^2+n_\perp^2\ga^2)^2}\lesssim \frac{1}{20}.
\end{equation}
The optimal case for the fulfillment of this inequality is $n_\perp\ga=K$. Then, for large $K$,
\begin{equation}\label{x_arg}
    x\approx K^{-2},
\end{equation}
and \eqref{wig_largem} is satisfied for
\begin{equation}\label{large_K}
    K\gtrsim5.
\end{equation}
If the inequality \eqref{wig_largem} holds, we can employ the approximate formulas (see, e.g., \cite{Bord.1,NIST})
\begin{equation}\label{BessAi}
    J_m(m\sqrt{1-x})\approx \Big(\frac{2}{m}\Big)^{1/3} \Ai\Big(\big(\frac{m}{2}\big)^{2/3}x\Big),\qquad J'_m(m\sqrt{1-x})\approx -\Big(\frac{2}{m}\Big)^{2/3} \Ai'\Big(\big(\frac{m}{2}\big)^{2/3}x\Big),\qquad m\gtrsim5.
\end{equation}
When the argument of the Airy function or its derivative is greater than $1/2$, the magnitudes of these functions decline exponentially to zero. Therefore, at the optimum, $n_\perp\ga=K$, the radiation of modes with
\begin{equation}
    m\gtrsim\frac{K^3}{\sqrt{2}}
\end{equation}
is exponentially suppressed. If, at fixed $m$, the quantity $n_\perp\ga$ deviates from its optimal value such that $x\gtrsim1/2$, then the average number of photons tends exponentially to zero. Hence, the main contribution to $dP$ comes from $n_\perp\ga$ close to $K$. It is an expected result since the main part of the radiation produced by the helical wiggler propagates at the angle $\ga\theta\approx K$ \cite{Bord.1}.

Let us find the number of photons produced in the mode $s$, $m$, $k_0$ with $k_\perp$ determined by \eqref{spectr_wigg}. For given $k_0$ and $m$, the projection $k_\perp$ is uniquely defined by Eq. \eqref{spectr_wigg}. However, the reverse is not true: for given $k_\perp$ and $m$, Eq. \eqref{spectr_wigg} gives the two values of the photon energy
\begin{equation}
    k^0_\pm=\frac{m\omega\ga^2}{1+K^2}\Big(1\pm\sqrt{1-\frac{(1+K^2)k_\perp^2}{m^2\omega^2\ga^2}}\Big).
\end{equation}
For example, for the optimal value $n_\perp\ga=K$,
\begin{equation}\label{k0pm_opt}
    k^0_+=\frac{2m\omega\ga^2}{1+2K^2},\qquad k_\perp=K\frac{2m\omega\ga}{1+2K^2}.
\end{equation}
However, in accordance with \eqref{spectr_wigg}, there is another one value of the photon energy corresponding to given $k_\perp$ and $m$:
\begin{equation}
    k^0_-=\frac{K^2}{1+2K^2}\frac{2m\omega\ga^2}{1+2K^2}\;\;\Rightarrow\;\;n_\perp\ga=K+K^{-1}.
\end{equation}
The distance between these peaks is
\begin{equation}
    k^0_+-k^0_-=\frac{2m\omega\ga^2}{1+K^2}\sqrt{1-\frac{(1+K^2)k_\perp^2}{m^2\omega^2\ga^2}}=\frac{2m\omega\ga^2}{(1+K^2)(1+2K^2)}\approx\frac{m\omega\ga^2}{K^4},
\end{equation}
where, in the second equality, we have taken $k_\perp$ from \eqref{k0pm_opt}. In that case,
\begin{equation}\label{twin_peaks}
    (k^0_+-k^0_-)/k_+^0\approx K^{-2}.
\end{equation}
Further, we suppose that these two peaks are sufficiently resolved and shall find the number of photons falling into one peak with $k_\perp$, $k^0_+$.

In order to estimate the number of photons, we multiply the magnitude of the average number of photons at the peak by the line width. The measure in \eqref{wigg_forw} is transformed in the usual manner
\begin{equation}\label{measure_k0}
    dk_3dk_\perp=\frac{k_0dk_0dk_\perp}{\sqrt{k_0^2-k_\perp^2}}=\frac{dk_0dk_\perp}{n_3}.
\end{equation}
The function $\sin^2(\pi Nx/2)/(\pi x)^2$ appearing in \eqref{wigg_forw} is equal to $N^2/4$ in its maximum. The effective width of its main peak is
\begin{equation}
    \frac{4\Sif(2\pi)}{\pi N}=:\frac{c_0}{N}\approx\frac{1.8}{N},
\end{equation}
where $\Sif(x)$ is the sine integral. We have defined here the effective width from the requirement that the total area under the main peak equals the maximum value of the function multiplied by the effective width. Taking the differential of the expression standing in the argument of $\de_N(x)$, we obtain
\begin{equation}\label{widths}
    \De k_0=\frac{c_0}{N}\frac{n_3\omega}{n_3-\ups_3},\qquad\De k_\perp=\frac{c_0}{N}\frac{n_3\omega}{n_\perp\ups_3}.
\end{equation}
Further, we assume that
\begin{equation}
    n_3-\ups_3=\frac{1+K^2-n_\perp^2\ga^2}{2\ga^2}>0.
\end{equation}
In particular, this inequality is fulfilled for $n_\perp\ga=K$. It follows from \eqref{widths} that
\begin{equation}
    \frac{\De k_0}{k_0}\approx\frac{c_0}{N}\frac{n_3}{m}\frac{1+K^2+n_\perp^2\ga^2}{1+K^2-n_\perp^2\ga^2}.
\end{equation}
The number of the wiggler sections $N$ and the harmonic number $m$ should be such that this ratio is less than \eqref{twin_peaks}. Substituting \eqref{BessAi}, \eqref{measure_k0}, and \eqref{widths} into \eqref{wigg_forw}, we find the average number of photons with the quantum numbers $s$, $m$ produced in the peak \eqref{spectr_wigg}:
\begin{equation}\label{phot_num}
    \De P=e^2\frac{c_0^2}{16}\frac{n_3(n_3-\ups_3)}{n_\perp^2\ups_3}\Big[\Big(\frac{2}{m}\Big)^{1/3}\Ai\Big(\big(\frac{m}{2}\big)^{2/3}x\Big) -\frac{sKn_\perp}{\ga(n_3-\ups_3)}\Big(\frac{2}{m}\Big)^{2/3} \Ai'\Big(\big(\frac{m}{2}\big)^{2/3}x\Big) \Big]^2.
\end{equation}
Notice that the number of photons does not depend on $N$.

Let us derive a simpler expression for the average number of twisted photons for $n_\perp\ga=K$. In this case, the second term in the square brackets in \eqref{phot_num} dominates. Therefore, taking into account \eqref{x_arg}, \eqref{large_K}, we have approximately
\begin{equation}
    \De P\approx e^2\frac{c_0^2}{8} K^2 \Big(\frac{2}{m}\Big)^{4/3} \Ai'\Big(\big(\frac{m}{2}\big)^{2/3}K^{-2}\Big)\approx e^2\frac{c_0^2}{8}\frac{(4/3)^{2/3}}{\Ga^{2}(1/3)}\frac{K^2}{m^{4/3}}\approx2\cdot10^{-3}\frac{K^2}{m^{4/3}},
\end{equation}
where it is assumed that $m\lesssim K^3/\sqrt{2}$. The number of photons in the peak \eqref{spectr_wigg} drops with increasing the harmonic number and, correspondingly, with increasing the quantum number $m$. This agrees with the known property of the wiggler radiation that the maximum of radiation spectrum is at the first harmonic \cite{Bord.1}. The efficiency of radiation of twisted photons grows quadratically with increasing the undulator strength parameter \eqref{dip_par} and virtually independent of the photon helicity.

So far we have investigated the forward radiation of an ideal helical wiggler, where the electron moves along the trajectory \eqref{traj_undul}, \eqref{circular_ondul} with $a_z=0$ and $a_x=-a_y=r_0$. Now we turn to the radiation at an angle to the undulator axis. For the same trajectory and the basis \eqref{basis1}, we have
\begin{equation}
    \dot{x}_3=1,\qquad\dot{x}_\pm=-\theta\pm i\frac{K}{\ga} e^{\pm i(\omega t-\vf)},
\end{equation}
in the leading order in $K/\ga$. As is well known \cite{Bord.1}, the most part of the wiggler radiation is concentrated in the cone with the opening angle $K/\ga$, $K/\ga\ll1$, and so we suppose that $\theta\lesssim K/\ga$ and $n_\perp\lesssim K/\ga$. On substituting the representation \eqref{Bessel_int1} of the Bessel functions into the integrals \eqref{I_n0} and saving only the leading terms in $K/\ga$, the following expression in the exponent arises
\begin{equation}
    -im\psi-ik_0\Big[t\frac{1+K^2+(n_\perp^2+\theta^2-2n_\perp\theta\cos\psi)\ga^2}{2\ga^2} -\frac{K}{\omega\ga}\theta\cos(\omega t-\vf) +\frac{K}{\omega\ga}n_\perp\cos(\omega t-\vf+\psi)\Big],
\end{equation}
where $\psi\in[-\pi,\pi]$ is in the integration variable. Introducing the notation,
\begin{equation}
\begin{gathered}
    \chi(\psi):=\frac{k_0K}{\omega\ga}\sqrt{n_\perp^2-2n_\perp\theta\cos\psi+\theta^2},\\
    \sin\de(\psi):=\frac{\theta-n_\perp\cos\psi}{\sqrt{n_\perp^2-2n_\perp\theta\cos\psi+\theta^2}},\qquad \cos\de(\psi):=\frac{n_\perp\sin\psi}{\sqrt{n_\perp^2-2n_\perp\theta\cos\psi+\theta^2}},
\end{gathered}
\end{equation}
we can expand the $T$ periodic functions in the integrand of $I_3$ in the Fourier series
\begin{equation}
    e^{i\chi\sin(\omega t-\vf+\de)}=\sum_{n=-\infty}^\infty e^{in(\omega t-\vf+\de)}J_n(\chi).
\end{equation}
Then the integrals over $t$ in \eqref{I_n0} are readily evaluated. The result can be cast into the form
\begin{equation}\label{I_n_wigg_hel}
\begin{split}
    I_3&=i^m\sum_{n=1}^\infty \int_{-\pi}^\pi d\psi e^{-im\psi}\de_N\big[\frac{\omega}{2}(a_n+b_n)(\cos\psi-\cos\xi_n)\big]e^{in(\de-\vf)}J_n ,\\
    I_\pm&=\pm i^m\frac{n_\perp}{s\mp n_3}\sum_{n=1}^\infty \int_{-\pi}^\pi d\psi e^{-i(m\mp1)\psi}\de_N\big[\cdots\big]e^{in(\de-\vf)} \Big(-\theta J_n\pm i\frac{K}{\ga}e^{\mp i\de}J_{n\mp1} \Big) ,
\end{split}
\end{equation}
where $\xi_n$, $a_n$, $b_n$ are defined in \eqref{anbn}, \eqref{xin},  $J_n\equiv J_n(\chi)$, and the argument of $\de_N(x)$ in the formula on the second line is the same as the argument of this functions on the first line. The terms with $n\leq0$ have been neglected since they are suppressed at large $N$.

\begin{figure}[!t]
\centering
a)\;\includegraphics*[align=c,width=0.6\linewidth]{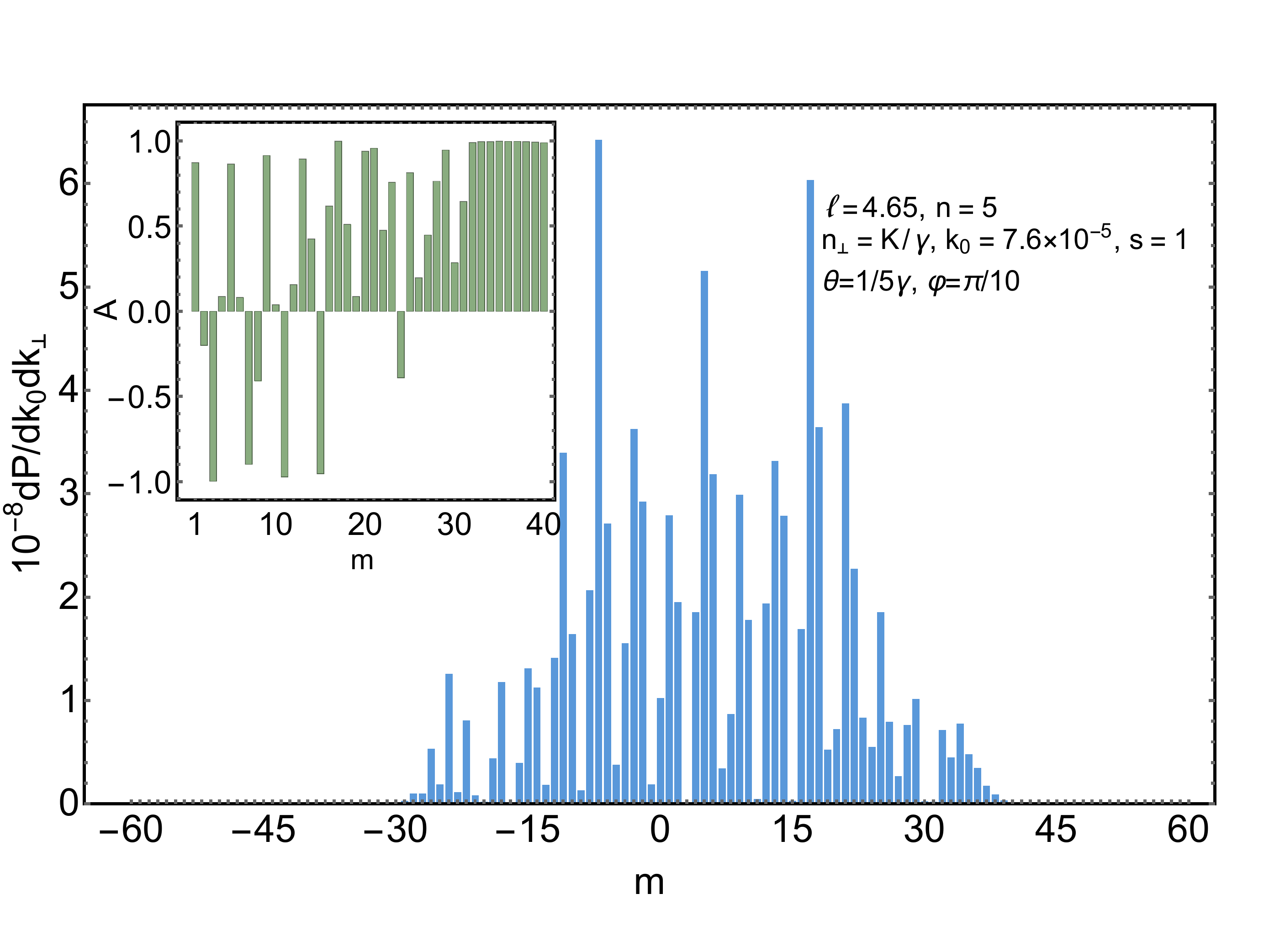}\\
b)\;\includegraphics*[align=c,width=0.4\linewidth]{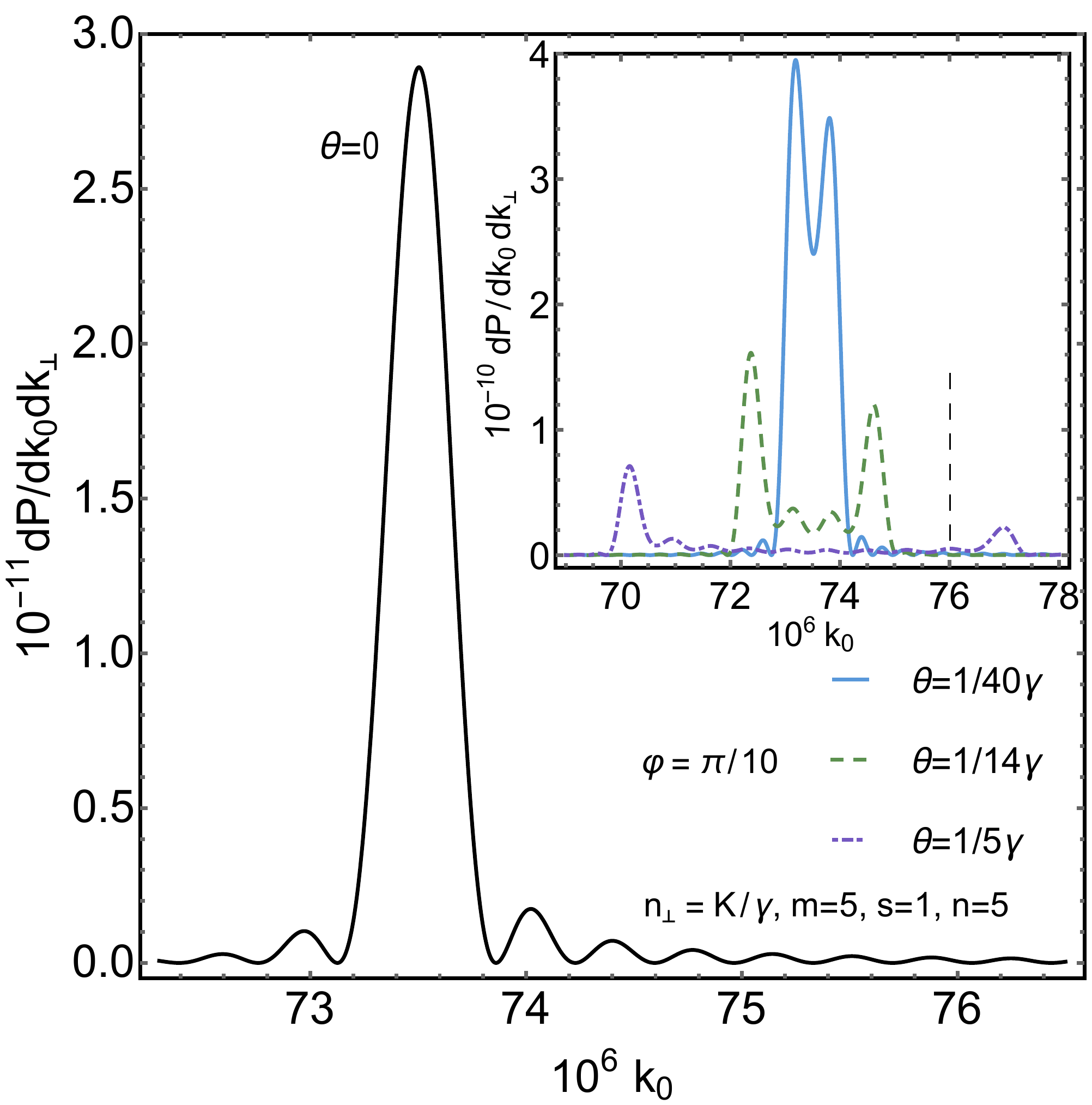}\;\;
c)\;\includegraphics*[align=c,width=0.39\linewidth]{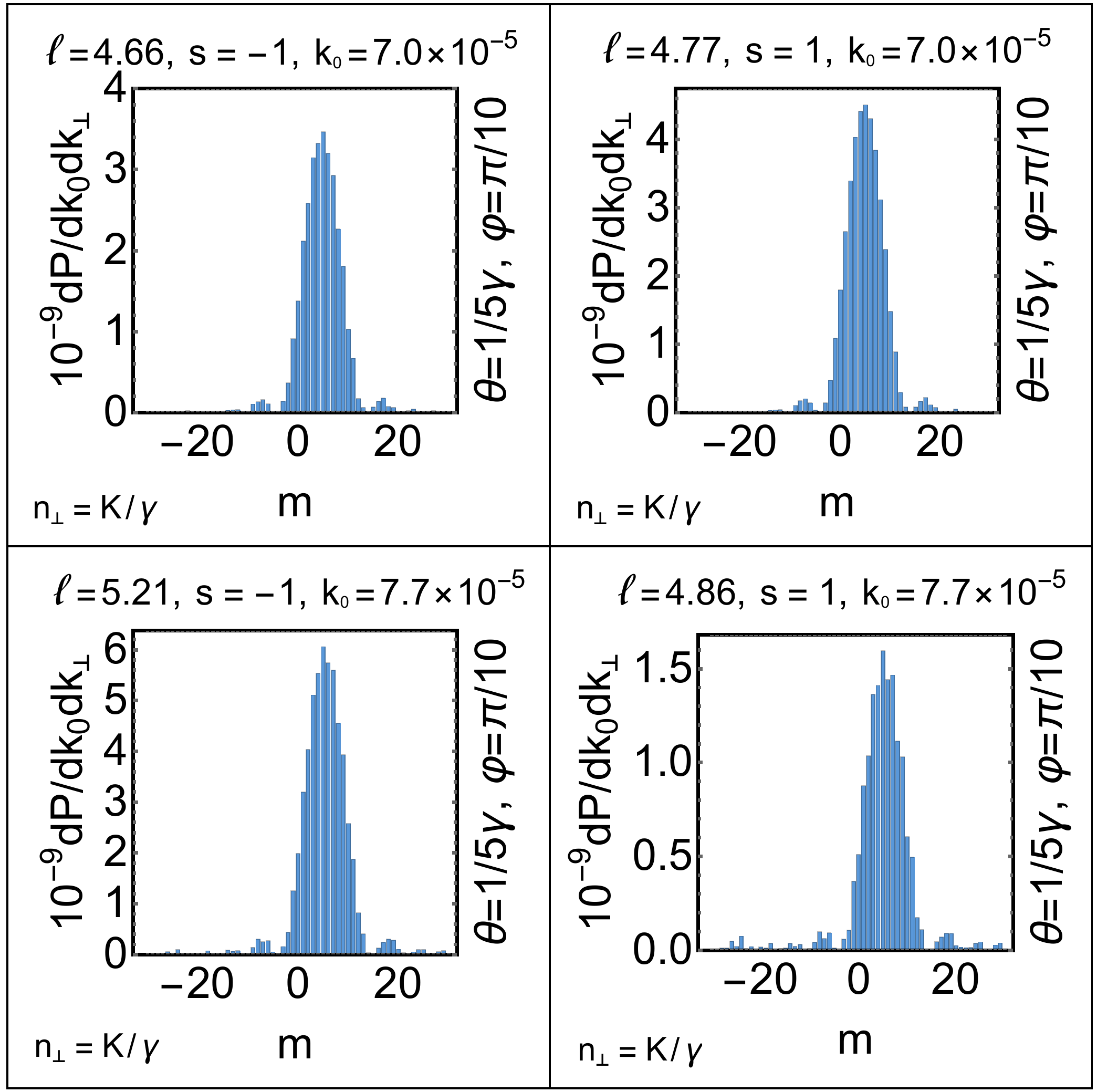}
\caption{{\footnotesize The radiation of twisted photons by the helical wiggler at the fifth harmonic. The trajectory of the electron is taken in the form \eqref{helic_wigg} with the undulator strength parameter $K=4$, $\omega=2\pi\be_\parallel/\la_0$, where $\la_0=1$ cm is the length of the undulator section, and $\ga=10^3$ is the Lorentz-factor of the electron. The number of the undulator sections $N=40$. The energy of photons is measured in the units of the rest energy of the electron $0.511$ MeV. (a) The distribution over $m$, the asymmetry, and the angular momentum projection per one photon. In accordance with \eqref{m_period}, the period of oscillations $T_m=4$. (b) The density of the average number of twisted photons against $k_0$ for the different observation angles. The position of the peak in the forward radiation and the boundaries of the spectral band in the inset are well described by \eqref{energy_ints}. The dashed vertical line in the inset depicts the photon energy used in the plot (a). (c) The density of the average number of twisted photons against $m$ and the angular momentum projection per one photon at the left, $\xi_n=\pi$, and right, $\xi_n=0$, peaks appearing in the distribution over the photon energy for $\theta=1/(5\ga)$ (see the inset in the plot (b)).}}
\label{hel_wig_plots}
\end{figure}

As in the dipole approximation, there are the three cases: the regular case, the weakly degenerate case, and the strongly degenerate one (the forward radiation). The last case has been already investigated. So we are left with the two cases. In deriving the analytic expression for the average number of twisted photons, we shall assume in the rest cases that $N$ is so large that we can remove all the functions standing at $\de_N(x)$ out of the integral sign and take them at those $\psi$'s where the argument of $\de_N(x)$ vanishes.

In the regular case, we have
\begin{equation}\label{points_regul}
    \psi=\pm\xi_n.
\end{equation}
The function $\de_N(x)$ can be replaced by the delta-function provided that
\begin{equation}
    \Big|\frac{(n_{\bot}-\theta\cos\xi_{n})n_{\bot} n}{n^{2}_{\bot}+\theta^{2}-2 n_{\bot}\theta\cos\xi_{n}}-m\Big|\ll\pi \sqrt{a_{n}b_{n}} N.
\end{equation}
Taking into account the contributions of the points \eqref{points_regul}, we arrive at
\begin{equation}
\begin{split}
    I_3=&2\sum_{n=1}^\infty \frac{\theta(a_n)\theta(b_n)}{\omega\sqrt{a_nb_n}}i^{m+n} e^{-in\vf}\cos\big(m\xi_n-n\de_n+\frac{\pi n}{2}\big) J_n(\chi_n) ,\\
    I_\pm=&\pm \frac{2n_\perp}{s\mp n_3}\sum_{n=1}^\infty \frac{\theta(a_n)\theta(b_n)}{\omega\sqrt{a_nb_n}}i^{m+n} e^{-in\vf} \Big\{-\theta\cos\big[(m\mp1)\xi_n-n\de_n+\frac{\pi n}{2}\big] J_n(\chi_n)\pm\\
    &\pm \frac{K}{\ga}\sin\big[(m\mp1)\xi_n-(n\mp1)\de_n +\frac{\pi n}{2}\big] J_{n\mp1}(\chi_n) \Big\} ,
\end{split}
\end{equation}
where $\de_n:=\de(\xi_n)$ and $\chi_n:=\chi(\xi_n)$. The photon energy belongs to the intervals \eqref{energy_ints}, which overlap starting with the harmonic number \eqref{n0over}. Now it is not difficult to obtain the average number of photons \eqref{probabil}. However, the resulting expression is rather huge, and we do not write it here. Notice that, for those harmonics where the energy intervals \eqref{energy_ints} do not overlap or one can neglect this overlapping, the average number of photons is independent of $\vf$ and has the form \eqref{m_depend_reg}. It is a periodic function of $m$ with the period \eqref{m_period}. If the harmonics overlap, then the both properties are violated. In particular, the average number of photons becomes a nontrivial function of $\vf$.

In the weakly degenerate case, $\xi_n=\{0,\pi\}$, which corresponds to $k_0\approx n\omega_+$ and $k_0\approx n\omega_-$, respectively. In the case when $\xi_n\approx0$,
\begin{equation}
    \psi\approx0,\qquad a_n\approx0,\qquad b_n\approx n(\omega_+\omega^{-1}_--1),
\end{equation}
and
\begin{equation}
    \de_N[\cdots]\approx\frac{\sin\big[\pi Nn(\omega_+\omega_-^{-1}-1)\psi^2/4\big]}{\omega\pi n(\omega_+\omega_-^{-1}-1)\psi^2/4}.
\end{equation}
Then,
\begin{equation}
    \int_{-\pi}^\pi d\psi\de_N[\cdots]\approx\omega^{-1}\sqrt{\frac{8N}{n(\omega_+\omega_-^{-1}-1)}},
\end{equation}
in the leading order in $1/N$. As for the other integrand functions in \eqref{I_n_wigg_hel}, we remove them out of the integral sign and set $\psi=0$ in their arguments. This is justified when
\begin{equation}
    \Big|\Delta\psi\Big(\frac{n_{\bot}n}{n_{\bot}-\theta}-m\Big)\Big|\ll 1, \qquad \Delta\psi\approx\Big(\frac{\pi n N}{4}(\omega_{+}\omega_{-}^{-1}-1)\Big)^{-1/2}.
\end{equation}
Thus, we come to
\begin{equation}
\begin{split}
    I_3&= \omega^{-1}\sqrt{\frac{8N}{n(\omega_+\omega_-^{-1}-1)}} i^{m+n}e^{-in\vf}J_n\Big(n\frac{\omega_+K}{\omega\ga}(\theta-n_\perp)\Big),\\
    I_\pm&=\pm \frac{n_\perp}{s\mp n_3}\omega^{-1}\sqrt{\frac{8N}{n(\omega_+\omega_-^{-1}-1)}} i^{m+n}e^{-in\vf} \big[-\theta J_n(\cdots)+\frac{K}{\ga}J_{n\mp1}(\cdots)\big].
\end{split}
\end{equation}
The arguments of all the Bessel functions are the same. After a little algebra, the average number of twisted photons \eqref{probabil} is reduced to
\begin{equation}
    dP=\frac{e^2}{\omega^2\ga^2}\frac{Nn_\perp K^2}{n(\omega_+\omega_-^{-1}-1)}\Big[\frac{1+K^2-\ga^2(\theta-n_\perp)^2}{2\ga K(\theta-n_\perp)}J_n(\cdots) +sJ'_n(\cdots) \Big]^2\frac{dk_3dk_\perp}{2\pi^2},
\end{equation}
and does not depend on $m$ in the range of applicability of the approximations made.

In the case, when $\xi_n\approx\pi$,
\begin{equation}
    \psi\approx\pi,\qquad a_n\approx n(1-\omega_-\omega^{-1}_+),\qquad b_n\approx 0,
\end{equation}
and
\begin{equation}
    \de_N[\cdots]\approx\frac{\sin\big[\pi Nn(1-\omega_-\omega_+^{-1})(\psi-\pi)^2/4\big]}{\omega\pi n(1-\omega_-\omega_+^{-1})(\psi-\pi)^2/4}.
\end{equation}
Proceeding in the same way as in the case just considered, we have
\begin{equation}\label{probabil_wigg_hel}
    dP=\frac{e^2}{\omega^2\ga^2}\frac{Nn_\perp K^2}{n(1-\omega_-\omega_+^{-1})}\Big[\frac{1+K^2-\ga^2(\theta+n_\perp)^2}{2\ga K(\theta+n_\perp)}J_n(\chi_n) +sJ'_n(\chi_n) \Big]^2\frac{dk_3dk_\perp}{2\pi^2},
\end{equation}
where
\begin{equation}
    \chi_n:=n\frac{\omega_-K}{\omega\ga}(\theta+n_\perp).
\end{equation}
The expression \eqref{probabil_wigg_hel} is also independent of the quantum number $m$. The applicability condition reads as
\begin{equation}
    \Big|\Delta\psi\Big(\frac{n_{\bot}n}{n_{\bot}+\theta}-m\Big)\Big|\ll 1, \qquad \Delta\psi\approx\Big(\frac{\pi n N}{4}(1-\omega_{-}\omega_{+}^{-1})\Big)^{-1/2}.
\end{equation}
Let us stress once again that the formulas above are obtained under the assumption that the number of the undulator sections $N$ is large. The accuracy of the analytical formulas and of the corresponding implications is increased with increasing $N$. The plots of the density of the average number of twisted photons are presented in Fig. \ref{hel_wig_plots}.

\subsubsection{Planar wiggler}

Now we consider the planar wiggler. We assume that the electron trajectory is of the form \eqref{traj_undul} with
\begin{equation}\label{planar_wiggl}
    r_x=0,\qquad r_y=\frac{\sqrt{2}\be_\parallel K}{\ga\omega}\sin(\omega t),\qquad r_z=-\frac{\be_\parallel K^2}{4\omega\ga^2}\sin(2\omega t),
\end{equation}
where $K\gg1$, but $K/\ga\ll1$ (see, for details, \cite{Bord.1}). For further analysis of the radiation of twisted photons, it is convenient to use the basis
\begin{equation}
    \spe_3=(\cos\de \sin\al,\sin\de,\cos\de\cos\al),\qquad \spe_1=(-\sin\de\sin\al,\cos\de,-\sin\de\cos\al),\qquad \spe_2=(-\cos\al,0,\sin\al),
\end{equation}
and introduce the notation
\begin{equation}\label{reangles}
    a:=\al \ga/K,\qquad d:=\de \ga/K,\qquad n_k:=n_\perp \ga/K.
\end{equation}
As long as the main part of the wiggler radiation is concentrated in the cone with the opening angle $\theta\ga\lesssim K$, the magnitudes of \eqref{reangles} are of the order of unity or less, while $\al$, $\be$, and $n_\perp$ are much less than unity. Notice that the detector lies in the orbit plane when $\al=0$.

In the leading order in $K/\gamma$, we have
\begin{equation}\label{velocities}
    \dot{x}_3=1,\qquad \dot{x}_\pm=\frac{K}{\ga}[\pm ia+d+\sqrt{2}\cos(\omega t)].
\end{equation}
Let us start with evaluation of the integral $I_3$ in \eqref{I_n0}. The integrals $I_\pm$ are found analogously. Substituting the integral representation \eqref{Bessel_int1} into $I_3$ and denoting $\tau:=\omega t$, the expression in the exponent in $I_3$ can be cast into the form
\begin{equation}
    -im\vf-i\frac{k_0K^2}{2\omega\ga^2}\big[\tau(K^{-2}+1+(d+n_k\sin\vf)^2+(a-n_k\cos\vf)^2) +\sin\tau(\cos\tau-2\sqrt{2}(d+n_k\sin\vf))\big],
\end{equation}
in the leading order in $K/\ga$. Then we represent the $2\pi$ periodic functions in the integrand of $I_3$ as a Fourier series
\begin{equation}\label{Four_ser}
    e^{-i\frac{k_0K^2}{2\omega\ga^2}\sin\tau(\cos\tau-2\sqrt{2}(d+n_k\sin\vf))}=\sum_{n=-\infty}^\infty c_n(\vf) e^{in\tau},
\end{equation}
where
\begin{equation}\label{cn}
    c_n(\vf)=\int_{-\pi}^\pi\frac{d\tau}{2\pi}e^{-in\tau-i\frac{k_0K^2}{2\omega\ga^2}\sin\tau(\cos\tau-2\sqrt{2}(d+n_k\sin\vf))} =J_n\Big(\frac{k_0K^2}{\omega\ga^2}\sqrt{2}(d+n_k\sin\vf),-\frac{k_0K^2}{4\omega\ga^2}\Big),
\end{equation}
and $J_n(x,y)$ is a generalized Bessel function of two arguments (see, e.g., \cite{Bord.1,Diden79,NikRit64,Dattol90,Dattol91}). Notice that $c_n(\vf)\in \mathbb{R}$. Substituting \eqref{velocities}, \eqref{Four_ser} into $I_3$ and neglecting the radiation produced at the entrance and exit points of the wiggler, we obtain
\begin{equation}\label{I3_wigg_pl}
    I_3=\sum_{n=-\infty}^\infty\int_{-\pi}^\pi d\vf e^{-im\vf}\de_N\Big[\frac{k_0K^2}{2\ga^2}(K^{-2}+1+(d+n_k\sin\vf)^2+(a-n_k\cos\vf)^2)-\omega n\Big]c_n(\vf).
\end{equation}
For large $N$, the contributions of the terms with $n\leq0$ are suppressed. Therefore, we retain only the terms with $n\in \mathbb{N}$ in what follows. As for $I_\pm$, similar calculations lead to
\begin{equation}\label{Ipm_wigg_pl}
    I_\pm=i\frac{s\pm n_3}{n_k}\sum_{n=1}^\infty\int_{-\pi}^\pi d\vf e^{-i(m\mp1)\vf}\de_N\big[\cdots\big][(\pm ia-d)c_n+\frac1{\sqrt{2}}(c_{n-1}+c_{n+1})],
\end{equation}
where the argument of $\de_N(x)$ is the same as in \eqref{I3_wigg_pl}. To the same accuracy that we are carrying out the calculations, we can put $n_3=1$ in \eqref{Ipm_wigg_pl}. This means that we can neglect $I_-$ for $s=1$ and $I_+$ for $s=-1$.

Let us introduce the notation (cf. \eqref{omega_pm})
\begin{equation}
    \omega_\pm:=\frac{2\omega\ga^2}{1+K^2+K^2(\sqrt{a^2+d^2}\mp n_k)^2},
\end{equation}
and
\begin{equation}
    \sin\vf_0:=\frac{-d}{\sqrt{a^2+d^2}},\qquad \cos\vf_0:=\frac{a}{\sqrt{a^2+d^2}}.
\end{equation}
Then the energy of radiated photons satisfies \eqref{energy_ints}. The energy intervals overlap starting with the harmonic number
\begin{equation}
    n_0=\frac{K^{-2}+1+(\sqrt{a^2+d^2}-n_k)^2}{4n_k\sqrt{a^2+d^2}}.
\end{equation}
The argument of $\de_N(x)$ can be written as
\begin{equation}\label{den_arg}
    \frac{2\omega n n_k\sqrt{a^2+d^2}[\cos\xi_n-\cos(\vf-\vf_0)]}{K^{-2}+1+a^2+d^2+n_k^2-2n_k\sqrt{a^2+d^2}\cos\xi_n},
\end{equation}
where $\xi_n$ is defined in \eqref{xin}, and $a_n$, $b_n$ appearing in the definition of $\xi_n$ are presented in \eqref{anbn}. We see from \eqref{den_arg} that, as in the case of the dipole approximation, there are the three cases: regular, weakly degenerate, and strongly degenerate. In the first two cases, $\de_N(x)$ removes the integral over $\vf$ for not very large $|m|$, and in the last case (the forward radiation) $\de_N(x)$ weakly depends on $\vf$ or is independent of it.

Before proceeding to the analysis of these three cases, we derive the approximate expression for $c_n(\vf)$ analogous to the approximate relations \eqref{BessAi}. In order to evaluate approximately $c_n(\vf)$ for $n\gtrsim5$ and $k_0\in n[\omega_-,\omega_+]$, we can employ the steepest descent method. The expression in the exponent in \eqref{cn} possesses the four stationary points in the strip $\re\tau\in[-\pi,\pi]$ that are placed symmetrically with respect to the real and imaginary axes. The contributions of these stationary points are not strongly exponentially suppressed only if these points approach closely the real axis. In this case, the stationary points become degenerate, viz., $\ddot{f}(\tau)\approx0$ in these points, where $f(\tau)$ is the expression in the exponent. Inasmuch as the integral is saturated in a small neighbourhood of the stationary points, we can develop the expression in the exponent as a Taylor series in $(\tau-\tau_0)$ in the vicinity of the point $\ddot{f}(\tau_0)=0$ and keep only the terms of the order $(\tau-\tau_0)^3$, inclusive. The condition $\ddot{f}=0$ leads to
\begin{equation}\label{statio_p}
    \cos\tau_0=\frac{d+n_k\sin\vf}{\sqrt{2}},\qquad \sin\tau_0=\pm\sqrt{1-\frac{(d+n_k\sin\vf)^2}{2}},
\end{equation}
and the exponential quantity in \eqref{cn} becomes
\begin{equation}
    -i\Big[n\tau_0-\frac{3k_0K^2}{4\omega\ga^2}\sin2\tau_0 +\big(n-\frac{k_0K^2}{2\omega\ga^2}(1+2\cos^2\tau_0)\big)(\tau-\tau_0) +\frac{k_0K^2}{\omega\ga^2}\sin^2\tau_0\frac{(\tau-\tau_0)^3}{3}\Big],
\end{equation}
in the neighbourhood of two points \eqref{statio_p} each. Having deformed the contour according to the steepest descent, the integral with respect to $\tau$ is performed in the infinite limits. Hence, the variable $\tau$ can be shifted safely by $\tau_0$. As a result, taking into account that
\begin{equation}
    \Ai(x)=\int\frac{dt}{2\pi}e^{-i(xt+t^3/3)},
\end{equation}
we obtain
\begin{equation}\label{cnphi}
\begin{split}
    c_n(\vf)&\approx2\Big(\frac{k_0K^2}{\omega\ga^2}\sin^2\tau_0\Big)^{-1/3} \cos\big[n\tau_0-\frac{3k_0K^2}{4\omega\ga^2}\sin2\tau_0\big]\Ai(B_n(\vf)),\\ B_n(\vf)&:=\frac{n-\frac{k_0K^2}{2\omega\ga^2}(1+2\cos^2\tau_0)}{\Big(\frac{k_0K^2}{\omega\ga^2}\sin^2\tau_0\Big)^{1/3}},
\end{split}
\end{equation}
where it is supposed that $\tau_0$ defined in \eqref{statio_p} is real and $\sin\tau_0\geq0$.

Let us begin with the regular case. The replacement of $\de_N(x)$ by the delta-function in the integrals \eqref{I3_wigg_pl}, \eqref{Ipm_wigg_pl} is justified when the peak width of $\de_N(x)$,
\begin{equation}\label{peak_width}
    \De\vf\approx\frac{1+a^2+d^2+n_k^2-2n_k\sqrt{a^2+d^2}\cos\xi_n}{2\pi nNn_k\sqrt{a^2+d^2}|\sin\xi_n|},
\end{equation}
is much smaller than the characteristic scale of variation of the rest integrand functions in \eqref{I3_wigg_pl}, \eqref{Ipm_wigg_pl} at the points
\begin{equation}
    \vf^\pm_n:=\vf_0\pm\xi_n.
\end{equation}
This requirement entails the restrictions
\begin{equation}\label{regul_cond1}
    |m|\De\vf\ll1,\qquad \frac{k_0K^2}{2\omega\ga^2}(a-n_k\cos\vf)^2\frac{n_k|\cos\vf|}{\sqrt{2}|\sin\tau_0|}\De\vf\ll1,
\end{equation}
where $\vf=\vf^\pm_n$. The second condition is the requirement of a small variation of $c_n(\vf)$ on the scale $\De\vf$ in the vicinity of the points $\vf^\pm_n$. Expressing $k_0$ in terms of $\xi_n$, we can write this condition as
\begin{equation}\label{regul_cond2}
    \frac{(a-n_k\cos\vf^\pm_n)^2}{2\pi nN\sqrt{a^2+d^2}}\frac{n_k|\cos\vf^\pm_n|}{\sqrt{2}|\sin\tau_0|}\ll|\sin\xi_n|.
\end{equation}
As we see, the conditions \eqref{regul_cond1}, \eqref{regul_cond2} are violated in the neighbourhood of the points $\sin\xi_n=0$, which corresponds to the weakly degenerate case. The both conditions \eqref{regul_cond1}, \eqref{regul_cond2} are also violated at small $\sqrt{a^2+d^2}$, which corresponds to the strongly degenerate case.

\begin{figure}[!t]
\centering
a)\;\includegraphics*[align=c,width=0.6\linewidth]{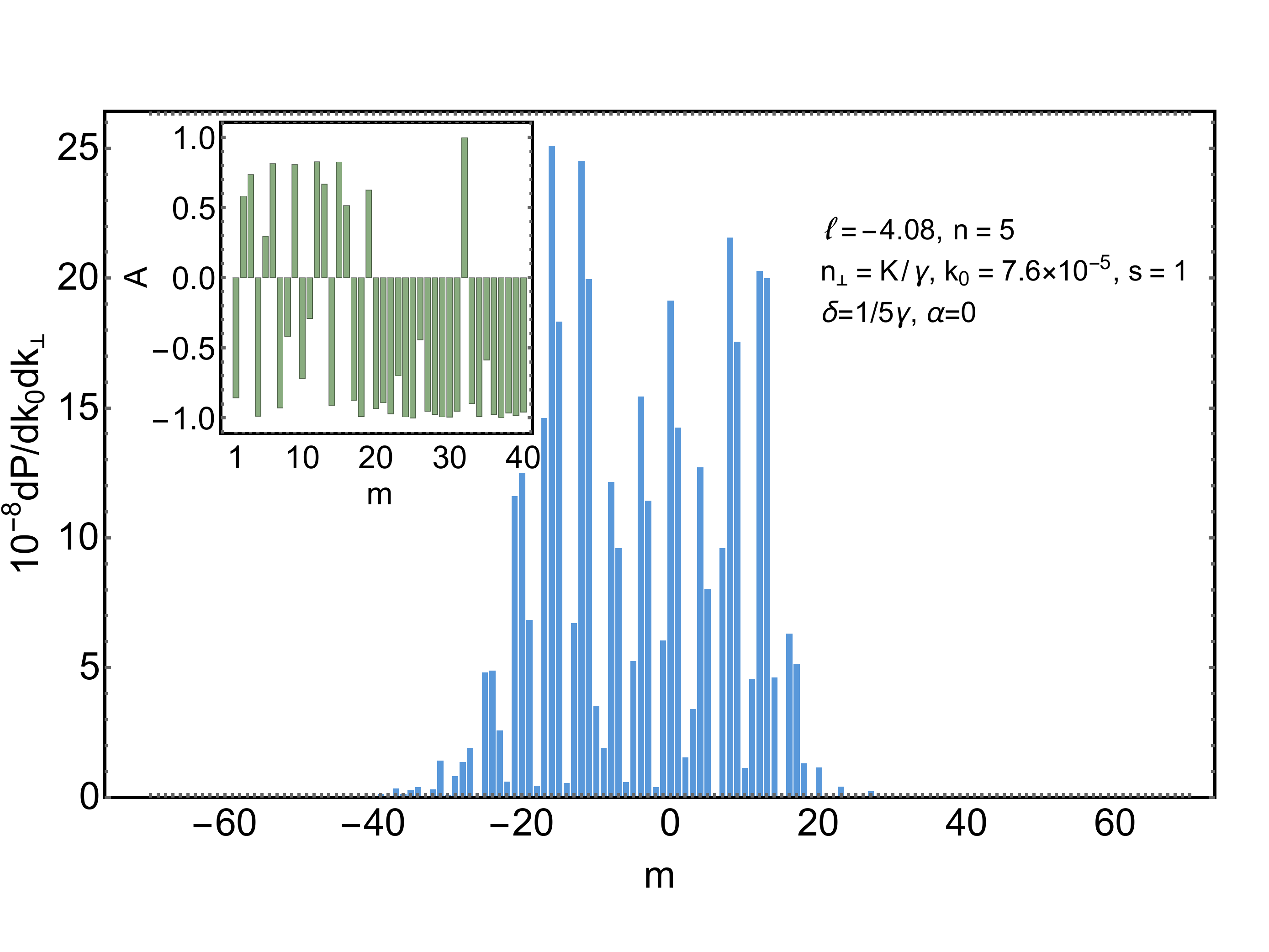}\\
b)\;\includegraphics*[align=c,width=0.4\linewidth]{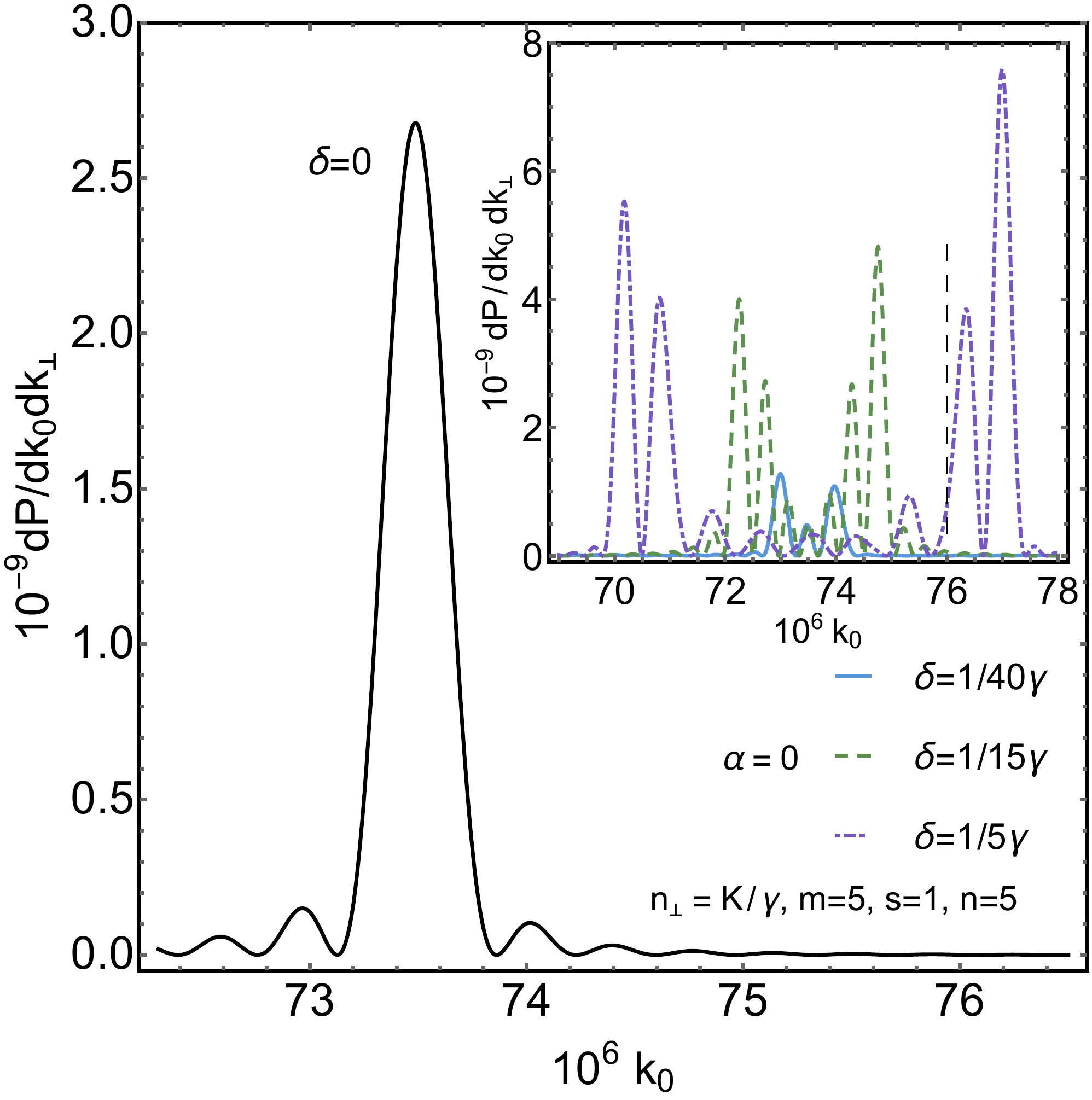}\;\;
c)\;\includegraphics*[align=c,width=0.39\linewidth]{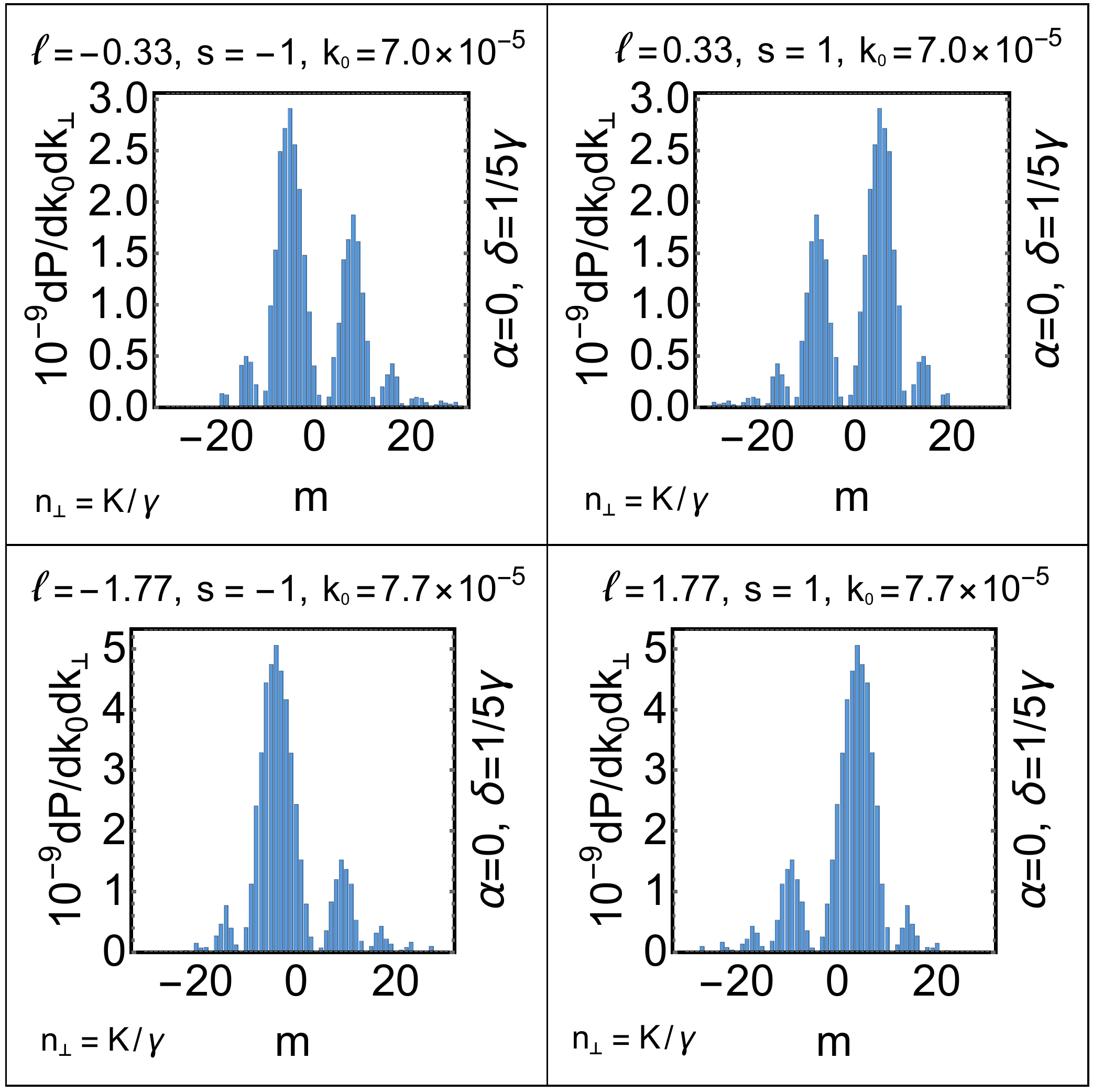}
\caption{{\footnotesize The radiation of twisted photons by the planar wiggler at the fifth harmonic. The trajectory of the electron is taken in the form \eqref{planar_wiggl} with the undulator strength parameter $K=4$, $\omega=2\pi\be_\parallel/\la_0$, where $\la_0=1$ cm is the length of the undulator section, and $\ga=10^3$ is the Lorentz-factor of the electron. The number of the undulator sections $N=40$. The energy of photons is measured in the units of the rest energy of the electron $0.511$ MeV. (a) The distribution over $m$, the asymmetry, and the angular momentum projection per one photon. In accordance with \eqref{m_period}, the period of oscillations $T_m=4$. (b) The density of the average number of twisted photons against $k_0$ for the different observation angles. The position of the peak in the forward radiation and the boundaries of the spectral band in the inset are well described by \eqref{energy_ints}. The dashed vertical line in the inset depicts the photon energy used in the plot (a). (c) The density of the average number of twisted photons against $m$ and the angular momentum projection per one photon at the left, $\xi_n=\pi$, and right, $\xi_n=0$, peaks appearing in the distribution over the photon energy for $\theta=1/(5\ga)$ (see the inset in the plot (b)). The presence of sufficiently high lateral peaks on the upper plots is a consequence of a comparatively large width \eqref{peak_width_sing} of the main peak of $\de_N(x)$. These lateral peaks disappear with increasing $N$. As long as the observation angle $\al=0$, the distributions over $m$ have the symmetry property \eqref{plane_symm}.}}
\label{plan_wig_plots}
\end{figure}

If the conditions \eqref{regul_cond1}, \eqref{regul_cond2} are fulfilled, then
\begin{equation}\label{I_wigg_pl}
\begin{split}
    I_3=&\sum_{n=1}^\infty\frac{\theta(a_n)\theta(b_n)}{\omega\sqrt{a_nb_n}}\big[e^{-i m\vf^+_n}c_n(\vf^+_n) +e^{-i m\vf^-_n}c_n(\vf^-_n)\big],\\
    I_\pm=& i\frac{s\pm n_3}{n_k}\sum_{n=1}^\infty\frac{\theta(a_n)\theta(b_n)}{\omega\sqrt{a_nb_n}}\Big\{ e^{-i (m\mp1)\vf^+_n} \big[(\pm ia-d)c_n+\frac1{\sqrt{2}}(c_{n-1}+c_{n+1})\big]_{\vf=\vf^+_n}+ \\
    &+e^{-i (m\mp1)\vf^-_n} \big[(\pm ia-d)c_n+\frac1{\sqrt{2}}(c_{n-1}+c_{n+1})\big]_{\vf=\vf^-_n}\Big\}.
\end{split}
\end{equation}
Notice that
\begin{equation}
    \pm ia-d=\pm i\sqrt{a^2+d^2}e^{\mp i\vf_0}.
\end{equation}
To obtain the average number of photons \eqref{probabil}, one ought to add up \eqref{I_wigg_pl} taking into account the factor $1/2$ in \eqref{probabil} and square the absolute value of the result. The explicit expression for $dP$ is rather bulky and we do not present it here but discuss some of its general properties. In the case when the intervals \eqref{energy_ints} do not overlap, the dependence of the average number of twisted photons on $m$ has the form \eqref{m_depend_reg} until the first condition in \eqref{regul_cond1} holds. If
\begin{equation}
    |m|\De\vf\gg1,\qquad |m|\gg\frac{k_0K^2}{2\omega\ga^2}(a-n_k\cos\vf)^2\frac{n_k|\cos\vf|}{\sqrt{2}|\sin\tau_0|},
\end{equation}
then $dP(m)$ tends exponentially to zero. For $a=0$, the average number of photons possesses the symmetry property \eqref{plane_symm} (see Fig. \ref{plan_wig_plots}).

In the weakly degenerate case, $\xi_n\approx\{0,\pi\}$, we assume that the main peak of $\de_N(x)$ is so sharp that it allows one to bring the integrand functions out of the integral and take them at the points $\vf=\vf^\pm_n$. The width of the peak of $\de_N(x)$ is of the order
\begin{equation}\label{peak_width_sing}
    \De\vf\approx\Big[\frac{1+(\sqrt{a^2+d^2}\pm n_k)^2}{\pi nNn_k\sqrt{a^2+d^2}}\Big]^{1/2},
\end{equation}
where the sign $\pm$ corresponds to $\xi_n=\{\pi,0\}$, respectively. The integral over $\vf$ is removed in the above mentioned sense under the assumption that the estimates \eqref{regul_cond1} are satisfied with $\vf=\vf^\pm_n$ and $\De\vf$ taken from \eqref{peak_width_sing}. The rest integral is of the form
\begin{equation}
    \int_{-\pi}^\pi d\vf\de_N[\cdots]\approx\int_{-\pi}^\pi d\vf\frac{\sin\big[\pi nN(\omega_+\omega_-^{-1}-1)(\vf-\vf_0)^2/4\big]}{\omega\pi n(\omega_+\omega_-^{-1}-1)(\vf-\vf_0)^2/4} \approx\omega^{-1}\sqrt{\frac{8N}{n(\omega_+\omega_-^{-1}-1)}},
\end{equation}
for $\xi_n=0$. Analogously, for $\xi_n=\pi$,
\begin{equation}
    \int_{-\pi}^\pi d\vf\de_N[\cdots]\approx \omega^{-1}\sqrt{\frac{8N}{n(1-\omega_-\omega_+^{-1})}}.
\end{equation}
Thus, for $\xi_n\approx0$, i.e., $k_0\rightarrow n\omega_+$,
\begin{equation}
\begin{split}
    I_3&=\omega^{-1}\sqrt{\frac{8N}{n(\omega_+\omega_-^{-1}-1)}}e^{-im\vf_0}c_n(\vf_0),\\
    I_\pm&=\omega^{-1}\sqrt{\frac{8N}{n(\omega_+\omega_-^{-1}-1)}}\frac{s\pm n_3}{n_k}e^{-im\vf_0} \Big[\mp\sqrt{a^2+d^2}c_n(\vf_0)+i\frac{e^{\pm i\vf_0}}{\sqrt{2}}\big(c_{n-1}(\vf_0)+c_{n+1}(\vf_0)\big) \Big].
\end{split}
\end{equation}
The average number of photons \eqref{probabil} becomes
\begin{equation}
    dP=\frac{e^2K}{n\ga}\frac{Nn_k\omega^{-2}}{\omega_+\omega_-^{-1}-1}\Big\{(c_{n-1}+c_{n+1})^2\cos^2\vf_0+\big[\sqrt{2}(n_k-\sqrt{a^2+d^2})c_n-(c_{n-1}+c_{n+1})\sin\vf_0 \big]^2 \Big\} \frac{dk_3dk_\perp}{4\pi^2}.
\end{equation}
For $\xi_n\approx\pi$, i.e., $k_0\rightarrow n\omega_-$, we deduce in the same way
\begin{equation}
    dP=\frac{e^2K}{n\ga}\frac{Nn_k\omega^{-2}}{1-\omega_-\omega_+^{-1}}\Big\{(c_{n-1}+c_{n+1})^2\cos^2\vf_0+\big[\sqrt{2}(n_k-\sqrt{a^2+d^2})c_n +(c_{n-1}+c_{n+1})\sin\vf_0 \big]^2 \Big\} \frac{dk_3dk_\perp}{4\pi^2},
\end{equation}
where one should set $\vf=\vf_0+\pi$ in all the arguments of $c_k(\vf)$. The expressions for the average number of photons obtained above are independent of $s$ and $m$ in the domain \eqref{regul_cond1}.

In the strongly degenerate case,
\begin{equation}\label{degen_cond}
    \frac{2\pi nNn_k\sqrt{a^2+d^2}}{1+(\sqrt{a^2+d^2}-n_k^2)^2}\ll1.
\end{equation}
Then, we can neglect the dependence of $\de_N(x)$ in \eqref{I3_wigg_pl}, \eqref{Ipm_wigg_pl} on $\vf$. We consider only the particular case of \eqref{degen_cond} when
\begin{equation}
    2\pi nN\sqrt{a^2+d^2}\ll1,
\end{equation}
i.e., when one can take $a=d=0$ in the integrals $I_3$, $I_\pm$ (the forward radiation). Let us introduce the notation
\begin{equation}\label{fmn0}
    f_{nm}(x,y):=i^m\int_{-\pi}^\pi d\vf e^{-im\vf}J_n(2x\sin\vf,y).
\end{equation}
Integrating term by term the expansion \cite{Bord.1,Diden79,NikRit64,Dattol90,Dattol91}
\begin{equation}
    J_n(2x\sin\vf,y)=\sum_{k=-\infty}^\infty J_k(y)J_{n-2k}(2x\sin\vf),
\end{equation}
we derive
\begin{equation}\label{fmn}
    f_{nm}(x,y)=\pi (1+(-1)^{n+m})\sum_{k=-\infty}^\infty J_k(y)J_{(n-m)/2-k}(x)J_{(n+m)/2-k}(x).
\end{equation}
The terms of this series tend rapidly to zero when the absolute value of the index of the Bessel function becomes greater than its argument. The integral over $\vf$ is written as
\begin{equation}\label{I3_wigg_pl1}
    I_3=i^{-m}\sum_{n=1}^\infty \de_N\Big[\frac{k_0K^2}{2\ga^2}(K^{-2}+1+n_k^2)-\omega n\Big]f_{nm}\Big(\frac{k_0K^2}{\sqrt{2}\omega\ga^2}n_k,-\frac{k_0K^2}{4\omega\ga^2}\Big).
\end{equation}
This expression is symmetric with respect to $m\rightarrow-m$ and does not vanish only when $m+n$ is an even number. As for the integrals $I_\pm$, let us define
\begin{equation}\label{fmnpm1}
    f^\pm_{nm}:=\frac{f_{n-1,m\mp1} +f_{n+1,m\mp1}}{\sqrt{2}}.
\end{equation}
Then
\begin{equation}\label{fmnpm}
    f^\pm_{nm}(x,y)=\pi (1+(-1)^{n+m})\sum_{k=-\infty}^\infty\frac{J_k(y)+J_{k-1}(y)}{\sqrt{2}}J_{(n\mp m)/2+1-k}(x)J_{(n\pm m)/2-k}(x).
\end{equation}
Using the above notation, we can write
\begin{equation}\label{Ipm_wigg_pl1}
    I_\pm=\mp\frac{s\pm n_3}{n_k} i^{-m} \sum_{n=1}^\infty \de_N\Big[\frac{k_0K^2}{2\ga^2}(K^{-2}+1+n_k^2)-\omega n\Big] f^\pm_{nm}\Big(\frac{k_0K^2}{\sqrt{2}\omega\ga^2}n_k,-\frac{k_0K^2}{4\omega\ga^2}\Big).
\end{equation}
This expression differs from zero only for even $m+n$. Summing \eqref{I3_wigg_pl1}, \eqref{Ipm_wigg_pl1} and neglecting the overlapping of the functions $\de_N(x)$ with different arguments, we obtain the average number of photons
\begin{equation}\label{dP_wigg_pl1}
    dP=e^2\sum_{n=1}^\infty\de^2_N\Big[\frac{k_0K^2}{2\ga^2}(K^{-2}+1+n_k^2)-\omega n\Big]\Big\{f_{nm}-\frac{n_3}{2n_k}(f^+_{nm}+f^-_{nm}) -\frac{s}{2n_k}(f^+_{nm}-f^-_{nm}) \Big\}^2n_\perp^3\frac{dk_3dk_\perp}{16\pi^2}.
\end{equation}
The expression for the average number of photons complies with the symmetry property \eqref{plane_symm} and is different from zero only for even $m+n$ (see Fig. \ref{forw_dip_plots}). The energy of photons $k_0$ in the arguments of $f_{nm}$ and $f^\pm_{nm}$ can be replaced by
\begin{equation}
    k_0=\frac{2n\omega\ga^2}{1+K^2+K^2n_k^2}.
\end{equation}
In this case, the dependence of $dP$ on the photon energy is determined only by the factor $\de^2_N(x)$.

The expression \eqref{dP_wigg_pl1} is not always useful for calculations for sufficiently large $m$ and $n$ as, in this case, one needs to take a large number of terms in the series \eqref{fmn}, \eqref{fmnpm}. In order to obtain the approximate expression for the average number of photons for large $m$ and $n$, one can make use of the approximate expression \eqref{cnphi} and find the integrals $I_3$, $I_\pm$ by the steepest descent method. Let,
\begin{equation}\label{S12}
\begin{split}
    S_{1nm}(\vf):=-i\big[m\vf -n(\tau_0-\frac{\pi}{2}) +\frac{3k_0K^2}{4\omega\ga^2}\sin2\tau_0\big]+\ln\Ai(B_n(\vf)),\\
    S_{2nm}(\vf):=-i\big[m\vf +n(\tau_0-\frac{\pi}{2}) -\frac{3k_0K^2}{4\omega\ga^2}\sin2\tau_0\big]+\ln\Ai(B_n(\vf)),
\end{split}
\end{equation}
be the rapidly varying expressions in the exponents in \eqref{cnphi}. The stationary points of these expressions are invariant under the reflection in the imaginary axis
\begin{equation}
    \vf\rightarrow-\vf^*.
\end{equation}
Moreover, the stationary points of $S_2$ are obtained from the stationary points of $S_1$ by the replacement
\begin{equation}
    \vf\rightarrow\vf+\pi.
\end{equation}
In evaluating the integral over $\vf$, we are interested in the stationary points $\vf^{nm}_{ext}$ nearest to the real axis and such that
\begin{equation}
    \im\vf^{nm}_{ext}\sgn m\leq0.
\end{equation}
When $|m|\ll n$, the expressions \eqref{S12} possess the extremum points at
\begin{equation}
    \dot{\tau}_0=-\frac{n_k\cos\vf}{\sqrt{2}\sin\tau_0}=0,
\end{equation}
i.e., at $\vf^{nm}_{ext}=\pm\pi/2$ in the strip $\re\vf\in[-\pi,\pi]$. In increasing $|m|$, these extremum points shift and move away from the real axis. It is these stationary points that give the leading contribution to the integral.

Let, for definiteness, $\vf^{nm}_{ext}$ be the stationary point of $S_{1nm}$ obtained by shifting from the stationary point $\pi/2$. Introduce the notation
\begin{equation}
    h_{nm}:=\Big(\frac{k_0K^2}{\omega\ga^2}\sin^2\tau_0\Big)^{-1/3} e^{-im\vf +in\tau_0-i\frac{3k_0K^2}{4\omega\ga^2}\sin2\tau_0} \Ai(B_n(\vf))\sqrt{\frac{2\pi}{-\ddot{S}_{1nm}}},
\end{equation}
where the principal branch of the square root is taken, the dot denotes the derivative with respect to $\vf$, and one should set $\vf=\vf^{nm}_{ext}$. Then the contribution coming from the two stationary points and the two exponents with the powers $S_{1,2}$ reads
\begin{equation}\label{fnmap}
    f_{nm}\approx i^m(1+(-1)^{m+n})(h_{nm}+(-1)^mh^*_{nm}).
\end{equation}
In virtue of the relation \eqref{fmnpm1}, we obtain
\begin{equation}\label{fnmap1}
    f^\pm_{nm}\approx i^{m\mp1}\frac{1+(-1)^{m+n}}{\sqrt{2}}\big[h_{n-1,m\mp1}+h_{n+1,m\mp1} +(-1)^{m+1}(h^*_{n-1,m\mp1}+h^*_{n+1,m\mp1})\big].
\end{equation}
As expected, the expressions \eqref{fnmap}, \eqref{fnmap1} are real-valued and different from zero only for even $m+n$. The approximate expression for $f_{nm}(x,y)$ can also be derived by applying the WKB method immediately to the double integral \eqref{cn}, \eqref{fmn0}. The evaluation of the stationary points reduces in this case to the solution of a cubic equation. When the variables $x$, $y$, $m$, and $n$ are of the same order, the general solution of this equation is rather awkward and leads to a huge expression for $f_{nm}$. Therefore, we do not present it here.

\section{Conclusion}

Let us sum up the results obtained in this paper.

First, we have derived the general formula for the average number of twisted photons produced by a classical source and establish some of its general properties. In particular, we have proved that the average number of twisted photons recorder by the detector obeys the symmetry property \eqref{plane_symm} when a charged particle moves along a planar trajectory, while the detector of twisted photons is placed in the orbit plane and projects the angular momentum onto the axis lying in this plane. We also have provided the pictorial representation of the general formula for the average number of twisted photons in terms of the radiation amplitude of plane-wave photons (see Figs. \ref{plane_symm_plots}, \ref{cylind_plots}). We have obtained the integral representations for the projection of the total angular momentum of twisted photons with given the energy, the longitudinal projection of momentum, and the helicity in terms of the trajectory of a charged particle.

Second, we have developed the general theory of radiation of twisted photons by undulators. We have derived the explicit formulas for the average number of twisted photons produced by the undulator and recorded by the detector located, in general, off the undulator axis and projecting the angular momentum onto the axis directed from the radiation point to the detector. These formulas are obtained for both the dipole and wiggler regimes of the undulator. We have established some general properties of the undulator radiation of twisted photons. It turns out that the forward radiation of an ideal right-handed helical undulator consists of the twisted photons with the projection of the total angular momentum $m=n$, where $n$ is the harmonic number. The radiation of an ideal left-handed helical undulator consists of the twisted photons with $m=-n$. We have shown that the forward radiation of the planar undulator obeys the selection rule that $n+m$ is in even number. As for the undulator radiation at an angle, we have found that, in particular, the average number of twisted photons is a periodic function of $m$ in a certain range of the quantum numbers $m$. We have checked the obtained analytical results by the numerical simulations and ascertained that they match each other. The accuracy of the analytical formulas is increased with increasing the number of the undulator sections $N$.

Thus, we may conclude that the general formula we have derived provides a reliable and effective tool for further studies of generation of the twisted photons by classical currents. As regards the possible generalizations not mentioned in the main text, it would be interesting to investigate the production of twisted gravitons by binary systems along the same lines.

\paragraph{Acknowledgments.}

We are thankful to V.G. Bagrov and D.V. Karlovets for fruitful conversations. This work is supported by the Russian Science Foundation (project No. 17-72-20013).

\appendix
\section{Some special functions}\label{Bessel_Prop}

It is useful to express the mode functions of the electromagnetic field in terms of the functions
\begin{equation}
    j_\nu(p,q):=\frac{p^{\nu/2}}{q^{\nu/2}}J_\nu(p^{1/2}q^{1/2}),
\end{equation}
where those branches of the multi-valued functions are taken that are real-valued and analytic for positive $p$ and $q$ (some properties of these functions can also be found in \cite{KazShip}). Then, for real $\nu$,
\begin{equation}
    j^*_\nu(p,q):=j_\nu(p^*,q^*).
\end{equation}
These functions possess the following properties (see, e.g., \cite{Wats.6})
\begin{equation}\label{recurr_rels}
\begin{gathered}
    2\frac{\partial}{\partial p}j_\nu(p,q)= j_{\nu-1}(p,q),\qquad2\frac{\partial}{\partial q}j_\nu(p,q)=-j_{\nu+1}(p,q),\\
    2\nu j_\nu(p,q)=pj_{\nu-1}(p,q)+qj_{\nu+1}(p,q),\\
    j_\nu(e^{i\pi}p,e^{i\pi}q)=e^{i\pi\nu}j_\nu(p,q),\\
    j_\nu(p,q)=\int_H\frac{dt}{2\pi i}t^{-\nu-1}e^{\frac12(pt-\frac{q}{t})},\quad\re p>0,
\end{gathered}
\end{equation}
where $H$ is the Hankel contour running from $-\infty$ a little bit lower than the real axis, encircling the origin, and then going to $-\infty$ a little bit higher than the real axis.

If $\nu=m\in \mathbb{Z}$, then $j_m(p,q)$ is an entire analytic function of complex variables $p$ and $q$, and the additional relations hold
\begin{equation}
    j_m(p,q)=(-1)^mj_{-m}(q,p),\qquad j_m(0,0)=\de_{m0}.
\end{equation}
In particular,
\begin{equation}
    j_m(x_+,x_-)=(-1)^mj_{-m}(x_-,x_+)=(-1)^mj^*_{-m}(x_+,x_-),
\end{equation}
where it is supposed in the last equality that $x_{1,2}\in\R$. Let $\De_\pm:=x_\pm-y_\pm$. Then the addition theorem takes place \cite{Wats.6}
\begin{equation}\label{add_thm}
    \sum_{m=-\infty}^\infty j_{\nu+m}(x_+,x_-)j_{m}(y_-,y_+)=\sum_{m=-\infty}^\infty j_{\nu+m}(x_+,x_-)j^*_{m}(y_+,y_-)=j_\nu(\De_+,\De_-).
\end{equation}
The integral representation \eqref{recurr_rels} is written as
\begin{equation}
    j_m(p,q)=\int_{|t|=1}\frac{dt}{2\pi i}t^{-m-1}e^{\frac12(pt-\frac{q}{t})},
\end{equation}
for any complex $p$ and $q$. In particular,
\begin{equation}\label{Bessel_int1}
    j_m(k_\perp x_+,k_\perp x_-)=\int_{-\pi}^\pi\frac{d\vf}{2\pi}e^{-im\vf+ik_\perp(x_1\sin\vf+x_2\cos\vf)} =i^m\int_{-\pi}^\pi\frac{d\vf}{2\pi}e^{-im\vf+ik_\perp(x_2\sin\vf-x_1\cos\vf)}.
\end{equation}

In describing the radiation of twisted photons by undulators in the dipole approximation, the function $G^m_N(a,b)$ arises (see \eqref{GmN}). We present some of its properties here. The function $G^m_N(a,b)$ is an entire analytic function of $a$ and $b$. It is obvious from \eqref{GmN} that
\begin{equation}
    G_N^m(a,b)=(-1)^mG_N^m(b,a)=(-1)^mG^{-m}_N(a,b).
\end{equation}
Furthermore, the following recurrence relation holds
\begin{equation}
    G_N^{m-1}(a,b)-G_N^{m+1}(a,b)=2i^{-m-1}\sin\Big(\frac{\pi N}{2}(b-a)-\frac{\pi m}{2}\Big)\frac{J_m\big(\frac{\pi N}{2}(b+a)\big)}{\pi (b+a)}+2i\frac{b-a}{b+a}G_N^{m}(a,b).
\end{equation}
For $N\rightarrow\infty$ and real $a$ and $b$ \cite{GrRy},
\begin{equation}\label{GmN_approx}
    G_N^{m}(a,b)\rightarrow\frac{i^{-m}\theta(a)\theta(b)}{2\pi\sqrt{ab}}\cos\Big(m\arccos\frac{b-a}{b+a}\Big)=\frac{i^{-m}\theta(a)\theta(b)}{2\pi\sqrt{ab}}T_m\Big(\frac{b-a}{b+a}\Big),
\end{equation}
where $T_m(x)$ are the Chebyshev polynomials of the first kind. The function $G_N^{m}(a,b)$ is well approximated by the expression on the right-hand side, with a relative error not exceeding $0.1$, when
\begin{equation}\label{a_b_m_ests}
    a\gtrsim\frac{5}{N},\qquad b\gtrsim\frac{5}{N},\qquad |m|\lesssim\frac{10}{7}N(a+b).
\end{equation}
For $|m|$ beyond this bound, $G_n^m(a,b)$ tends exponentially to zero. If $a\ll5/N$ and $b$ satisfies the estimate \eqref{a_b_m_ests}, then
\begin{equation}\label{GmN_bound}
    G_N^m(a,b)\approx G_N^m(0,b)\approx\frac{i^{-m}}{\pi}\sqrt{\frac{N}{2b}},
\end{equation}
for
\begin{equation}\label{m_est_a0}
    |m|\lesssim(10\pi bN)^{1/3}.
\end{equation}
If $|m|$ is greater than the above estimate, then $G_N^m(0,b)$ tends exponentially to zero. For $a\ll5/N$ and $b\ll5/N$, we have
\begin{equation}\label{GmN00}
    G_N^m(a,b)\approx G_N^m(0,0)=\frac{N}{2}\de_{m0}.
\end{equation}

\end{document}